\documentstyle[prb,aps,multicol,eqsecnum]{revtex}

\begin{document}
\draft

\title{ Quantum melting of magnetic long-range order near orbital
        degeneracy.\\
        Classical phases and Gaussian fluctuations }

\author { Andrzej M. Ole\'{s}\cite{AMO} }
\address{ Institute of Physics, Jagellonian University,
          Reymonta 4, PL-30059 Krak\'ow, Poland \\
          and Max-Planck-Institut f\"ur Festk\"orperforschung,
          Heisenbergstrasse 1, D-70569 Stuttgart, 
          Federal Republic of Germany }
\author { Louis Felix Feiner }
\address{ Institute for Theoretical Physics, Utrecht University,
          Princetonplein 5, NL-3584 CC Utrecht, The Netherlands \\
          and Philips Research Laboratories, Prof. Holstlaan 4,
          NL-5656 AA Eindhoven, The Netherlands }
\author { Jan Zaanen }
\address{ Lorentz Institute for Theoretical Physics, Leiden University,
          P.O.B. 9506, NL-2300 RA Leiden, The Netherlands }

\date{25 March 1999}
\maketitle

\begin{abstract}
We address the role played by orbital degeneracy in strongly correlated
transition metal compounds. Specifically, we study the effective spin-orbital
model derived for the $d^9$ ions in a three-dimensional perovskite lattice,
as in KCuF$_3$, where at each site the doubly degenerate $e_g$ orbitals
contain a single hole. The model describes the superexchange interactions
that depend on the pattern of orbitals occupied and shows a nontrivial
coupling between spin and orbital variables at nearest neighbor sites.
We present the ground state properties of this model, depending on the 
splitting between the $e_g$ orbitals $E_z$, and the Hund's rule coupling in 
the excited $d^8$ states, $J_H$. The classical phase diagram consists of six 
magnetic phases which all have different orbital ordering: two 
antiferromagnetic (AF) phases with G-AF order and either $x^2-y^2$ or 
$3z^2-r^2$ orbitals occupied, two phases with mixed orbital (MO) patterns 
and A-AF order, and two other MO phases with either C-AF or G-AF order. 
All of them become degenerate at the multicritical point 
$M\equiv (E_z,J_H)=(0,0)$. Using a generalization of linear spin-wave theory 
we study both the transverse excitations which are spin-waves and 
spin-and-orbital-waves, as well as the longitudinal (orbital) excitations. 
The transverse modes couple to each other, and the spin-and-orbital-wave 
turns into a soft mode near the $M$ point. Therefore, quantum corrections to 
the long-range-order parameter are drastically increased near the orbital 
degeneracy, and classical order is suppressed in a crossover regime between 
the G-AF and A-AF phases in the $(E_z,J_H)$ plane. This behavior is 
reminiscent of that found in frustrated spin models, and we conclude that 
orbital degeneracy provides a new and physically realizable mechanism which 
stabilizes a spin liquid ground state due to inherent frustration of 
magnetic interactions. We also point out that such a disordered magnetic 
phase is likely to be realized at low $J_H$ and low electron-phonon coupling, 
as in LiNiO$_2$.
\end{abstract}

\begin{multicols}{2} 

\section{ New mechanism of frustration near orbital degeneracy }
\label{sec:orbitals}

Quite generally, strongly correlated electron systems involve orbitally
degenerate states,\cite{Ima98} such as $3d$ ($4d$) states in transition metal
compounds, and $4f$ ($5f$) states in rare-earth compounds. Yet, the orbital
degrees of freedom are ignored in most situations and the common approach
is to consider a single correlated orbital per atom
which leads to spin degeneracy alone. Indeed, most of the current studies of
strongly correlated electrons deal with models of nondegenerate orbitals. The
problems discussed recently include mechanisms of ferromagnetism in the
Hubbard model,\cite{ferro} hole propagation and quasiparticles in the $t-J$
model,\cite{Dag94} and magnetic states of the Kondo lattice.\cite{Man97}
Of course, in many actually existing compounds the orbital degeneracy is
removed by the crystal field, and a single-orbital approach is valid {\it per
se\/}. Also, from a fundamental point of view it is often possible to argue
that orbital degeneracy is qualitatively irrelevant, and that a single-orbital
approach can capture the generic mechanisms operative in the presence of
strong correlations.

However, neither of these arguments applies for a class of insulating
strongly correlated transition metal compounds, where the crystal field
leaves the $3d$ orbitals explicitly degenerate and thus the type of occupied
orbitals is not known {\it a priori\/}, while the magnetic interaction
between the spins of neighboring transition metal ions depends on which
orbitals are occupied. In this particular class of Mott-Hubbard insulators
(MHI) the orbital degrees of
freedom acquire a separate existence in much the same way as the spins do.
Thereby, the degeneracy of $t_{2g}$ orbitals is of less importance, as the
magnetic superexchange and the coupling to the lattice are rather weak.
A more interesting situation occurs when $e_g$ orbitals are partly occupied,
which results in stronger magnetic interactions, and strong
Jahn-Teller (JT) effect. Typical examples of such ions are:
Cu$^{2+}$ ($d^9$ configuration, one hole in $e_g$-orbitals), low-spin
Ni$^{3+}$ ($d^7$ configuration, one electron in $e_g$-orbitals), as well as
Mn$^{3+}$ and Cr$^{2+}$ ions (high-spin $d^4$ configuration, one $e_g$
electron). The simplest model, relevant for $d^9$ transition metal ions,
which is also the subject of the present paper, was introduced
by Kugel and Khomskii more than two decades ago,\cite{Kug73} but its
mean-field (MF) phase diagram was analyzed only recently.\cite{Fei97}
It describes magnetic superexchange interactions between spins $S=1/2$,
and the accompanying orbital superexchange interactions.

One might argue that the (classical) orbital degeneracy is not easy to
realize in such systems, as the electron-phonon coupling will lead to the
conventional collective JT instability. In fact, it can be shown that the JT
instability is enhanced by the orbital pattern once this has been established
as the result of effective interactions:\cite{Kug73,Kho97,crete} the lattice
has to react to the symmetry lowering in the orbital sector, which can only
increase the stability of a given magnetic state. So the lattice follows
rather than induces the orbital order, and therefore, as was pointed out in
the early work by Kugel and Khomskii,\cite{Kug73,Kug82} in the orbitally
degenerate MHI one has to consider in first instance the purely electronic
problem. This is supported by the results of recent band structure
calculations using the local density approximation (LDA) with the electron
interactions treated in Hartree-Fock approximation, the so-called LDA+U
method, which permits both orbitals and spins to polarize while keeping the
accurate treatment of the electron-lattice coupling of LDA intact.
These calculations reproduce the observed orbital ordering in
KCuF$_3$ (Ref. \onlinecite{Lie95}) and in LaMnO$_3$ (Ref. \onlinecite{Ani97}),
even when the lattice distortions are suppressed, while allowing the lattice
to relax only yields an energy gain which is minute in comparison with the
energies involved in the orbital ordering.

Effects of orbital degeneracy are expected as soon as crystal-field
splittings become small. Such situations are frequently encountered in
rare-earth systems, where they lead to the so-called singlet-triplet models
discussed in the seventies,\cite{Hsi72} while in the $3d$ oxides only a small
number of so-called Kugel-Khomskii (KK) systems \cite{Kug82} have been
recognized that actually exhibit orbital effects.\cite{Kho97}
As pointed out by Kugel and Khomskii,\cite{Kug73} in such situations the
superexchange interactions have a more complex form than in spin-only models
and one expects that also in some other Mott-Hubbard (or charge-transfer)
insulators new magnetic phases might arise due to the competition of various
magnetic and orbital interactions. Some examples of such a competition of
magnetic interactions are encountered in the heavy fermion
systems,\cite{Man97,Cox87} and in the manganites where the phase diagrams
show a particular frustration of magnetic
interactions.\cite{Miz95,Ish96,Shi97,Fei99}

Even more interesting behavior is expected for the doped systems, as the
competition between the magnetic, orbital, and kinetic energy is then
described by $t-J$ Hamiltonians of a novel type, which exhibit qualitatively
different excitation spectra due to the underlying orbital
degeneracy.\cite{Zaa93}
A few examples of such models have already been discussed in the literature,
such as the triplet $t-J$ model,\cite{Zaa92} the low-spin defects in a $S=1$
background,\cite{Dag96} or a new $t-J$ model for the manganites.\cite{Mul96}
Whether such models are realistic enough is not yet clear, as for example in
the manganites there are experimental\cite{Oki97} and theoretical \cite{Mil95}
indications that the double-exchange model which includes only the spin
degrees of freedom is insufficient to understand the transport properties
under doping. Recent work \cite{Shi97,Fei99,Tak98,Bri99} strongly suggests
that an extension of the $t-J$ and double-exchange models which include fully
the orbital physics should be studied instead.

In this paper we shall consider only the insulating situation, where one can
integrate out the $d-d$ excitations and derive an effective low-energy
Hamiltonian. This approach is justified by the large on-site Coulomb
interaction $U$, being the largest energy scale in MHI.
A low-energy Hilbert space splits off, spanned by {\it spin and orbital\/}
configuration space, with superexchangelike couplings between both spin
and orbital local degrees of freedom. The orbital sector carries a discrete
symmetry and the net outcome is that the clock-like orbital degrees of
freedom get coupled into the $SU(2)$ spin problem. The resulting low-energy
Hamiltonian is called a {\it spin-orbital model\/}. Here we focus
on the simplest situation with two nearly degenerate partially filled $e_g$
orbitals, and completely filled $t_{2g}$ orbitals, as encountered in KCuF$_3$
and related systems.\cite{Kug82} These are JT-distorted cubic crystals,
three-dimensional (3D) analogues of the cuprate superconductors.\cite{Web88}
In the high-$T_c$ cuprates, orbital degeneracy would occur if the Cu-O bonds
which involve apical oxygens were squeezed such
as to recover the cubic symmetry of the perovskite lattice. Of course, such
a degeneracy of $e_g$ orbitals is far from being realized in the
actual high-$T_c$ materials, and in their parent compounds.\cite{Kho91,Gra92}

If only one correlated orbital is present, the system
may be described by the effective single-band Hubbard model (typically with
more extended hopping), as in the cuprate superconductors.\cite{Fei96}
In this simplest case the effective model at half-filling is the Heisenberg
model with antiferromagnetic (AF) superexchange.\cite{And59} This changes
when more than one $3d$ orbital is partly occupied. For example, we show in
Sec. \ref{sec:model} that virtual excitations involving $d^8$ local triplet
states become possible in the case of degenerate $e_g$ orbitals, and this
leads to additional ferromagnetic (FM) interactions. The origin of these new
interactions was first discussed by Kugel and Khomskii\cite{Kug73} and by
Cyrot and Lyon-Caen\cite{Cyr75} who pointed out that the strongest
superexchange constant results from the excitation to the lowest energy
triplet state in the degenerate Hubbard model.

The model proposed by Kugel and Khomskii explains qualitatively the observed
magnetic ordering in KCuF$_3$ as being due to an orbital ordering which gives
planes of perpendicularly oriented orbitals, and the magnetic coupling becomes
then FM according to the Goodenough-Kanamori rules.\cite{Goo63} As mentioned
above, such a state was indeed found in the band structure calculations of
Liechtenstein {\it et al.}\cite{Lie95} using the LDA+U method. An analogous
orbital order is responsible for ferromagnetism in the planar FM insulator
K$_2$CuF$_4$.\cite{Kug82} In the colossal magnetoresistance parent compound
LaMnO$_3$, where the $e_g$ orbitals contain one electron instead of one hole,
a similar orbital ordering occurs,\cite{Kho97,Ish96} although the situation
there is more complex due to the presence of $t_{2g}$ spins, so that the
resulting superexchange is not between spins $S=1/2$ but between total spins
$S=2$.\cite{Fei99} Another example of degenerate orbitals is found in
V$_2$O$_3$, with the orbital ordering studied by Castellani, Natoli and
Ranninger in a series of papers.\cite{Cas78} In fact, their prediction that
the transition into the AF insulator is accompanied by the onset of orbital
ordering was experimentally verified only recently.\cite{Bao97} However,
this case is still open, as recent electronic structure calculations suggest
that doubly degenerate orbitals are occupied by two electrons in the
high-spin state and the orbital degree of freedom plays no role.\cite{Ezh99}

In any of the above situations the orbital ordering breaks the translational
symmetry and represents an analogon of spin antiferromagnetism in orbital
space. So, {\em classically\/} orbital ordering is expected to occur quite
generally whenever one encounters $e_g$ orbitals containing either one hole
or one electron, with important consequences for the magnetism. This
immediately raises a number of questions about what happens in the
{\em quantum regime\/}. Will orbital long-range order (LRO) be robust or
will it give way to an {\em orbital liquid\/}, as proposed by Ishihara,
Yamanaka and Nagaosa (Ref. \onlinecite{Nag97})? In either case, what are the
consequences of the enlarged phase space and the associated additional
channels for quantum fluctuations for the magnetism: can magnetic LRO
survive or will it be replaced by a {\em spin liquid\/}?

Quantum disordered phases are of great current interest. Spin disorder is
well known to occur in one-dimensional (1D) and quasi 1D quantum spin
systems, and the best example is the 1D Heisenberg model, where the famous
exact solution found by Bethe many years ago \cite{Bet31} showed that the
quantum fluctuations prevent true AF LRO, giving instead a slow decay of
spin correlations. A similar situation is encountered in spin ladders with
an even number of legs, which have a spin gap and purely short-range
magnetic order.\cite{Whi94,Ric96} This is one of the realizations of a
spin-liquid ground state due to purely short-range spin correlations. In
the limit of a two-dimensional (2D) Heisenberg model the spin disorder is
replaced by a ground state with AF LRO.

It is well known that frustrated magnetic interactions may lead to spin
disordered states in two dimensions. However, in order to achieve this, i.e.,
to prevent 2D macroscopic spin systems from behaving classically and to make
quantum mechanics take over instead, the frustration of the interactions
must be sufficiently severe. This shows that global SU(2) by itself is not
symmetric enough to defeat classical order in $D>1$ and one has to change
the magnetic interactions in such a way that they lead to sufficiently
strong quantum fluctuations.
So far, this strategy has been shown to lead to spin disorder in (quasi) 2D
systems in three different situations:
  (i) Frustrating a 2D square lattice by adding longer-range AF interactions,
      as in $J_1-J_2$ and $J_1-J_2-J_3$ models, gives a high degeneracy of
      the classical sector, and a disordered state is found for particular
      values of the magnetic interactions.\cite{Cha88,Chu91} This mechanism
      involves fine-tuning of parameters and therefore such systems are
      hard to realize in nature.
 (ii) In the bilayer Heisenberg model two planes are coupled by interlayer AF
      superexchange $J_{\perp}$ which generates zero-dimensional fluctuations.
      This leads to a crossover to the disordered ground state of an
      incompressible spin liquid above a certain critical value of
      $J_{\perp}$.\cite{Mil94,Chu95} Also this mechanism is hard to
      realize experimentally.
(iii) In contrast, a spin disordered state can be obtained in nature by
      reducing the number of magnetic bonds in a 2D square lattice. The
      model of CaV$_4$O$_9$ studied by Taniguchi {\it et al.}\cite{Tan95} is
      a 1/5 depleted square lattice, which gives a plaquette resonating
      valence bond (PRVB) ground state for realistic interactions, and
      a spin gap which agrees with experimental observations.\cite{Ued96}
A common feature of these systems is a crossover between different magnetic
ground states, either between two different patterns of LRO, as in case (i),
or simply between the ordered and disordered states, which results in all
three situations in a tendency towards the formation of spin singlets on
the bonds with the strongest AF superexchange. One may further note that in
these spin-only models very specific patterns of magnetic interactions are
required already in two dimensions to prevent the system to order classically,
while up to now it has proven impossible to realize a spin liquid in three
dimensions.

In the present paper we address two fundamental questions for the Heisenberg
antiferromagnet (HAF) extended to include the orbital degrees of freedom in
orbitally degenerate MHI:
 (i) Which {\em classical\/} states with magnetic LRO do exist in the
       neighborhood of orbital degeneracy?
(ii) Are those forms of classical order
       always stable against {\em quantum\/} fluctuations?
We will show that the orbitally degenerate MHI represent a class of
systems in which spin disorder occurs due to frustration of {\em spin and
orbital\/} superexchange couplings.
This frustration mechanism is different from that operative in pure spin
systems, and suppresses the magnetic LRO in the ground state {\em even in
three dimensions\/}.

As explained above, the low-energy behavior of such systems is described by
a spin-orbital model. We will show that within the framework of such a
spin-orbital model the occurrence of spin disorder may be regarded
as resulting from a competition between various classical ordered phases,
each one with a simultaneous symmetry breaking
in spin and orbital space. The qualitatively new aspect is that the magnetic
interactions follow the orbital pattern, and thus these systems tend to
"self-tune" to (critical) points of high classical degeneneracy.
We show explicitly that in the vicinity of such a multicritical point
classical order is highly unstable with respect to quantum fluctuations.
As a result, a qualitatively new quantum spin liquid with strong orbital
correlations is expected. We believe that a 3D state of this type is realized
in LiNiO$_2$.

The paper is organized as follows. The spin-orbital model for $d^9$
transition metal ions, such as Cu$^{2+}$ in KCuF$_3$, is derived in Sec.
\ref{sec:model} using the correct multiplet structure of Cu$^{3+}$ excited
configurations. We solve this model first in the MF approximation and 
present the resulting classical phases and the accompanying orbital 
orderings in Sec. \ref{sec:mfa}. The elementary excitations obtained within 
an extension of the linear spin-wave theory (LSW) are presented in Sec. 
\ref{sec:magnons}, where we demonstrate that two transverse modes are 
strongly coupled to each other. This leads to soft modes next to the 
classical transition lines, and to the collapse of LRO due to diverging 
quantum corrections, as shown in Sec. \ref{sec:rpa}. We summarize the 
results and present our conclusions in Sec. \ref{sec:spinliquid}.

\section{ The spin-orbital model }
\label{sec:model}

Our aim is to construct the effective low-energy Hamiltonian for a 3D
perovskite-like lattice. The original charge-transfer multiband model, as
considered for instance for the cuprates, includes the hybridization elements
between the $3d$ orbitals of transition metal ions and the $2p$ orbitals
of oxygen ions.\cite{Fei96}
If the Coulomb elements at the $3d$ orbitals and the charge-transfer energy
between the $3d$ and $2p$ orbitals are large, this model can be transformed
into an effective spin-fermion model. For example, this transformation
performed for the three-band model gives an effective Hamiltonian with
localized spins at the Cu sites which interact by superexchange
interactions, while the doped carriers interact with them by a Kondo-like
coupling.\cite{Zaa88} In the limit of undoped compounds, one is thus left
with a model which describes interacting transition metal ions.

The simplest form of (superexchange) interaction, namely a spin model, is
obtained for the case of nondegenerate $d$ orbitals, whereas orbital
degeneracy gives a spin-orbital model acting in a larger Hilbert space
defined by both spin and orbital degrees of freedom at each transition
metal site. Having in mind the strongly correlated late transition
metal oxides, we consider specifically the case of one hole per unit cell in
the $3d^9$ configuration, characterized in the absence of JT-distortion by
two degenerate $e_g$ orbitals: $x^2-y^2\sim |x\rangle$ and
$(3z^2-r^2)/\sqrt{3}\sim |z\rangle$. The derivation is, however, more
general and applies as well to the low-spin $d^7$ configuration; in the case
of the early transition metal oxides the $d^1$ case would involve the
$t_{2g}$ orbitals instead.

The holes in the undoped compound which corresponds to the $d^9$
configuration of transition metal ions, as in La$_2$CuO$_4$ or KCuF$_3$, are
fairly localized.\cite{notecov} Hence, we take as a starting point the
following Hamiltonian which describes $d$-holes on transition metal ions,
\begin{equation}
\label{hband}
H_{e_g} = H_{kin} + H_{int} + H_z,
\end{equation}
and consider the kinetic energy, $H_{kin}$, and the electron-electron
interactions, $H_{int}$, within the subspace of the $e_g$ orbitals (the 
$t_{2g}$ orbitals are filled by electrons, do not couple to $e_g$ orbitals 
due to the hoppings via oxygens, and hence can be neglected). The last term 
$H_z$ describes the crystal-field splitting of the $e_g$ orbitals.

Due to the shape of the two $e_g$ orbitals $|x\rangle$ and $|z\rangle$,
their $d-p$ hybridization in the three cubic directions is unequal, and
is different between them, so that the effective hopping elements are
direction dependent and different for $|x\rangle$ and $|z\rangle$.
The only nonvanishing hopping in the $c$-direction connects two $|z\rangle$
orbitals, while the elements in the $(a,b)$ planes fulfill the Slater-Koster
relations,\cite{Sla54} as presented before by two of us.\cite{Zaa93} Taking
the hopping $t$ along the $c$-axis as a unit, the kinetic energy is given by,
\begin{eqnarray}
\label{hkin}
H_{kin}&=&{t\over 4}\sum_{\langle ij\rangle{\parallel}}
\left[ 3 d^{\dagger}_{ix\sigma}d^{}_{ix\sigma}
 + (-1)^{{\vec \delta}\cdot {\vec y}}\sqrt{3}
         (d^{\dagger}_{iz\sigma}d^{}_{ix\sigma}\right.  \nonumber \\
&+&\left. H.c.)+d^{\dagger}_{iz\sigma}d^{}_{iz\sigma}\right]
+t\sqrt{\beta} \sum_{\langle ij\rangle{\perp}}
          d^{\dagger}_{iz\sigma}d^{}_{iz\sigma},
\end{eqnarray}
where $\langle ij\rangle{\parallel}$ and $\langle ij\rangle{\perp}$ stand
for the bonds between nearest neighbors within the $(a,b)$-planes, and
along the
$c$-axis, respectively, and $\beta=1$ in a cubic system. The $x-z$ hopping 
in the $(a,b)$ planes depends on the phases of the $x^2-y^2$ orbitals along 
$a$- and $b$-axis, respectively, included in the factors
$(-1)^{{\vec \delta}\cdot {\vec y}}$ in Eq. (\ref{hkin}).

The electron-electron interactions are described by the on-site terms,
\begin{eqnarray}
\label{hint}
&H_{int}&=(U+\case{1}{2}J_H)
          \sum_{i\alpha}n_{i\alpha  \uparrow}n_{i\alpha\downarrow}
 + (U-J_H)\sum_{i\sigma}n_{ix\sigma}n_{iz\sigma}
                                                              \nonumber \\
&+&(U-\case{1}{2}J_H)\sum_{i\sigma}n_{ix\sigma}n_{iz\bar{\sigma}}
 -\case{1}{2}J_H \sum_{i\sigma}d^{\dagger}_{ix \sigma}d^{}_{ix\bar{\sigma}}
                               d^{\dagger}_{iz\bar{\sigma}}d^{}_{iz \sigma}
                                                              \nonumber \\
&+&\case{1}{2}J_H\sum_{i}
            ( d^{\dagger}_{ix  \uparrow}d^{\dagger}_{ix\downarrow}
              d^{       }_{iz\downarrow}d^{       }_{iz  \uparrow}
            + d^{\dagger}_{iz  \uparrow}d^{\dagger}_{iz\downarrow}
              d^{       }_{ix\downarrow}d^{       }_{ix  \uparrow} ),
\end{eqnarray}
with $U$ and $J_H$ standing for the Coulomb and Hund's rule exchange
interaction,\cite{noteuj} respectively, and $\alpha=x,z$. For convenience,
we used the simplified notation $\bar{\sigma}=-\sigma$. This Hamiltonian
describes correctly the multiplet structure of $d^8$ (and $d^2$)
ions,\cite{Gri71} and is rotationally invariant in the orbital
space.\cite{Ole83} The wave functions have been assumed to be real which
gives the same element $J_H/2$ for the exchange interaction and for the
pair hopping term between the $e_g$ orbitals, $|x\rangle$ and $|z\rangle$.

In fact, we adopted here the most natural units for the elements of the
Coulomb interaction, with the energy of the central $|^1E\rangle$ doublet
being equal to $U$. By definition this energy does not depend on the Hund's
exchange element $J_H$, as we show below, and is thus the measure of the
average excitation energy in the
$d^9_id^9_j\rightarrow d^{10}_id^8_j$ transition.
The interaction element $J_H$ stands for
the singlet-triplet splitting in the $d^8$ spectrum (Fig. \ref{virtual}) and
is just twice as big as the exchange element $K_{xz}$ used usually in quantum
chemistry.\cite{Gra92} The typical energies for the Coulomb and exchange
elements can be found using constrained-occupation local-density functional
theory.\cite{Hyb89} Unfortunately, such calculations have been performed
only for a few compounds so far. For La$_2$CuO$_4$, a parent compound of
superconducting cuprates, one finds $U=7.77$ eV and $J_H=2.38$ eV;\cite{Gra92}
other estimations of $U$ based on the experimental data report values $6<U<8$
eV for cuprates and nickelates.\cite{Zaa90} This results in the ratio
$J_H/U\simeq 0.3$ which we take as a representative value for the strongly
correlated late transition metal oxides. The values of intersite hopping $t$,
being an effective parameter, are more difficult to estimate. As a
representative value for La$_2$CuO$_4$ one might take $t\approx 0.65$ eV,
which results in the superexchange interaction between the $|x\rangle$
orbitals in $(a,b)$ planes, $J_{(a,b)}=(9/4)t^2/U\simeq 0.13$ eV,\cite{Esk93}
in good agreement with the experimental value.\cite{expJ} Similar values of
the effective $t$ are expected also in the other transition metal oxides, and
thus we can safely assume that at the filling of one hole per ion the
ionic Hamiltonian (\ref{hband}) describes an insulating state, and
that the effective magnetic interactions can be derived in the strongly
correlated regime of $t\ll U$.

The last term in Eq. (\ref{hband}) stands for the crystal field which lifts
the degeneracy of the two $e_g$ orbitals and breaks the symmetry in
the orbital space,
\begin{equation}
\label{hz}
H_{z}=\sum_{i\sigma}(\varepsilon_xn_{ix\sigma}+\varepsilon_zn_{iz\sigma}),
\end{equation}
if $\varepsilon_x\neq \varepsilon_z$. It acts as a magnetic field in the
orbital space, and together with the parameter $\beta$ in
$H_{kin}$ (\ref{hkin}) quantifies the deviation in the electronic structure
from the ideal cubic local point group.

In the atomic limit, i.e., at $t=0$ and $E_z=0$, one has orbital degeneracy
next to spin degeneracy. This gives four basis states per site, as each hole
may occupy either orbital, $|x\rangle$ or $|z\rangle$, and either spin state,
$\sigma=\uparrow$ or $\sigma=\downarrow$. The system of $N$ $d^9$ ions thus
has a large degeneracy $4^N$, which is, however, removed by the effective
interactions between each pair of nearest neighbor ions $\{i,j\}$ which
originate from virtual transitions to the excited states,
$d^9_id^9_j\rightleftharpoons d^{10}_id^8_j$, due to hole hopping. Hence,
we derive the effective spin-orbital model following Kugel and
Khomskii,\cite{Kug73} starting from the Hamiltonian in the atomic limit,
$H_{at}=H_{int}+H_z$, and treating $H_{kin}$ as a perturbation. However,
in the present study we include the {\em full multiplet structure\/} of the
excited states within the $d^8$ configuration which gives corrections of the
order of $J_H$ compared with the earlier results of Refs.
\onlinecite{Kug73,Kug82}.

Knowing the multiplet structure of the $d^8$ intermediate states, the
derivation of the effective Hamiltonian can be done in various ways. The
most straightforward but lengthy procedure is a generalization of the
canonical transformation method used before for the Hubbard\cite{Cha77} and
the three-band\cite{Zaa88} model. A significantly shorter derivation is
possible, however, using the cubic symmetry and starting with the
interactions along the $c$-axis. Here the derivation simplifies tremendously
as one finds only effective interactions which result from the hopping of
holes between the directional $|z\rangle$ orbitals, as shown in Fig.
\ref{virtual}. Next the interactions in the remaining directions can be
generated by the appropriate rotations to the other cubic axes $a$ and $b$,
and applying the symmetry rules for the hopping elements between the
$e_g$ orbitals.\cite{Sla54} The derivation of the spin-orbital model
is given in more detail in Appendix \ref{sec:derivation}.

Depending on whether the initial state is $|z\rangle_i|x\rangle_j$ or
$|z\rangle_i|z\rangle_j$, the intermediate $d_i^{10}d_j^8$ configuration
resulting from the hole-hop $|z\rangle_i \rightarrow |z\rangle_j$, involves
on the $d^8$ site either the interorbital states, the triplet $^3A_2$ and
the singlet $^1E_{\theta}$, or the two singlets built from the states with
doubly occupied orbitals, $^1E_{\varepsilon}$ and $^1A_1$. Of course, the
spins have to be opposite in the latter case, while in the former case also
parallel spin configurations contribute in the triplet channel. Apart from
a constant term, this atomic problem is equivalent to that of the $d^2$
configuration, and thus one might consider instead the spectrum of $d^2$
ions. The eigenstates within the $e_g$ subspace are:
  (i) triplet $|^3A_2\rangle$,
 (ii) interorbital singlet $|^1E_{\epsilon}\rangle$, and
(iii) bonding and antibonding singlets, $|^1E_{\theta}\rangle$ and
      $|^1A_1\rangle$, with double occupancies of both orbitals, where
      bonding/antibonding refers to pair hopping term $\propto J_H$ between
      $|x\rangle$ and $|z\rangle$ orbital.
The energies of the states $|^3A_2\rangle$ and $|^1E_{\epsilon}\rangle$ are
straighforwardly obtained using ${\vec S}_{ix}\cdot {\vec S}_{iz}=+1/4$ and
                                ${\vec S}_{ix}\cdot {\vec S}_{iz}=-3/4$,
for $S=1$ and $S=0$ states, respectively. The remaining two singlet
energies are found by diagonalizing a $2\times 2$ problem in the subspace
of doubly occupied states. Hence, the resulting spectrum is,\cite{noted8}
\begin{eqnarray}
\label{specd8}
 E( ^3A_2          ) &=& U - J_H,    \nonumber \\
 E( ^1E_{\epsilon} ) &=& U      ,    \nonumber \\
 E( ^1E_{\theta}   ) &=& U + \case{1}{2}J_H
        - \case{1}{2}J_H\left[ 1-(E_z/J_H)^2 \right]^{1/2}, \nonumber \\
 E( ^1A_1          ) &=& U + \case{1}{2}J_H
        + \case{1}{2}J_H\left[ 1-(E_z/J_H)^2 \right]^{1/2},
\end{eqnarray}
where $E_z=\varepsilon_x-\varepsilon_z$. At $E_z=0$ it consists of
equidistant states, with a distance of $J_H$ between the triplet
$|^3A_2\rangle$ and the degenerate singlets $|^1E_{\theta}\rangle$ and
$|^1E_{\epsilon}\rangle$ (which form of course an orbital doublet),
as well as between the above singlets and the top singlet $|^1A_1\rangle$.
We emphasize that the simplified Hubbard-like form of electron-electron
interactions (\ref{hint}) which uses two parameters, $U$ and $J_H$, in this
case is an {\em exact representation\/} of the Coulomb interaction in
the $t_{2g}^6e_g^2$ configuration as obtained in the theory of
multiplet spectra, and one finds a one-to-one correspondence between the
energies calculated above, and those found with the Racah parameters $A$,
$B$, and $C$,\cite{Gri71}
\begin{eqnarray}
\label{racah}
  E( ^3A_2 ) &=& A - 8B     ,    \nonumber \\
  E(  ^1E  ) &=& A      + 2C,    \nonumber \\
  E( ^1A_1 ) &=& A + 8B + 4C.
\end{eqnarray}
Thus, the parameters used by us are $U=A+2C$ and $J_H=8B+2C$.\cite{noteuj}
We normalize the energies by the Coulomb interaction $U$, and introduce
\begin{equation}
\label{jh}
\eta\equiv J_H/U
\end{equation}
as an energy unit for the Hund's rule exchange interaction. This gives the
excitation energies which correspond to the {\em local excitations\/}
$d^9_id^9_j\rightarrow d^{10}_id^8_j$ on a given bond $(ij)$,
\begin{eqnarray}
\label{ourd8}
 \varepsilon( ^3A_2          ) &=& 1 - \eta,    \nonumber \\
 \varepsilon( ^1E_{\epsilon} ) &=& 1        ,    \nonumber \\
 \varepsilon( ^1E_{\theta}   ) &=& 1 + \case{1}{2}\eta
                      - \case{1}{2}\eta\left[ 1-(E_z/J_H)^2 \right]^{1/2},
                                                 \nonumber \\
 \varepsilon( ^1A_1          ) &=& 1 + \case{1}{2}\eta
                      + \case{1}{2}\eta\left[ 1-(E_z/J_H)^2 \right]^{1/2},
\end{eqnarray}
shown in Fig. \ref{msd8}. We note that the deviation from the equidistant
spectrum at $E_z=0$ becomes significant only for $|E_z|/J_H>1$. Taking the
realistic parameters of the cuprates,\cite{Gra92} one finds for La$_2$CuO$_4$
with $E_z=0.64$ eV that $E_z/J_H\simeq 0.54$, a value representative for
systems that are already far from orbital degeneracy. Since we are
interested here in what happens close to orbital degeneracy, this
allows us to neglect the $E_z$ dependence of the energies of the excited
$d^8$ states, and use the atomic spectrum (\ref{racah}) in the derivation
presented in Appendix \ref{sec:derivation}.

Following the above procedure, we have derived the effective Hamiltonian
${\cal H}$ in spin-orbital space,
\begin{equation}
\label{somcu}
{\cal H} = {\cal H}_J + {\cal H}_{\tau},
\end{equation}
where the superexchange part ${\cal H}_J$ can be most generally written
as follows (a simplified form was discussed recently in Ref.
\onlinecite{Fei97}),
\begin{eqnarray}
{\cal H}_J&=&\sum_{\langle ij\rangle}\left\{
 - \frac{t^2}{\varepsilon(^3A_2)}
   \left(\vec{S}_i\cdot\vec{S}_j+\frac{3}{4}\right)
   {\cal P}_{\langle ij\rangle}^{\zeta\xi}\right.              \nonumber \\
&+&\left. \frac{t^2}{\varepsilon(^1E_\epsilon)}
   \left(\vec{S}_i\cdot\vec{S}_j-\frac{1}{4}\right)
   {\cal P}_{\langle ij\rangle}^{\zeta\xi}\right.              \nonumber \\
&+&\left. \left[\frac{t^2}{\varepsilon(^1E_\theta)}
        +\frac{t^2}{\varepsilon(^1A_1)}\right]
   \left(\vec{S}_i\cdot\vec{S}_j-\frac{1}{4}\right)
   {\cal P}_{\langle ij\rangle}^{\zeta\zeta}\right\}.
\label{somj}
\end{eqnarray}
Here $\vec{S}_i$ refers to a spin $S=1/2$ at site $i$, and
${\cal P}_{\langle ij\rangle}^{\alpha\beta}$ are projection operators
on the orbital states for each bond,
\begin{eqnarray}
\label{porbit}
{\cal P}_{\langle ij\rangle}^{\zeta\xi}&=&
(\case{1}{2}+\tau^c_i)(\case{1}{2}-\tau^c_j)+
(\case{1}{2}-\tau^c_i)(\case{1}{2}+\tau^c_j),             \nonumber \\
{\cal P}_{\langle ij\rangle}^{\zeta\zeta}&=&
2(\case{1}{2}-\tau^c_i)(\case{1}{2}-\tau^c_j).
\end{eqnarray}
They are either parallel ($P_{i\zeta}=\frac{1}{2}-\tau^c_i$) to the
direction of the bond $\langle ij\rangle$ on site $i$, and perpendicular
($P_{j\xi}=\frac{1}{2}+\tau^c_j$) on the other site $j$, or parallel on both
sites, respectively, and are constructed with the following
orbital operators associated with the three cubic axes ($a$, $b$, $c$),
\begin{eqnarray}
\label{orbop}
\tau^{a}_i & = & -\case{1}{4}(\sigma^z_i - \sqrt{3}\sigma^x_i ),
                                                            \nonumber \\
\tau^{b}_i & = & -\case{1}{4}(\sigma^z_i + \sqrt{3}\sigma^x_i ),
                                                            \nonumber \\
\tau^c_i & = & \case{1}{2} \sigma^z_i.
\end{eqnarray}
The $\sigma$'s are Pauli matrices acting on the orbital pseudo-spins
$|x\rangle ={\scriptsize\left( \begin{array}{c} 1\\ 0\end{array}\right)},\;
 |z\rangle ={\scriptsize\left( \begin{array}{c} 0\\ 1\end{array}\right)}$.
Hence, we find a Heisenberg Hamiltonian for the spins, coupled into an
orbital problem. While the spin problem is described by the continuous
symmetry group $SU(2)$, the orbital problem is clock-model-like, i.e., there
are three directional orbitals: $3x^2-r^2$, $3y^2-r^2$, and $3z^2-r^2$, but
they are not independent. The orbital basis consists of one directional
orbital and its orthogonal counterpart, and we have chosen here
$|z\rangle\equiv 3z^2-r^2$ and $|x\rangle\equiv x^2-y^2$ orbitals.

In general, the energies of these two orbital states, $|x\rangle$ and
$|z\rangle$, are different, and thus the complete effective Hamiltonian of
the $d^9$ model (\ref{somcu})
includes as well the crystal-field term (\ref{hz}) which we write as
\begin{equation}
\label{somez}
{\cal H}_{\tau} = - E_z \sum_i \tau^c_i.
\end{equation}
Here $E_z$ is a crystal field which acts as a "magnetic field" for the
orbital pseudospins, and is loosely associated with an uniaxial pressure
along the $c$-axis. The $d^9$ spin-orbital model (\ref{somcu}) depends thus
on two parameters: (i) the crystal field splitting $E_z$, and (ii) the
Hund's rule exchange $J_H$.

While the first two terms in (\ref{somj}) cancel for the magnetic
interactions in the limit of $\eta\to 0$, the last term favors AF spin
orientation. Although the form (\ref{somj}) might in principle be used for
further analysis, we prefer to make an expansion of the excitation energies
$\varepsilon_n$ in the denominators of Eq. (\ref{somj}) in terms of $J_H$,
and use $\eta=J_H/U$ (\ref{jh}) as a parameter which quantifies the Hund's
rule exchange. This results in the following form of the effective exchange
Hamiltonian in the $d^9$ model (\ref{somcu}),\cite{Fei97,notebugs}
\begin{eqnarray}
\label{somjexp}
{\cal H}_J\! &\simeq& \! J\sum_{\langle ij\rangle} \left[
      2\left( {\vec S}_i\cdot{\vec S}_j -\frac{1}{4} \right)
       {\cal P}_{\langle ij\rangle}^{\zeta\zeta}
     -{\cal P}_{\langle ij\rangle}^{\zeta\xi  }\right]        \nonumber \\
 &-&\! J\eta\sum_{\langle ij\rangle} \left[ {\vec S}_i\cdot{\vec S}_j
       \left(\! {\cal P}_{\langle ij\rangle}^{\zeta\zeta}
       \! +\! {\cal P}_{\langle ij\rangle}^{\zeta\xi  }\!\right)
   +\frac{3}{4}{\cal P}_{\langle ij\rangle}^{\zeta\xi  }
   -\frac{1}{4}{\cal P}_{\langle ij\rangle}^{\zeta\zeta}\right].\nonumber \\
\end{eqnarray}

The first term in Eq. (\ref{somjexp}) describes the AF superexchange
$\propto J=t^2/U$ (where $t$ is the hopping between $|z\rangle$ orbitals
along the $c$-axis), and is obtained when the splittings between different
excited $d^8$ states $\sim J_H$ (Fig. \ref{msd8}) are neglected. As we show
below, in spite of the AF superexchange $\propto J$, {\it no {\rm LRO} can
stabilize in a system described by the spin-orbital model (\ref{somcu}) in
the limit $\eta\to 0$ at orbital degeneracy $(E_z=0)$} because of
the presence of the frustrating orbital interactions which gives a highly
degenerate classical ground state. We emphasize that even in the limit of
$J_H\to 0$ the present Kugel-Khomskii model {\em does not obey\/}
SU(4) symmetry, essentially because of the directionality of the $e_g$
orbitals.  Therefore, such an idealized SU(4)-symmetric model\cite{Fri99}
does not correspond to the realistic situation of degenerate $e_g$ orbitals
and is expected to give different answers concerning the interplay of spin
and orbital ordering in cubic crystals.

Taking into account the multiplet splittings, we obtain [second line of
(\ref{somjexp})] again a Heisenberg-like Hamiltonian for the spins coupled
into an orbital problem, with a reduced interaction $\propto J\eta$. It is
evident that the new terms support FM rather than AF spin interactions for
particular orbital orderings. This net FM superexchange originates from the
virtual transitions which involve the triplet state $|^3A_2\rangle$, which
has the lowest energy and thus gives the strongest effective coupling.
We remark in passing that the FM channel is additionally
enhanced for $d^4$ ions when the virtual excitations to double
occupancies in $e_g$ orbitals happen in the presence of partly filled
$t_{2g}$ orbitals, as realized in the manganites.\cite{Shi97,Fei99}

The important feature of the spin-orbital model (\ref{somcu}) is that the
{\it actual magnetic interactions depend on the orbital pattern\/}. This
follows essentially from the hopping matrix elements in $H_{kin}$
(\ref{hkin}) being different between a pair of $|x\rangle$ orbitals, between
a pair of different orbitals (one $|x\rangle$ and one $|z\rangle$ orbital),
and between a pair of $|z\rangle$ orbitals, respectively, and depending on
the bond direction either in the $(a,b)$ planes, or along the
$c$-axis.\cite{Zaa93} We show in Sec. \ref{sec:mfa} that this leads to a
particular competition between magnetic and orbital interactions, and the
resulting phase diagram contains a rather large number of classical
phases, stabilized for different values of $E_z$ and $J_H$.

\section{ Mean-field phase diagram }
\label{sec:mfa}

\subsection{ Anisotropy of antiferromagnetic interactions }

We start the analysis of the $d^9$ spin-orbital (or Kugel-Khomskii) model 
(\ref{somcu})--(\ref{somjexp}) by analyzing the MF solution obtained by 
replacing the scalar products $\vec{S}_i\cdot\vec{S}_j$ by the Ising term, 
$S^z_iS^z_j$. The MF Hamiltonian may be written for the more general 
situation where the interaction has uniaxial
anisotropy along the $c$-direction in the 3D lattice as follows,
\begin{eqnarray}
\label{somcumf}
{\cal H}_{\rm MF}\! &\simeq&\! \sum_{\langle ij\rangle} J_{\alpha}\left[
      2\left( S_i^zS_j^z -\case{1}{4} \right)
       {\cal P}_{\langle ij\rangle}^{\zeta\zeta}
     -{\cal P}_{\langle ij\rangle}^{\zeta\xi  }\right]        \nonumber \\
 &-&\! \eta\sum_{\langle ij\rangle} J_{\alpha}\left[ S_i^zS_j^z
       \left(\!{\cal P}_{\langle ij\rangle}^{\zeta\zeta}
       +{\cal P}_{\langle ij\rangle}^{\zeta\xi  }\!\right)
       +\case{3}{4}{\cal P}_{\langle ij\rangle}^{\zeta\xi  }
       -\case{1}{4}{\cal P}_{\langle ij\rangle}^{\zeta\zeta}\right]
                                                              \nonumber \\
&-& E_z \sum_i \tau^c_i,
\end{eqnarray}
where $J_a=J_b=J$, and $J_c=J\beta$. For $\beta>1$ the nearest-neighbor bonds
$\langle ij\rangle$ $\parallel c$ are shorter, while for $\beta<1$ these
bonds are longer than the bonds within the $(a,b)$ planes. In the limit of
$\beta\to 0$ the bonds along the $c$-axis may be neglected and the model
reduces to a 2D model, representative for the magnetic interactions between
Cu ions within the CuO$_2$ planes of the high-temperature superconductors.

The presence of AF spin interactions $\propto J$ suggests magnetic
superstructures with staggered magnetization, and we considered several
possibilities, with two- and four-sublattice 3D structures, giving rise to
G-AF and A-AF phases, AF 1D chains
coupled ferromagnetically, and others. The MF Hamiltonian contains
as well an AF {\em interaction between orbital variables\/},
$\sim J\tau^{\alpha}_i\tau^{\alpha}_j$, which suggests that it might be
energetically more favorable to alternate the orbitals in a certain regime of
parameters, and pay thereby part of the magnetic energy. This illustrates the
essence of the {\em frustration\/} of the magnetic interactions present
in the spin-orbital model (\ref{somcu}), as discussed in Sec.
\ref{sec:orbitals}. Therefore, for any classical state the orbitals occupied
by the holes have to be optimized, and we allowed mixed orbitals (MO),
\begin{equation}
\label{mixing}
|i\mu\sigma\rangle=\cos\theta_i|iz\sigma\rangle+\sin\theta_i|ix\sigma\rangle,
\end{equation}
with the values of the mixing angles $\{\theta_i\}$ being variational
parameters to be found from the minimization of the classical energy.

The superexchange in (\ref{somcumf}) depends strongly on the orbital state.
At large positive $E_z$, where the crystal field strongly favors
$|x\rangle$-occupancy over $|z\rangle$-occupancy, one expects that
$\theta_i=\pi/2$ in Eq. (\ref{mixing}), and the holes occupy $|x\rangle$
orbitals on every site. In this case the spins do not interact in the
$c$-direction (see Fig. \ref{virtual}), and there is also no orbital energy
contribution. Hence, the $(a,b)$ planes will decouple magnetically, while
within each plane the superexchange is AF and equal to $9J/4$ along $a$ and
$b$. These interactions stabilize a 2D antiferromagnet, called further AFxx.
The resulting 2D N\'eel state with decoupled $(a,b)$ planes along the
$c$-direction is the well-known classical ground state of the high-$T_c$
superconductors, La$_2$CuO$_4$ and YBa$_2$Cu$_3$O$_6$.\cite{Kam94}
In contrast, if $E_z<0$ and $|E_z|$ is large, $|E_z|/J\gg 1$, then
$\theta_i=0$ in Eq. (\ref{mixing}), and the holes occupy $|z\rangle$
orbitals. The spin system has then strongly anisotropic AF superexchange,
being $4J$ between two $|z\rangle$ orbitals along the $c$-axis, and $J/4$
between two $|z\rangle$ orbitals in the $(a,b)$ planes, respectively. The
corresponding 3D N\'eel state with holes occupying $|z\rangle$ orbitals is
called AFzz. The spin and orbital order in both AF phases is shown
schematically within the $(a,b)$ planes in Fig. \ref{allmfa}.

\subsection{ Antiferromagnetic states in the 3D model }

Assuming an AF classical order in all three directions, the so-called G-AF
state, it is thus obvious that for large $|E_z|$ one finds either the AFxx
or the AFzz phase, depending on whether $E_z>0$ or $E_z<0$, with the
following energies normalized per one site,
\begin{eqnarray}
\label{mfaf}
E_{\rm AFxx}&=&-3J\left(1-{\eta\over 4}\right)-{1\over 2}E_z, \nonumber \\
E_{\rm AFzz}&=&- J\left(1+{\eta\over 4}\right)
          -2J\beta\left(1-{\eta\over 2}\right)+ {1\over 2}E_z.
\end{eqnarray}
The AFxx and AFzz phases are degenerate in a 3D system ($\beta=1$) along
the line $E_z=0$, while decreasing $\beta$ moves the degeneracy to
negative values of $E_z$, namely to $E_z=-2J(1-\beta)(1-{\eta \over 2})$.

However, for intermediate values of $|E_z|$ one should allow for mixed
orbitals. Following the argument above about the AF nature of the orbital
interaction, we assume alternating orbitals at two sublattices, A and B.
The alternation should allow the orbitals to compromise between being
identical (optimizing the magnetic energy) and being orthogonal (optimizing
the orbital energy). This is realized by choosing in Eq. (\ref{mixing}) the
angles alternating between the sublattices: $\theta_i=+\theta$ for $i\in A$,
and $\theta_j=-\theta$ for $j\in B$, respectively,
\begin{eqnarray}
\label{orbmoffa}
|i\mu\sigma\rangle&=&\cos\theta|iz\sigma\rangle+\sin\theta|ix\sigma\rangle,
                                                        \nonumber \\
|j\mu\sigma\rangle&=&\cos\theta|jz\sigma\rangle-\sin\theta|jx\sigma\rangle.
\end{eqnarray}
The calculation of the energy can be performed either by evaluating the
average values of the operator variables $\{\tau_i^{\alpha}\}$, or by taking
the average values of the orbital projection operators $\{P_{i\alpha}\}$ as
given in Eq. (\ref{fullij}). Using the two-sublattice orbital ordering
(\ref{orbmoffa}), one finds for the bonds $\langle ij\rangle\parallel (a,b)$,
\begin{eqnarray}
\label{orbavab}
\langle P_{i\xi}P_{j\zeta}+P_{i\zeta}P_{j\xi}\rangle&=&
\case{1}{8}(7-4\cos^2 2\theta),
                                                        \nonumber \\
\langle 2P_{i\zeta}P_{j\zeta}\rangle&=&
\case{1}{8}(1-2\cos 2\theta)^2,
\end{eqnarray}
and for the bonds $\langle ij\rangle\parallel c$,
\begin{eqnarray}
\label{orbavc}
\langle P_{ix}P_{jz}+P_{iz}P_{jx}\rangle&=&
\case{1}{2}(1- \cos^2 2\theta),
                                                        \nonumber \\
\langle 2P_{iz}P_{jz}\rangle&=&
\case{1}{2}(1+\cos 2\theta)^2.
\end{eqnarray}
The classical energy per site as a function of $\theta$ is then given by
\begin{eqnarray}
\label{enemoaaat}
E(\theta)&=&-\frac{J}{4}(1+\frac{\eta}{2})(7-4\cos^2 2\theta)
                                                   \nonumber\\
         &-&\frac{J}{4} (1-\frac{\eta}{2})(1-2\cos 2\theta)^2
                                                   \nonumber\\
         &-&\frac{J}{2}\beta(1+\frac{\eta}{2})(1-\cos^2 2\theta)
                                                   \nonumber\\
         &-&\frac{J}{2}\beta(1-\frac{\eta}{2})(1+\cos 2\theta)^2
                                                   \nonumber\\
         &+&\frac{1}{2}E_z\cos 2\theta.
\end{eqnarray}
This has a minimum at
\begin{equation}
\label{orbmoaaa}
\cos 2\theta=-{{(1-\frac{\eta}{2})(1-\beta)+\frac{1}{2}\varepsilon_z\over
               (2+\beta)\eta}},
\end{equation}
where $\varepsilon_z=E_z/J$, if $\eta\neq 0$, and provided that
$|\cos 2\theta|\le 1$ (a similar condition applies to all the other states
with MO considered below). So, as long as
$2J(\beta-1)-3J(\beta+1)\eta \le E_z \le 2J(\beta-1)+J(5+\beta)\eta$,
there is genuine MO order, while upon reaching the smaller (larger)
boundary value for $E_z$, the orbitals go over smoothly into
$|z\rangle$ ($|x\rangle$), i.e. one retrieves the AFzz (AFxx) phase.
Taking the magnetic ordering in the three cubic directions [$abc$]
as a label to classify the classical phases with MO (\ref{orbmoffa}),
we call the phase obtained in the regime of genuine MO order
MO{\scriptsize AAA}, with classical energy given by
\begin{eqnarray}
\label{enemoaaa}
E_{\rm MO{\scriptsize AAA}}&=&-\left(2+\beta+\frac{3}{4}\eta\right)J
                                                            \nonumber \\
&-&J {\left[ (2-\eta)(1-\beta)+\varepsilon_z \right]^2\over
           4(2+\beta)\eta}.
\end{eqnarray}
Upon increasing $J_H$, the FM interactions occur which increase the energy
of the AF phases in three dimensions by the term $\frac{3}{4}\eta$ per site
in Eqs. (\ref{mfaf}) (a similar increase of energy occurs also in the
MO{\scriptsize AAA} phase in the region of its existence). This indicates
frustration of magnetic interactions and opens a potential possibility that
other classical phases with FM order along particular directions might be
more stable. We have found a few classical phases when the spins order
ferromagnetically either in particular planes, or along one spatial
direction, and this magnetic order coexists with MO occupied by holes.

For example, the angles in Eq. (\ref{mixing}) can be chosen in such a way
that at least one of the orbitals on two neighboring sites is perpendicular
to the bond direction, e.g. is like $y^2-z^2$ type for a bond along the
$a$-axis. In such a case, the AF superexchange vanishes, and one finds
instead a weaker FM interaction, in agreement with the Goodenough-Kanamori
rules.\cite{Goo63} By this mechanism Kugel and Khomskii\cite{Kug73} proposed
an alternating orbital order to explain the FM planes observed in KCuF$_3$.
Following this argument, let us assume FM order within $(a,b)$ planes, and
the same form (\ref{orbmoffa}) as above for the alternating orbitals at the
two sublattices, $A$ and $B$.
As alternating orbitals can only be arranged to be perpendicular to the
bonds in at most two spatial directions, such an arrangement for the
$(a,b)$ planes forces the orbitals to have nonzero lobes along $c$.
This results in sizable AF superexchange for the bonds
$\langle ij\rangle$ parallel to $c$, which will order the spins
antiferromagnetically in the $c$ direction.
The orbitals may either repeat or stagger along the $c$-axis, and both
states give the same mean-field energy. Taking the magnetic ordering in the
three cubic directions [$abc$] as a label to classify the classical phases
with MO (\ref{orbmoffa}), we call this ground state the MO{\scriptsize FFA}
phase.
With the help of Eqs. (\ref{orbavab}) and (\ref{orbavc}) one obtains the
following classical energy as a function of $\theta$,
\begin{eqnarray}
\label{enemoffat}
E(\theta)&=&-\frac{J}{4}(1+\eta)(7-4\cos^2 2\theta)        \nonumber \\
         &-&\frac{J}{2}\beta(1+\frac{\eta}{2})(1-\cos^2 2\theta)  \nonumber \\
         &-&\frac{J}{2}\beta(1-\frac{\eta}{2})(1+\cos 2\theta)^2  \nonumber \\
         &+&\frac{1}{2}E_z\cos 2\theta ,
\end{eqnarray}
with a minimum at
\begin{equation}
\label{thetamoffa}
\cos 2\theta={\beta(1-\frac{\eta}{2})-\frac{1}{2}\varepsilon_z
              \over 2+(2+\beta)\eta},
\end{equation}
where again the MO exist as long as $|\cos 2\theta|\le 1$. Using Eqs.
(\ref{enemoffat}) and (\ref{thetamoffa}) one finds that the classical
energy of the MO{\scriptsize FFA} phase is given by
\begin{equation}
\label{enemoffa}
E_{\rm MO{\scriptsize FFA}}=-\frac{J}{4}\left(11-7\eta\right)
  -\frac{J}{2}{[\beta(1-\frac{\eta}{2})-\frac{1}{2}\varepsilon_z]^2
               \over 2+(2+\beta)\eta}.
\end{equation}

As a special case, let us consider first degenerate orbitals ($E_z=0$) in a
3D system ($\beta=1$). Eq. (\ref{thetamoffa}) simplifies in this case to
$\cos 2\theta= (1-\frac{\eta}{2})/(2+3\eta)$. A particularly simple
result is found at $\eta=0$ where $\cos 2\theta= 1/2$, i.e., $\theta=\pi/6$,
and the orbitals stagger like $x^2-z^2$ and $y^2-z^2$, as shown in Fig.
\ref{allmfa}. This staggering was proposed by Kugel and Khomskii as a ground
state of KCuF$_3$;\cite{Kug82} of course, this state is not realized for the
realistic parameters with $\eta\simeq 0.3$, but the optimized orbitals
with $\theta$ given by
(\ref{thetamoffa}) are not so far from this idealized picture.

The energy of the MO{\scriptsize FFA} phase is degenerate with that of the AF
phases at the classical degeneracy point, $M\equiv (E_z/J,\eta)=(0,0)$, and
this phase becomes more stable at $\eta>0$, and $E_z/J\simeq 0$. The magnetic
energy is gained due to relatively strong AF interactions on the bonds
$\langle ij\rangle\parallel c$, and weak FM interactions in the planes
$(a,b)$, perpendicular to the preferred directionality of the MO
(\ref{mixing}) along the $c$-direction, while the orbital energy is gained
due to orbital alternation within the $(a,b)$ planes. Such orbital ordering
remains stable with decreasing $E_z<0$, while two similar states with the
staggering either within the $(b,c)$ or the $(a,c)$ planes, are more stable
for $E_z>0$. Following our convention, these two degenerate MO states stable
at $E_z>0$ are called MO{\scriptsize AFF} and MO{\scriptsize FAF} (see Fig.
\ref{allmfa}), respectively. However, the MO involve in this case the
directional orbital $|\zeta\rangle$ along the AF bonds (i.e.,
$|\zeta_a\rangle \sim 3x^2-r^2$ for MO{\scriptsize AFF} or
$|\zeta_b\rangle \sim 3y^2-r^2$ for MO{\scriptsize FAF},
respectively), and the corresponding orthogonal orbital, $|\xi\rangle$.
Therefore, since the symmetry-breaking field acts on $|z\rangle$ orbitals,
the angles in the two sublattices cannot be exactly equivalent in this case,
unlike in the MO{\scriptsize FFA} phase, and we adopted an ansatz,
\begin{eqnarray}
\label{orbmoaff}
|i\sigma\rangle&=&\cos\theta_+|i\xi  \sigma\rangle
                 +\sin\theta_+|i\zeta\sigma\rangle,         \nonumber \\
|j\sigma\rangle&=&\cos\theta_-|i\xi  \sigma\rangle
                 -\sin\theta_-|i\zeta\sigma\rangle,
\end{eqnarray}
where $i\in A$, $j\in B$, and $\theta_{\pm}>0$ for the two
sublattices. Introducing for convenience the new angles,
$\phi=\frac{1}{2}(\theta_++\theta_-)$, and $\delta=\theta_+-\theta_-$, one
finds the following conditions for the energy minimum of the classical
MO{\scriptsize AFF} phase,
\begin{eqnarray}
\label{thetamoaff1}
\cos 2\phi  &=&
-\case{1}{4}\left\{\left[(1+{\beta})(2-\eta)
            +\varepsilon_z\right]\cos\delta\right.    \nonumber \\
 &+&\left.\sqrt{3}\varepsilon_z\sin\delta\right\}
   \left[1+\beta+(1+2\beta)\eta\right]^{-1} ,                   \\
\label{thetamoaff2}
\tan 2\delta&=&
+\case{1}{2}\sqrt{3}\left[(1+{\beta})(2-\eta)+\varepsilon_z\right]
                    \varepsilon_z  \nonumber \\
&\times&\left\{4\left[1+\beta+(1+2\beta)\eta\right]
        +\left[(1+{\beta})(2-\eta)+\varepsilon_z\right]^2\right. \nonumber \\
  &-&\left.\case{3}{4}\varepsilon_z^2\right\}^{-1},
\end{eqnarray}
and the energy is given by
\begin{eqnarray}
\label{enemoaff}
&E&_{\rm MO{\scriptsize AFF}}=-\frac{J}{4}\left[
              7(1+\eta)+2\beta(1+\cos\delta)\right]           \nonumber \\
&-&\frac{J}{32}{\left[[(1+{\beta})(2-\eta)+\varepsilon_z]\cos\delta
            +\sqrt{3}\varepsilon_z\sin\delta\right]^2 \over
            1+\beta+(1+2\beta)\eta} .                         \nonumber \\
\end{eqnarray}

Finally, one may consider states in which magnetic energy is gained in
the $c$-direction due to MO with a small admixture of $|z\rangle$ into
orbitals of predominantly $|x\rangle$-character, i.e.,
$\sin\theta_i=1-\epsilon$ in Eq. (\ref{mixing}).
As such a state is a modification of the AFxx phase, the two sublattices
in the $(a,b)$ planes are again physically equivalent, and it suffices to
introduce a single angle $\theta$ to characterize this state. Apart from
(large) energy contributions due to AF order on the bonds in the $(a,b)$
planes, the expansion of the ground state energy contains also (small)
terms depending on the spin order in the $c$-direction,
$\langle S_i^zS_j^z\rangle_{\parallel c}$,
\begin{equation}
\label{mixc}
E=\left(1+\cos2\theta\right)\left(1+\cos2\theta-\eta\right)
  \langle S_i^zS_j^z\rangle_{\parallel c}+const,
\end{equation}
which prefers FM order as long as $(1+\cos2\theta)<\eta$. The reason is
that the AF superexchange is a fourth order effect $\sim\epsilon^4$, while
the FM interactions $\propto\eta$ are second order, $\sim\epsilon^2$, and
give a lower energy $E$ as long as the $|z\rangle$ occupancy is small enough.
Following our convention, we call the resulting state the MO{\scriptsize AAF}
phase, with the mixing angle given by
\begin{equation}
\label{orbmoaaf}
\cos2\theta=-{{1-\frac{\eta}{2}+\frac{1}{2}\varepsilon_z
             \over \beta(1+\eta)+2\eta}},
\end{equation}
and the classical energy by
\begin{eqnarray}
\label{enemoaaf}
E_{\rm MO{\scriptsize AAF}}&=&-\left(2+\frac{3}{4}\eta\right)J  \nonumber \\
&-&\frac{1}{2}\beta(1+\eta)
-J{\left( 2-\eta+\varepsilon_z\right)^2\over 2[\beta(1+\eta)+2\eta]}.
\end{eqnarray}
Therefore, only when the average population of the $|z\rangle$ orbitals,
$\sim\cos^2\theta$, increases sufficiently, one can find a transition to
the AF phase with mixed orbitals, MO{\scriptsize AAA}, discussed above.

By making several other choices of orbital mixing and classical magnetic
order, we have verified that no other commensurate ordering with up to four 
sublattices can be stable in the present situation. Although some other 
phases could be found, they were degenerate with the above phases only at 
the $M$ point, and otherwise had higher energies. Thus, we obtain the
classical phase diagram of the 3D spin-orbital model (\ref{somcu}) by
comparing the energies of the six above phases for various values of two
parameters, $\{E_z/J,J_H/U\}$: two AF phases with two sublattices and pure
orbital character (AFxx and AFzz), three A-AF phases with four sublattices
(MO{\scriptsize FFA} and two degenerate phases: MO{\scriptsize AFF} and 
MO{\scriptsize AFF}), one C-AF phase (MO{\scriptsize AAF}), and one G-AF 
phase with MO's (MO{\scriptsize AAA}). While the orbital mixing is unstable 
at $\eta=0$, the generic sequence of classical phases at finite $\eta$ and 
decreasing $E_z/J$ is: AFxx, MO{\scriptsize AAF}, MO{\scriptsize AAA}, 
MO{\scriptsize AFF}, MO{\scriptsize FFA}, and AFzz, and the magnetic order 
is tuned together with the gradually increasing $|z\rangle$ character of 
the occupied orbitals.

The result for cubic symmetry ($\beta=1$) is presented in Fig. \ref{mfa3d}, 
where one finds all six phases, but the MO{\scriptsize AAA} phase does 
stabilize only in a very restricted regime of parameters with $J_H/U<0.1$, 
before MO{\scriptsize AFF} takes over. Only the first of the above 
transitions is a continuous one, and the $|z\rangle$ amplitude
$\sim\cos^2\theta$ increases smoothly from zero and removes the built-in
degeneracy of the 2D AFxx phase with respect to the magnetic order along
the $c$-direction. All the other transition lines in Fig. \ref{mfa3d} are
associated with jumps in the magnetic and in orbital patterns. We emphasize
that all the considered phases with magnetic LRO are degenerate at the point 
$M$, with classical energy of $-3J$. In fact $M$ is an infinite-order 
quantum critical point, since not only may the spins be chosen to be FM in
certain planes, whence the orbitals have to be tuned to compensate the loss
of the magnetic energy by the orbital energy contributions, as realized in
all MO phases, but also may the orbitals be rotated freely when the spins
are AF in all three directions.We note, however, that the magnetic terms are
essential, and in a purely disordered spin system, with
$\langle S^z_iS^z_j\rangle=0$, a higher energy of $-21J/8$ is found even
with the optimal choice of orbitals with $\cos 2\theta=0$.

We also investigated the phase diagrams for the case of modified hopping
along the $c$-direction ($\beta \neq 1$). One finds that increased hopping
($\beta=1.414$) in the $c$-direction stabilizes the MO phases, and in
particular the MO{\scriptsize AFF} (MO{\scriptsize FAF}) phase [Fig.
\ref{beta}(a)]. By contrast, the MO phases are stable in a narrower range of
$E_z$ for a fixed value of $J_H/U$, if the hopping along the $c$-direction
is decreased below $\beta=1$ [an example of $\beta=0.707$ is shown in Fig.
\ref{beta}(b)]. The decreased stability of the MO{\scriptsize AFF} phase
promotes in this case the AF order with MO in the MO{\scriptsize AAA} phase.
The latter phase is stable only in a relatively narrow range of $E_z$, and
only for small enough $J_H/U$; an increase of $J_H/U$ favors instead FM
order along the $c$-direction. We also note that the orbital mixing sets in
for the MO{\scriptsize AAA} phase (\ref{orbmoaaa}) only at a smaller value
of $E_z$ than in the MO{\scriptsize AAF} phase (\ref{orbmoaaf}).
Interestingly, the point of high degeneracy of the classical states exists
{\em independently of the value of\/} $\beta$, and moves for $\beta\neq 1$
to $E_z=-2J(1-\beta)$. This demonstrates the generic nature of the internal
frustration of spin and orbital interactions in the model, and the crystal
field term just plays here a compensating role for the missing (or enhanced)
magnetic interactions within the $(a,b)$ planes.

Independently of the value of $\beta$, the spin-orbital model (\ref{somcu})
has a universal feature: different classical spin structures become
degenerate at the critical lines in Figs. \ref{mfa3d}-\ref{mfa2d}. This is
also encountered in frustrated 2D magnetic lattices described by simple
Heisenberg Hamiltonians,\cite{Chu91} and may thus be regarded as a
signature of frustration. However, unlike in the purely spin models,
in the present case (\ref{somcu}), the {\em sign\/} of the interactions
changes because of the coupling to the orbital sector, and this
{\em reduces the effective dimensionality\/} for the AF interactions
$\sim J$, with the 3D system behaving like a quasi-1D antiferromagnet.

\subsection{ Phase diagram of a 2D model }

As a special case, we considered the limit of $\beta\to 0$ which gives a 2D
spin-orbital model. The two AF phases with either $|x\rangle$ or $|z\rangle$
orbitals occupied, AFxx and AFzz, are degenerate at $E_z=-2J$. This asymmetry
reflects the large difference between the superexchange interactions for
$|x\rangle$ and $|z\rangle$ orbitals within the $(a,b)$ planes of a 2D system
which has to be compensated by the orbital energy (\ref{somez}).

As the presence of FM planes $\parallel c$-axis is crucial for the ordering
in the MO{\scriptsize AFF} phase (see Fig. \ref{allmfa}), this phase
disappears, while the remaining two phases with AF order within $(a,b)$
planes, MO{\scriptsize AAA} and MO{\scriptsize AAF}, collapse into a single
MO{\scriptsize AA} phase. Hence, one finds in two dimensions a classical
phase diagram with only four phases, which are stable with decreasing $E_z$
and at finite $\eta$ in the following order: AFxx, MO{\scriptsize AA},
MO{\scriptsize FF}, and AFzz (Fig. \ref{mfa2d}). The 2D phase diagram shows
in particular that strong AF superexchange in the $c$-direction is not the
stabilizing factor of the MO{\scriptsize FFA} phase in the 3D model, but
instead these phases are stable due to the orbital interactions which
enforce the orbital alternation shown in Fig. \ref{allmfa}.

For the realistic parameters of La$_2$CuO$_4$ the Cu $d_{x^2-y^2}$ and
$d_{3z^2-r^2}$ orbitals are split, and $E_z\simeq 0.64$ eV.\cite{Gra92}
This material belongs together with Nd$_2$CuO$_4$ to the class of cuprates
with weakly coupled CuO$_2$ planes, and one finds in the present treatment
a 2D AFxx state, as observed in neutron experiments.\cite{Kas98} If however
the orbital splitting is small in a 2D situation, the orbital ordering 
couples strongly to the lattice, as the hybrids with alternating phasing on 
two sublattices are formed according to Eqs. (\ref{orbmoaff}) The net result 
is a quadrupolar distortion as indicated in Fig. \ref{disto}. In fact, using 
these arguments Kugel and Khomskii predicted\cite{Kug73} the existence of 
such a structural distortion in the MO{\scriptsize FF} phase of a quasi-2D 
compound K$2$CuF$_4$. This prediction was confirmed experimentally a few 
years later.\cite{Ito76}

The MO{\scriptsize FF} phase of K$_2$CuF$_4$ is magnetically polarized, has 
no quantum fluctuations, and is thus well described in a classical theory.
In the next sections we concentrate ourselves on the 3D case, where the
quantum fluctuations are strong and destabilize the classical magnetic 
ordering in a particular regime of parameters.

\section{ Elementary excitations }
\label{sec:magnons}

\subsection{General formalism}
\label{sec:genrpa}

The presence of the orbital degrees of freedom in the Hamiltonian
(\ref{somcu}) results in excitation spectra that are qualitatively different
from those of the HAF with a single spin-wave mode. As we have discussed in
the limit of $J_H=0$, the transverse excitations are twofold:
{\em spin-waves\/} and {\em spin-and-orbital waves\/}.\cite{Fei98}
In addition to these two modes there are also {\em longitudinal\/} (purely
orbital) excitations, and thus one finds three elementary excitations for
the present spin-orbital model (\ref{somcu}).\cite{Fei97,Fei98,Bri98} This
gives therefore the same number of modes as found in a 1D SU(4) symmetric
spin-orbital model in the Bethe ansatz method.\cite{Sut75,Fri99}
We emphasize that this feature is a consequence of the dimension (equal
to 15) of the $so(4)$ Lie algebra of the local operators, as explained 
below, and is not related to the global symmetry of the Hamiltonian.
Here we present the analysis of the realistic $d^9$ spin-orbital model
for the 3D simple cubic (i.e., perovskite-like) lattice, using linear
spin-wave theory,\cite{Tak89,Aue94} generalized such as to make it
applicable to the present situation.

Before we introduce the excitation operators, it is convenient to rewrite
the spin-orbital model (\ref{somcu}) in a different representation which
uses a four-dimensional space,
$\{|x\! \uparrow\rangle$, $|x\! \downarrow\rangle$,
  $|z\! \uparrow\rangle$, $|z\! \downarrow\rangle\}$, instead of a direct 
product of the spin and orbital spaces. Hence, we introduce operators which
define purely spin excitations in individual orbitals,
\begin{equation}
S^{+}_{ixx}=d^{\dagger}_{ix\uparrow}d^{}_{ix\downarrow}, \hskip 1.0cm
S^{+}_{izz}=d^{\dagger}_{iz\uparrow}d^{}_{iz\downarrow},
\label{splus}
\end{equation}
and operators for simultaneous spin-and-orbital excitations,
\begin{equation}
K^{+}_{ixz}=d^{\dagger}_{ix\uparrow}d^{}_{iz\downarrow}, \hskip 1.0cm
K^{+}_{izx}=d^{\dagger}_{iz\uparrow}d^{}_{ix\downarrow}.
\label{oplus}
\end{equation}
The corresponding $S^z_{i\alpha\alpha}$ and $K^z_{i\alpha\beta}$ operators
are defined as follows,
\begin{eqnarray}
S^{z}_{ixx}&=&\case{1}{2}(n_{ix\uparrow}-n_{ix\downarrow}), \nonumber \\
S^{z}_{izz}&=&\case{1}{2}(n_{iz\uparrow}-n_{iz\downarrow}),           \\
K^{z}_{ixz}&=&\case{1}{2}(d^{\dagger}_{ix  \uparrow}d_{iz  \uparrow}
                        -d^{\dagger}_{ix\downarrow}d_{iz\downarrow}),
                                                            \nonumber \\
K^{z}_{izx}&=&\case{1}{2}(d^{\dagger}_{iz  \uparrow}d_{ix  \uparrow}
                        -d^{\dagger}_{iz\downarrow}d_{ix\downarrow}).
\label{szet}
\end{eqnarray}

The Hamiltonian (\ref{somcu}) contains also purely orbital interactions which
can be expressed using the following orbital-flip ($T_{i\alpha\beta}$)
and orbital-polarization ($n_{i-}$) operators,
\begin{eqnarray}
T_{ixz}&=&\case{1}{2}(d^{\dagger}_{ix  \uparrow}d_{iz  \uparrow}
                    +d^{\dagger}_{ix\downarrow}d_{iz\downarrow}),
                                                            \nonumber \\
T_{izx}&=&\case{1}{2}(d^{\dagger}_{iz  \uparrow}d_{ix  \uparrow}
                    +d^{\dagger}_{iz\downarrow}d_{ix\downarrow}),
                                                            \nonumber \\
n_{i-}&=&\case{1}{2}(d^{\dagger}_{ix  \uparrow}d_{ix  \uparrow}\! +\!
                    d^{\dagger}_{ix\downarrow}d_{ix\downarrow}\! -\!
                    d^{\dagger}_{iz  \uparrow}d_{iz  \uparrow}\! -\!
                    d^{\dagger}_{iz\downarrow}d_{iz\downarrow} ).
\label{ozet}
\end{eqnarray}
In order to simplify the notation, we also introduce sum operators for
the spin-and-orbital and purely orbital operators,
\begin{eqnarray}
\label{top}
K^{+}_{i}&=&K^{+}_{ixz}+K^{+}_{izx},      \nonumber \\
K^{z}_{i}&=&K^{z}_{ixz}+K^{z}_{izx},      \nonumber \\
T    _{i}&=&T    _{ixz}+T    _{izx} .
\end{eqnarray}
The full set of local operators at a site $i$ constitute an $so(4)$ Lie
algebra. While the spin operators (\ref{splus}) fulfill of course for $x$
and $z$ separately the usual $su(2)$ commutation relations, they also form
collectively a subalgebra of $so(4)$, and the same holds for the
spin-and-orbital operators (\ref{oplus}). However, as we will see below,
for the calculation of the excitations one also needs commutators between
spin and spin-and-orbital operators, so that one cannot avoid considering
the full Lie-algebra structure of $so(4)$, discussed in Appendix
\ref{sec:commute}.

The number of collective modes in a particular phase may be determined as 
follows. The $so(4)$ Lie algebra consists of three Cartan operators, i.e., 
operators diagonal on the local eigenstates of the symmetry-broken phase 
under consideration (e.g. $S^z_{ixx}$, $S^z_{izz}$, and $n_{i-}$ in the AFxx 
phase), plus 12 non-diagonal operators turning the eigenstates into one 
another (like $S^{+}_{ixx}$ and $S^{+}_{izz}$ in AFxx). Out of those twelve 
operators, six connect two excited states (like $S^{+}_{izz}$ in AFxx), 
and are physically irrelevant (at the RPA level), because they give only 
rise to 'ghost' modes, modes for which the spectral function vanishes 
identically. The remaining six operators connect the local ground state with 
an excited state, three of them describing an excitation and three a 
deexcitation, and only these six operators are physically relevant. Out of 
the three excitations (deexcitations), two are transverse, i.e., change the 
spin, and one is longitudinal, i.e., does not affect the spin. For a 
classical phase with $L$ sublattices one therefore has $4L$ transverse and 
$2L$ longitudinal operators per unit cell. Since the spin-orbital Hamiltonian
(\ref{somcu}) does not couple transverse and longitudinal operators, this 
yields also $4L$ transverse and $2L$ longitudinal modes. Because of 
time-reversal invariance they all occur in pairs with opposite frequencies, 
$\pm \omega^{(n)}_{\vec k}$. 

Finally, the $SU(2)$ spin invariance of the Hamiltonian guarantees that the 
transverse operators raising the spin are decoupled from those lowering the 
spin, and that they are described by the same set of equations of motion, so 
that the transverse modes are pairwise degenerate. Such a simplification does 
not occur in the longitudinal sector. So, in conclusion, in an $L$-sublattice 
phase there are $L$ doubly-degenerate positive-frequency transverse modes and 
$L$ nondegenerate positive-frequency longitudinal modes, accompanied by the 
same number of negative-frequency modes. This may be compared with the 
well-known situation in the HAF, where there is, with only spin operators 
involved, only one (not two) doubly-degenerate positive-frequency (transverse)
mode in the two-sublattice N\'{e}el state.

For the actual evaluation it is convenient to decompose the superexchange
terms in the spin-orbital Hamiltonian (\ref{somcu}),
\begin{equation}
{\cal H}_J={\cal H}_{\parallel}+{\cal H}_{\perp},
\label{somlong}
\end{equation}
into two parts which depend on the bond direction:

(i) for the bonds $\langle ij\rangle\parallel (a,b)$,
\begin{eqnarray}
\label{hpara}
{\cal H}_{\parallel}&=&\case{1}{4}J\sum_{\langle ij\rangle{\parallel}}
  \left[(1-\case{1}{2}\eta)
        (3{\vec S}_{ixx}+{\vec S}_{izz}+\lambda_{ij}\sqrt{3}{\vec K}_{i})
  \right.                                            \nonumber \\
 & &\left. \hskip 1.2cm
 \cdot (3{\vec S}_{jxx}+{\vec S}_{jzz}+\lambda_{ij}\sqrt{3}{\vec K}_{j})
  -2\eta {\vec S}_{i}\cdot {\vec S}_{j}
  \right.                                            \nonumber \\
 & &\left. \hskip .5cm +(1+2\eta)(n_{i-}+\lambda_{ij}\sqrt{3}T_{i})
                                 (n_{j-}+\lambda_{ij}\sqrt{3}T_{j})
  \right.                                            \nonumber \\
 & &\left. \hskip .5cm -(3+\eta)\right],
\end{eqnarray}
where $\lambda_{ij}=(-1)^{{\vec\delta}{\vec y}}$ with ${\vec y}$ being a unit
vector in the $b$-direction, and

(ii) for the bonds $\langle ij\rangle\perp (a,b)$, i.e., along the
$c$-axis,
\begin{eqnarray}
\label{hperp}
{\cal H}_{\perp}&=&J\sum_{\langle ij\rangle{\perp}}
  \left[(4-2\eta){\vec S}_{izz}\cdot{\vec S}_{jzz}
  -\eta ({\vec S}_{ixx}\cdot{\vec S}_{jzz}
  \right.                                            \nonumber \\
 &+&\left. {\vec S}_{izz}\cdot{\vec S}_{jxx})
  +(1+2\eta)n_{i-}n_{j-}-\case{1}{4}(3+\eta)\right].
\end{eqnarray}
Here and in the following Sections we consider a 3D model with $\beta=1$.
We note that the orbital interactions (\ref{orbop}) are quite different in
$H_{\parallel}$ and $H_{\perp}$; propagating spin-and-orbital
excitations are possible only within the $(a,b)$ planes, where they are
coupled to the spin excitations, while in the $c$-direction only pure
spin excitations and pure orbital excitations occur, which are decoupled
from one another. This breaking of symmetry between $H_{\parallel}$ and
$H_{\perp}$ is a consequence of the choice of basis as $|x\rangle$ and
$|z\rangle$ orbitals.

In the following Sections we consider transverse and longitudinal
excitations in the various symmetry-broken states. The transverse 
excitations, i.e., spin-waves and spin-and-orbital-waves, are calculated 
using the spin-changing operators which make a transition to a state 
realized in a classical phase at a given site $i$; for example for the 
AFxx phase these operators are for $i$ in the A (spin-up) sublattice,
\begin{equation}
S^+_{ixx}=d^{\dagger}_{ix\uparrow}d^{}_{ix\downarrow},    \hskip 1cm
K^+_{ixz}=d^{\dagger}_{ix\uparrow}d^{}_{iz\downarrow}.
\label{excopt}
\end{equation}
The longitudinal excitations without spin-flip are most conveniently
obtained starting from spin-dependent orbital excitation operators,
\begin{equation}
T_{ixz\sigma}=d^{\dagger}_{ix\sigma}d^{}_{iz\sigma},      \hskip 1cm
T_{izx\sigma}=d^{\dagger}_{iz\sigma}d^{}_{ix\sigma}.
\label{excop}
\end{equation}
The commutation relations for these operators are presented in Appendix
\ref{sec:commute}.

\subsection{Antiferromagnetic AFxx phase}
\label{sec:afxxrpa}

The nature and dispersion of elementary excitations in the spin-orbital model
(\ref{somcu}) can be conveniently studied in the leading order of the $1/S$
expansion using the Green function formalism. We note, however, that
equivalent results for the AFxx and AFzz phases can be obtained using 
instead an expansion around a classical saddle point with Schwinger 
bosons.\cite{Aue94}

We start from the equations of motion for the Green functions generated by
the excitation operators (\ref{excopt}) written in the energy
representation,\cite{Zub60,Hal72}
\begin{eqnarray}
\label{gfafxx1}
E\langle\langle     S_{ixx}^+|...\rangle\rangle &=&
{1\over 2\pi}\langle [S_{ixx}^+,...]\rangle +
 \langle\langle [S_{ixx}^+,H]|...\rangle\rangle,          \\
\label{gfafxx2}
E\langle\langle     K_{ixz}^+|...\rangle\rangle &=&
{1\over 2\pi}\langle [K_{ixz}^+,...]\rangle +
 \langle\langle [K_{ixz}^+,H]|...\rangle\rangle,
\end{eqnarray}
where the average of the commutator on the right hand side, e.g.
$\langle [S_{ixx}^+,S_{jxx}^-]\rangle$, is evaluated in the classical
ground state. The excitation operators were chosen as leading to the local
states $|ix\!\uparrow\rangle$ realized at one of the sublattices in the 
ground state of the AFxx phase. As usually, the commutators in Eqs. 
(\ref{gfafxx1}) and (\ref{gfafxx2}) generate higher-order Green functions. 
In contrast to the HAF, it does not suffice to consider the spin-flip
Green function $\langle\langle S_{ixx}^+|...\rangle\rangle$, as the
spin-flips may also occur together with an accompanying orbital-flip,
as described by $\langle\langle K_{ixz}^+|...\rangle\rangle$.

We derived the equations of motion for the Green functions generated by
the set of operators $\{S_{ixx}^+,K_{ixz}^+,S_{jxx}^+,K_{jxz}^+\}$, where 
$i\in A$ and $j\in B$, and used the random-phase approximation (RPA) for 
spinlike operators which linearizes the equations of motion by a decoupling 
procedure.\cite{Zub60,Hal72} Thereby, the operators which have nonzero 
expectation values in the considered classical state give finite 
contributions, e.g. for the first spin-flip Green function one uses
\begin{equation}
\langle\langle S_{ixx}^+S_{mxx}^z|...\rangle\rangle\simeq
\langle S_{mxx}^z\rangle \langle\langle S_{ixx}^+|...\rangle\rangle,
\label{rpas}
\end{equation}
and a similar formula for the mixed spin-and-orbital excitation described by
$\langle\langle K_{ixz}^+|...\rangle\rangle$,
\begin{equation}
\langle\langle K_{ixz}^+S_{mxx}^z|...\rangle\rangle\simeq
\langle S_{mxx}^z\rangle \langle\langle K_{ixz}^+|...\rangle\rangle .
\label{rpak}
\end{equation}
It is crucial that the decoupled operators have different site indices, and
thus the decoupling procedure preserves the local commutation rules given
in Appendix \ref{sec:commute}. Instead, if one uses products of spin and
orbital operators, e.g., $K_{ixz}^+=S_{ixx}^+\sigma_i^+$, one is tempted
to decouple these operators locally\cite{Cas78,Kha97} which would violate
the algebraic structure of the $so(4)$ Lie algebra.

In the present case of the AFxx phase one uses the respective N\'eel state
average values,
\begin{eqnarray}
\langle S_{ixx}^z\rangle&=&-\langle S_{jxx}^z\rangle=\case{1}{2}, \\
   \langle n_{i-}\rangle&=& \langle n_{j-}\rangle=\case{1}{2},
\label{avx}
\end{eqnarray}
where $i\in A$ and $j\in B$, and $A$ and $B$ are the two sublattices in a 2D
lattice for the AFxx phase. All the remaining averages vanish, as this phase
has a pure $|x\rangle$-orbital character at every site, which simplifies
significantly the equations of motion which result from the RPA procedure.

The translational invariance of the N\'eel state implies that the
transformed Green functions are diagonal in the reduced Brillouin zone (BZ).
As in the HAF, the Fourier transformed functions are defined for the Green
functions which describe the spin dynamics on a given sublattice, either $A$
or $B$. For instance, the pure spin-flip Green functions are transformed as
follows,
\begin{eqnarray}
\langle\langle S_{{\vec k}xx}^+|...\rangle\rangle_A&=&
\frac{1}{\sqrt{N}}\sum_{i\in A}e^{i{\vec k}{\vec R}_i}
\langle\langle S_{ixx}^+|...\rangle\rangle_A,  \nonumber \\
\langle\langle S_{{\vec k}xx}^+|...\rangle\rangle_B&=&
\frac{1}{\sqrt{N}}\sum_{j\in B}e^{i{\vec k}{\vec R}_j}
\langle\langle S_{jxx}^+|...\rangle\rangle_B,
\label{fourier}
\end{eqnarray}
where $N$ is the number of sites in one sublattice. Hence, the problem of
finding the elementary excitations of the considered spin-orbital model
(\ref{somcu}) reduces to the diagonalization of a $4\times 4$ dynamical
matrix at each ${\vec k}$-point, as given in Appendix \ref{sec:dynama}.

The symmetric positive and negative eigenvalues
$\pm\omega_{\vec k}^{(n)}$, with $n=1,2$, solved from the matrix in Eq.
(\ref{gfeq}) may be written in the following form for the AFxx phase,
\begin{eqnarray}
\label{afsw}
[\omega_{\vec k}^{(n)}]^2&=&J^2\left(\lambda_{x}^2+\tau_{x}^2
-Q_{x\vec k}^2-R_{\vec k}^2-2P_{x\vec k}^2\right)
                                                           \nonumber \\
&\pm &J^2\left[ (\lambda_{x}^2-\tau_{x}^2)^2
-2(\lambda_{x}^2-\tau_{x}^2)(Q_{x\vec k}^2-R_{\vec k}^2) \right.
                                                           \nonumber \\
&-&\left. 4(\lambda_{x}-\tau_{x})^2P_{x\vec k}^2
+(Q_{x\vec k}^2+R_{\vec k}^2+2P_{x\vec k}^2)^2 \right.
                                                           \nonumber \\
&-&\left. 4(Q_{x\vec k}R_{\vec k}-P_{x\vec k}^2)^2\right]^{1/2}.
\end{eqnarray}
Here the quantities $\lambda_{\alpha}$ and $\tau_{\alpha}$ play the role
of local potentials and follow from the model parameters, $E_z$ and $J_H$,
\begin{eqnarray}
\label{afxxlambda}
\lambda_x&=&\case{9}{2}-3\eta,                               \\
\label{afxxtau}
\tau_x&=&\case{7}{2}-4\eta-2-\eta+\varepsilon_z.
\end{eqnarray}
The remaining terms are ${\vec k}$-dependent, and depend on
\begin{eqnarray}
\label{gammap}
\gamma_{+}(\vec k)&=&\case{1}{2}(\cos k_x+\cos k_y),                 \\
\label{gammam}
\gamma_{-}(\vec k)&=&\case{1}{2}(\cos k_x-\cos k_y),                 \\
\label{gammaz}
\gamma_{z}(\vec k)&=&\cos k_z.
\end{eqnarray}
The quantities $Q_{x\vec k}$ and $P_{x\vec k}$ for the AFxx phase take the
form,
\begin{eqnarray}
\label{afxxq}
Q_{x\vec k}&=&(\case{9}{2}-3\eta)\gamma_{+}(\vec k),              \\
\label{afxxp}
P_{x\vec k}&=&\case{1}{2}\sqrt{3}(3-\eta)\gamma_{-}(\vec k),
\end{eqnarray}
while the last dispersive term,
\begin{equation}
R_{\vec k}=\case{3}{2}\gamma_{+}(\vec k) ,
\label{rdef2}
\end{equation}
carries no index and remains identical for both AF phases (AFxx and AFzz). We 
emphasize that the coupling between the spin-wave and spin-and-orbital-wave
excitations occurs due to the terms $\propto P_{x\vec k}$, as seen from Eq.
(\ref{gfeq}). It vanishes in the planes of $k_x=\pm k_y$, but otherwise plays
an important role, as discussed in Sec. \ref{sec:rpa}. In the limit of large
$E_z\to\infty$, Eq. (\ref{afsw}) reproduces the spin-wave excitations for a 
2D antiferromagnet with an AF superexchange interaction of
$J(\frac{9}{4}-\frac{3}{2}\eta)$,
\begin{equation}
\label{limitafxx}
\omega_{\vec k}^{(1)}=J\left(\case{9}{2}-3\eta\right)
                       [1-\gamma_{+}^2(\vec k)]^{1/2},
\end{equation}
while the dispersion of the high-energy spin-and-orbital excitation,
$\omega_{\vec k}^{(2)}\simeq E_z$, becomes negligible.
As explained above, both modes are doubly degenerate.

Consider now the orbital (excitonic) excitations generated by the orbital-flip
operators (\ref{excop}). They are found by considering the equations of motion,
\begin{eqnarray}
\label{gfafxxl1}
E\langle\langle T_{i\alpha\beta \uparrow  }|...\rangle\rangle &=&
{1\over 2\pi}\langle [T_{i\alpha\beta \uparrow  },...]\rangle    
+ \langle\langle [T_{i\alpha\beta \uparrow  },H]|...\rangle\rangle, 
        \nonumber  \\ \\
\label{gfafxxl2}
E\langle\langle T_{i\alpha\beta \downarrow}|...\rangle\rangle &=&
{1\over 2\pi}\langle [T_{i\alpha\beta \downarrow},...]\rangle    
+  \langle\langle [T_{i\alpha\beta \downarrow},H]|...\rangle\rangle, 
        \nonumber  \\
\end{eqnarray}
and the commutators are calculated using the rules (\ref{excom}).
In general, one finds four different excitation operators at each site.
However, making a Fourier transformations as for the transverse operators
(\ref{fourier}), one may show that only two operators per sublattice suffice
to describe the modes in an antiferromagnet. The structure of the respective
RPA dynamical matrix is given in Appendix \ref{sec:dynama}. The orbital
excitations which follow from Eq. (\ref{gfeqor}) are in general given by
\begin{equation}
\label{genorb}
\zeta_{\vec k}= J \; 
\left[u_{\alpha}(u_{\alpha}\pm 2 \rho_{\alpha\vec k})\right]^{1/2},
\end{equation}
yielding two, in general nondegenerate, positive-frequency modes.
In the AFxx phase one finds,
\begin{eqnarray}
\label{afxxu}
u_x&=&\varepsilon_z-3\eta,                                \\
\label{afxxrho}
\rho_{x\vec k}&=&\case{3}{2}\eta\gamma_{+}(\vec k).
\end{eqnarray}

It is important to realize that the propagation of longitudinal excitations,
being equivalent to a finite dispersion of longitudinal modes, becomes
possible only at $\eta>0$. This follows from the multiplet structure of the
excited $d^8$ states, which allows a spin-flip between the orbitals in the
$|^1E_{\theta}\rangle$ and in the $S^z=0$ component of the
$|^3A_2\rangle$-state
only if $J_H\neq 0$, as illustrated in Fig. \ref{orbex}. The processes
$\sim t_{xz}$ are not shown, as they would lead to a final state shown in
Fig. \ref{orbex}(b), i.e., to a propagation of a spin-and-orbital excitation
which was already considered above. In contrast, the relevant longitudinal
orbital excitation in the symmetry-broken state implies that the exciton
has the same spin as imposed by the N\'eel state of the background; this
state is shown in Fig. \ref{orbex}(c). Therefore, in a perfect N\'eel state
without FM interactions due to $\eta\neq 0$, only local orbital
excitations are possible. These local excitations cost no energy in the limit
of $\varepsilon_z\to 0$ which demonstrates again the frustration of magnetic
interactions at the classical degeneracy point, $\varepsilon_z=\eta=0$.

An example of the excitation spectra is shown in Fig. \ref{modesxx} for the
main directions in the 2D BZ, with $X=(\pi,0)$ and $S=(\pi/2,\pi/2)$.
Near the $\Gamma$ point one finds a (doubly-degenerate) Goldstone mode
$\omega_{\vec k}^{(1)}$ with dispersion $\sim k$ at ${\vec k}\to 0$, as in 
the HAF, and a second (doubly-degenerate) transverse mode at higher energy,
$\omega_{\vec k}^{(2)}\simeq \omega_0+ak^2$.
Near $\Gamma$ the Goldstone mode is essentially purely spin-wave, the
second mode purely spin-and-orbital wave. With increasing ${\vec k}$ these
modes start to mix due to the $P_{x{\vec k}}$ term along the $\Gamma-X$
direction. This is best illustrated by the intensity measured in the
neutron scattering experiments, which see only the spin-wave component
in each transverse mode, as explained in more detail in Appendix
\ref{sec:neutrons}. The intensity $\chi({\vec q})$ moves from one mode
to the other along the $\Gamma-X$ direction in the 2D BZ
(Fig. \ref{modesxx}), demonstrating that indeed the lowest (highest)
mode is predominantly spin-wave-like (spin-and-orbital-wave-like) before
the anticrossing point, while this is reversed after the anticrossing of
the two modes. Thus, we make here a specific prediction that {\it two
spin-wave-like modes could be measurable in certain parts of the 2D
BZ}, in particular in the vicinity of an anticrossing,
if only an AFxx phase was realized for parameters not too distant
from the classical degeneracy point. Unfortunately, for the realistic
parameters for the cuprates,\cite{Gra92} one finds $E_z/J\simeq 10$ which
makes the spin-and-orbital excitation and the changes of the spin-wave
dispersion hardly visible in neutron spectroscopy.

The orbital (longitudinal) excitations are found for the parameters of Fig.
\ref{modesxx} at a finite energy, being of the same order of magnitude as
the energy of the spin-and-orbital excitation, $\omega^{(2)}_{\vec k}$. The
weak dispersion of these modes follows from the spin-flip processes in the
{\em excited\/} states, as explained in Fig. \ref{orbex} and discussed above.
We emphasize that the orbital mode has a gap and {\it does not couple\/} to
any spin excitation. At the classical degeneracy point $M$ the orbital mode
falls to zero energy and is dispersionless, expressing that the orbital
can be changed locally without any cost in energy.

\subsection{Antiferromagnetic AFzz phase}
\label{sec:afzzrpa}

The transverse excitations in the AFzz phase are determined by considering the
complementary set of Green functions to that given in Eqs. (\ref{gfafxx1})
and (\ref{gfafxx2}),
\begin{eqnarray}
\label{gfafzz1}
E\langle\langle     S_{izz}^+|...\rangle\rangle &=&
{1\over 2\pi}\langle [S_{izz}^+,...]\rangle +
 \langle\langle [S_{izz}^+,H]|...\rangle\rangle,                 \\
\label{gfafzz2}
E\langle\langle     K_{izx}^+|...\rangle\rangle &=&
{1\over 2\pi}\langle [K_{izx}^+,...]\rangle +
 \langle\langle [K_{izx}^+,H]|...\rangle\rangle,
\end{eqnarray}
with the excitations to the local $|iz\uparrow\rangle$ states.
As usually, the average of the commutator on the right hand side is next
evaluated in the classical ground state. After obtaining the RPA equations,
we thus use the following nonvanishing averages,
\begin{eqnarray}
\langle S_{izz}^z\rangle&=&-\langle S_{jzz}^z\rangle=\case{1}{2}, \\
   \langle n_{i-}\rangle&=& \langle n_{j-}\rangle=-\case{1}{2},
\label{avz}
\end{eqnarray}
in the AFzz phase. This leads again to the general form (\ref{gfeq}), with
all the elements except for $R_{\vec k}$ replaced by,
\begin{eqnarray}
\label{afzzlambda}
\lambda_z&=&\case{1}{2}-\eta+2(2-\eta),                                  \\
\label{afzztau}
\tau_z&=&-\case{1}{2}-\eta+2(1-2\eta)-\varepsilon_z,                     \\
\label{afzzq}
Q_{z\vec k}&=&(\case{1}{2}-\eta)\gamma_{+}(\vec k)
            +2(2-\eta)\gamma_{z}(\vec k), \\
\label{afzzp}
P_{z\vec k}&=&\case{1}{2}\sqrt{3}(1-\eta)\gamma_{-}(\vec k) .
\end{eqnarray}
Thus, the transverse excitations have the same form (\ref{afsw}) as in the
AFxx phase, but the above quantities (\ref{afzzlambda})--(\ref{afzzp}) have
to be used.

In the limit of large $E_z\to-\infty$ one finds the spin-wave
for a 3D anisotropic antiferromagnet with strong superexchange equal to
$2J(2-\eta)$ along the $c$-axis, and weak superexchange
$\frac{1}{4}J(1-2\eta)$ within the $(a,b)$-planes,
\begin{eqnarray}
\label{limitafzz}
\omega_{\vec k}^{(1)}&=&J\left\{
       \left[(\case{1}{2}-\eta)+2(2-\eta)\right]^2\right.  \nonumber \\
   & &-\left.\left[(\case{1}{2}-\eta)\gamma_{+}(\vec k)
            +2(2-\eta)\gamma_{z}\right]^2\right\}^{1/2},
\end{eqnarray}
while the spin-and-orbital excitation, $\omega_{\vec k}^{(2)}\simeq -E_z$,
is dispersionless. Again, both these transverse modes are doubly degenerate.
The orbital excitations in the AFzz phase are found using
the equations of motion of the form (\ref{gfafxxl1}) and (\ref{gfafxxl2})
which lead to Eq. (\ref{genorb}) with,
\begin{eqnarray}
\label{afzzu}
u_z&=&-\varepsilon_z-3\eta,                                         \\
\label{afzzrho}
\rho_{z,\vec k}&=&-\case{3}{2}\eta\gamma_{+}(\vec k),
\end{eqnarray}
and we find again zero-energy nondispersive modes at $\varepsilon_z=\eta=0$.

The representative excitation spectrum for the AFzz phase is shown in Fig.
\ref{modeszz}. We use the 3D BZ for a $bcc$ lattice with the
standard notation: $W=(\pi,\pi/2,0)$, $L=(\pi/2,\pi/2,\pi/2)$ and
$K=(3\pi/4,3\pi/4,0)$. The transverse modes have qualitatively the same
behavior as in the 2D AFxx phase, and one finds a Goldstone mode
$\omega_{\vec k}^{(1)}$ at the $\Gamma$ point which is spin-wave-like,
accompanied by a finite energy spin-and-orbital mode $\omega_{\vec k}^{(2)}$.
The first one is linear, while the second changes quadratically with
increasing ${\vec k}$. The dispersion in the $\Gamma-X$ direction is, however,
only $\sim 0.7J$, while in the AFxx phase a large dispersion of $\sim 2.5J$
was found (Fig. \ref{modesxx}). This demonstrates the very large difference
between the superexchange in the $(a,b)$-planes in the two AF phases.

Here one should bear in mind, that in a strongly anisotropic antiferromagnet,
such as the AFzz phase, the dispersion of the spin-wave mode in the
$(k_x,k_y)$ plane is roughly $(2 J_{ab} J_c)^{1/2}$, so actually enhanced by
$(J_c/2J_{ab})^{1/2}$ compared with the planar exchange constant. In fact,
there is also strong mixing between spin wave and spin-and-orbital wave along
$\Gamma-X$, depressing at the $X$-point $\omega^{(1)}_X$ by no less than
$0.5 J$ from its pure spin-wave value. The mixing effect is also visible in
the relatively large neutron intensity of the second mode. By contrast,
the transverse excitations are rather pure all along the $W-L$ direction
[where the neutron intensity $\chi({\vec q})$ is larger], except in the
regime where $\omega^{(1)}_{\vec k}\simeq\omega^{(2)}_{\vec k}$ and the
neutron intensity is distributed between the modes. However, owing to the
abruptness of the anticrossing, the range where the modes have
simultaneously appreciable intensity is very narrow, and their energetic
proximity then makes it likely that they would be measured as a single
broad maximum.

The (longitudinal) orbital excitation is found at the $X$ and $L$ points
at the same energy as that of a {\it local\/} excitation from $|z\rangle$
to $|x\rangle$ orbital (see Fig. \ref{modeszz}). It depends only on the
energy difference between the orbitals, and has a weak dispersion by the
same mechanism as described above for the AFxx phase (Fig. \ref{orbex}).

\end{multicols} 
\widetext

\subsection{Mixed-orbital FFA phase}
\label{sec:moffarpa}

The excitation operators which couple to the local states in a symmetry-broken
phase with mixed orbitals are linear combinations of the operators considered
in Secs. \ref{sec:afxxrpa} and \ref{sec:afzzrpa}. It is therefore convenient
to make a unitary transformation of the Hamiltonian (\ref{somcu}) to new
orbitals defined as follows for $i\in A$ or $i\in D$ sublattice,
\begin{equation}
\label{newstatei}
\left( \begin{array}{c}
 |i\mu\rangle   \\
 |i\nu\rangle
\end{array} \right) =
\left(\begin{array}{cc}
 \cos\theta &  \sin\theta \\
-\sin\theta &  \cos\theta
\end{array} \right)
\left( \begin{array}{c}
 |iz\rangle   \\
 |ix\rangle
\end{array} \right) ,
\end{equation}
and for $j\in B$ or $j\in C$ sublattice,
\begin{equation}
\label{newstatej}
\left( \begin{array}{c}
 |j\mu\rangle   \\
 |j\nu\rangle
\end{array} \right) =
\left(\begin{array}{cc}
 \cos\theta & -\sin\theta \\
 \sin\theta &  \cos\theta
\end{array} \right)
\left( \begin{array}{c}
 |jz\rangle   \\
 |jx\rangle
\end{array} \right) .
\end{equation}
With these definitions and by choosing the angle $\theta$ at the value which 
minimizes the classical energy (\ref{thetamoffa}), we guarantee that 
$|i\mu\rangle$ and $|j\mu\rangle$, respectively, are at each site the orbital 
state realized in the classical MO{\scriptsize FFA} phase, which is G-type 
with respect to the orbital ordering, while $|i\nu\rangle$ and $|j\nu\rangle$ 
are the excited state, so that one can readily define the excitation operators
pertinent to the symmetry-broken ground state of this phase. Thus the spin, 
spin-and-orbital, and orbital operators in terms of the new orbital states 
$\{|\mu\rangle,|\nu\rangle\}$ defined by Eqs. (\ref{newstatei}) and 
(\ref{newstatej}) are
\begin{eqnarray}
\label{newspin1}
{\cal K}_{i\alpha\beta }^+&=&
      |i\alpha  \uparrow\rangle\langle i\beta \downarrow|  ,        \\
\label{newspin2}
{\cal K}_{i\alpha\beta }^z&=&\case{1}{2}
     (|i\alpha  \uparrow\rangle\langle i\beta   \uparrow|
     -|i\alpha\downarrow\rangle\langle i\beta \downarrow|) ,        \\
\label{newspin3}
{\cal T}_{i-           }  &=&\case{1}{2}\sum_{\sigma}
     (|i\mu       \sigma\rangle\langle i\nu       \sigma|
     +|i\nu       \sigma\rangle\langle i\mu       \sigma|) ,        \\
\label{newspin4}
{\cal N}_{i-           }  &=&\case{1}{2}\sum_{\sigma}
     (|i\mu       \sigma\rangle\langle i\mu       \sigma|
     -|i\nu       \sigma\rangle\langle i\nu       \sigma|) .
\end{eqnarray}
The new operators: ${\vec {\cal K}}_{i\alpha\beta}$, ${\cal T}_i$ and
${\cal N}_{i-}$ fulfill the same commutation rules as the nontransformed
operators: ${\vec K}_{i\alpha\beta }$, $T_i$, and $n_{i-}$, respectively;
they are given in Appendix \ref{sec:commute}. To simplify the notation we
also introduce total spin and spin-and-orbital operators,
\begin{eqnarray}
\label{newsping1}
{\vec {\cal S}}_{i}&=&{\vec {\cal S}}_{i\mu\mu}+{\vec {\cal S}}_{i\nu\nu}, \\
\label{newsping2}
{\vec {\cal K}}_{i}&=&{\vec {\cal K}}_{i\mu\nu}+{\vec {\cal K}}_{i\nu\mu}.
\end{eqnarray}

The Hamiltonian (\ref{somcu}) has to be transformed by the inverse
transformations to those given by Eqs. (\ref{newstatei}) and
(\ref{newstatej}). For the bonds $\langle ij\rangle\parallel (a,b)$ with
$i\in A$ and $j\in B$ one finds,
\begin{eqnarray}
\label{hparalong}
{\cal H}_{\parallel}&=&\case{1}{4}J\sum_{\langle ij\rangle{\parallel}}
  \left\{(1-\case{1}{2}\eta)\left[
        ((2-\cos 2\theta){\vec {\cal S}}_{i\mu\mu}
        +(2+\cos 2\theta){\vec {\cal S}}_{i\nu\nu}
        +\sin 2\theta{\vec {\cal K}}_i)
  \right.\right.                                            \nonumber \\
 & &\left.\left.\hskip 2.5cm
  \times((2-\cos 2\theta){\vec {\cal S}}_{j\mu\mu}
        +(2+\cos 2\theta){\vec {\cal S}}_{j\nu\nu}
        -\sin 2\theta{\vec {\cal K}}_i)
  \right.\right.                                            \nonumber \\
 & &\left.\left.\hskip 1.2cm
       +3(\sin 2\theta({\vec {\cal S}}_{i\mu\mu}-{\vec {\cal S}}_{i\nu\nu})
         +\cos 2\theta {\vec {\cal K}}_{i})
         (\sin 2\theta({\vec {\cal S}}_{j\mu\mu}-{\vec {\cal S}}_{j\nu\nu})
         +\cos 2\theta {\vec {\cal K}}_{j})
  \right.\right.                                            \nonumber \\
 &+&\left.\left. \!\! \lambda_{ij}\sqrt{3}\left(
        ((2-\cos 2\theta){\vec {\cal S}}_{i\mu\mu}
        +(2+\cos 2\theta){\vec {\cal S}}_{i\nu\nu}
        +\sin 2\theta{\vec {\cal K}}_i)
     (\sin 2\theta({\vec {\cal S}}_{j\mu\mu}\! -\! {\vec {\cal S}}_{j\nu\nu})
        -\cos 2\theta{\vec {\cal K}}_j)
  \right.\right.\right.                                     \nonumber \\
 & &\left.\left.\left.\hskip 0.4cm
       -(\sin 2\theta({\vec {\cal S}}_{i\mu\mu}-{\vec {\cal S}}_{i\nu\nu})
        +\cos 2\theta{\vec {\cal K}}_i)
        ((2-\cos 2\theta){\vec {\cal S}}_{j\mu\mu}
        +(2+\cos 2\theta){\vec {\cal S}}_{j\nu\nu}
        -\sin 2\theta{\vec {\cal K}}_j)\right)\right]
  \right.                                                   \nonumber \\
 &+&\left.  \case{1}{2}\eta\left[
        (\cos 2\theta({\vec {\cal S}}_{i\mu\mu}-{\vec {\cal S}}_{i\nu\nu})
        -\sin 2\theta{\vec {\cal K}}_i)
        (\cos 2\theta({\vec {\cal S}}_{j\mu\mu}-{\vec {\cal S}}_{j\nu\nu})
        +\sin 2\theta{\vec {\cal K}}_j)
  \right.\right.                                            \nonumber \\
 & &\left.\left.\hskip .3cm
      -3(\sin 2\theta({\vec {\cal S}}_{i\mu\mu}-{\vec {\cal S}}_{i\nu\nu})
        +\cos 2\theta{\vec {\cal K}}_i)
        (\sin 2\theta({\vec {\cal S}}_{j\mu\mu}-{\vec {\cal S}}_{j\nu\nu})
        -\cos 2\theta{\vec {\cal K}}_j)
  \right.\right.                                            \nonumber \\
 & &\left.\left. - \lambda_{ij}\sqrt{3}\left(
        (\cos 2\theta({\vec {\cal S}}_{i\mu\mu}-{\vec {\cal S}}_{i\nu\nu})
        -\sin 2\theta{\vec {\cal K}}_i)
        (\sin 2\theta({\vec {\cal S}}_{j\mu\mu}-{\vec {\cal S}}_{j\nu\nu})
        -\cos 2\theta{\vec {\cal K}}_j)
  \right.\right.\right.                                     \nonumber \\
 & &\left.\left.\left.\hskip 1.2cm
       +(\sin 2\theta({\vec {\cal S}}_{i\mu\mu}-{\vec {\cal S}}_{i\nu\nu})
        +\cos 2\theta{\vec {\cal K}}_i)
        (\cos 2\theta({\vec {\cal S}}_{j\mu\mu}-{\vec {\cal S}}_{j\nu\nu})
        -\sin 2\theta{\vec {\cal K}}_j)\right)\right]
  - 2\eta {\vec {\cal S}}_{i}{\vec {\cal S}}_{j}
  \right.                                                   \nonumber \\
 &+&\left. (1+2\eta)\left[
        (\cos 2\theta{\vec {\cal N}}_i-\sin 2\theta{\vec {\cal T}}_i)
        (\cos 2\theta{\vec {\cal N}}_j+\sin 2\theta{\vec {\cal T}}_j)
  \right.\right.                                            \nonumber \\
 & &\left.\left.    \hskip 1.0cm
      -3(\sin 2\theta{\vec {\cal N}}_i+\cos 2\theta{\vec {\cal T}}_i)
        (\sin 2\theta{\vec {\cal N}}_j-\cos 2\theta{\vec {\cal T}}_j)
  \right.\right.                                            \nonumber \\
 & &\left.\left. -\lambda_{ij}\sqrt{3}\left(
        (\cos 2\theta{\vec {\cal N}}_i-\sin 2\theta{\vec {\cal T}}_i)
        (\sin 2\theta{\vec {\cal N}}_j-\cos 2\theta{\vec {\cal T}}_j)
  \right.\right.\right.                                     \nonumber \\
 & &\left.\left.\left.    \hskip 1.0cm
       +(\sin 2\theta{\vec {\cal N}}_i+\cos 2\theta{\vec {\cal T}}_i)
        (\cos 2\theta{\vec {\cal N}}_j+\sin 2\theta{\vec {\cal T}}_j)
         \right)\right]-\left(3+\eta\right)\right\},
\end{eqnarray}
while for the bonds $\langle ij\rangle\perp (a,b)$ it takes the form
\begin{eqnarray}
\label{hperplong}
{\cal H}_{\perp}\!&=\!&J\!\sum_{\langle ij\rangle{\perp}}
   \! \left\{(1-\case{1}{2}\eta)
        ((1+\cos 2\theta){\vec {\cal S}}_{i\mu\mu}
        +(1-\cos 2\theta){\vec {\cal S}}_{i\nu\nu}
        -\sin 2\theta{\vec {\cal K}}_i)                     
 \right.                                                   \nonumber \\
 & &\left. \hskip 2.1cm \times ((1+\cos 2\theta){\vec {\cal S}}_{j\mu\mu}
        +(1-\cos 2\theta){\vec {\cal S}}_{j\nu\nu}
        -\sin 2\theta{\vec {\cal K}}_j)
 \right.                                                   \nonumber \\
 &-&\left. \case{1}{4}\eta \left[
        ((1-\cos 2\theta){\vec {\cal S}}_{i\mu\mu}
        +(1+\cos 2\theta){\vec {\cal S}}_{i\nu\nu}
        +\sin 2\theta{\vec {\cal K}}_i)                     
 \right.\right.                                             \nonumber \\
 & &\left.\left. \hskip .5cm \times ((1+\cos 2\theta){\vec {\cal S}}_{j\mu\mu}
        +(1-\cos 2\theta){\vec {\cal S}}_{j\nu\nu}
        -\sin 2\theta{\vec {\cal K}}_j)
  \right.\right.                                            \nonumber \\
 & &\left.\left. \hskip .5cm
       +((1+\cos 2\theta){\vec {\cal S}}_{i\mu\mu}
        +(1-\cos 2\theta){\vec {\cal S}}_{i\nu\nu}
        -\sin 2\theta{\vec {\cal K}}_i)                     
 \right.\right.                                             \nonumber \\
 & &\left.\left. \hskip .5cm \times ((1-\cos 2\theta){\vec {\cal S}}_{j\mu\mu}
        +(1+\cos 2\theta){\vec {\cal S}}_{j\nu\nu}
        +\sin 2\theta{\vec {\cal K}}_j)\right]
  \right.                                                   \nonumber \\
 &+&\left. (1+2\eta)
        (\cos 2\theta{\vec {\cal N}}_i-\sin 2\theta{\vec {\cal T}}_i)
        (\cos 2\theta{\vec {\cal N}}_j-\sin 2\theta{\vec {\cal T}}_j)
        -\case{1}{4}(3+\eta)\right\}.
\end{eqnarray}
Finally, the transformed orbital-anisotropy term reads
\begin{equation}
\label{kk3long}
{\cal H}_{\tau} = E_z \sum_i
        (\cos 2\theta{\vec {\cal N}}_i-\sin 2\theta{\vec {\cal T}}_i).
\end{equation}

\begin{multicols}{2}

The transverse excitations may be found starting from the relevant raising 
operators that lead to the local state $|i\mu\uparrow\rangle$ realized in 
one of the sublattices, analogous to those introduced for the AFxx phase 
(\ref{excopt}), i.e., the set
$\{{\cal S}_{i\mu\mu}^+,{\cal K}_{i\mu\nu}^+,
{\cal S}_{j\mu\mu}^+,{\cal K}_{j\mu\nu}^+,
{\cal S}_{k\mu\mu}^+,{\cal K}_{k\mu\nu}^+,
{\cal S}_{l\mu\mu}^+,{\cal K}_{l\mu\nu}^+\}$, where
$i\in A$, $j\in B$, $k\in C$, and $l\in D$;
they lead as usual to the orbitals $\{|i\mu\rangle, |j\mu\rangle\}$
(\ref{orbmoffa}) realized in the MO{\scriptsize FFA} phase,
\begin{eqnarray}
\label{gfmoffa1}
E\langle\langle {\cal S}_{i\mu\mu}^+|...\rangle\rangle &=&
{1\over 2\pi}\langle [{\cal S}_{i\mu\mu}^+,...]\rangle +
 \langle\langle [{\cal S}_{i\mu\mu}^+,H]|...\rangle\rangle,         \\
\label{gfmoffa2}
E\langle\langle {\cal K}_{i\mu\nu}^+|...\rangle\rangle &=&
{1\over 2\pi}\langle [{\cal K}_{i\mu\nu}^+,...]\rangle +
 \langle\langle [{\cal K}_{i\mu\nu}^+,H]|...\rangle\rangle.
\end{eqnarray}
We applied the same RPA procedure as in Secs. \ref{sec:afxxrpa} and
\ref{sec:afzzrpa} in order to determine the Green function equations in the
${\vec k}$-space. The longitudinal excitations can be obtained from
operators similar to those used in the AFxx and AFzz phases (\ref{excop}),
\begin{equation}
\label{excop1}
{\cal T}_{i\mu\nu\uparrow}=
 d^{\dagger}_{i\mu\uparrow}d^{}_{i\nu\uparrow}, \hskip 1.0cm
{\cal T}_{i\nu\mu\uparrow}=
 d^{\dagger}_{i\nu\uparrow}d^{}_{i\mu\uparrow},
\end{equation}
for the $(a,b)$ planes with the $\uparrow$-spins, and the corresponding
${\cal T}_{i\mu\nu\downarrow}$ and ${\cal T}_{i\nu\mu\downarrow}$ for the
$(a,b)$ planes with the $\downarrow$-spins. The commutation operators for
these operators are analogous to those presented in Appendix
\ref{sec:commute} and may be easily obtained. The resulting dynamical
matrices for both transverse and longitudinal excitations are given in
Appendix \ref{sec:dynama}; their numerical diagonalization gave the modes
presented below. There are four doubly-degenerate positive-frequency 
transverse modes, and four non-degenerate positive-frequency longitudinal 
modes, consistent with the MO{\scriptsize FFA} phase having four sublattices.

An example of the transverse and longitudinal modes in the
MO{\scriptsize FFA} phase is presented in Fig. \ref{modesffa}. The modes are
shown in the respective BZ which corresponds to the magnetic unit cell of
the MO{\scriptsize FFA} phase: The 2D part along $\Gamma-X-S-\Gamma$ is
identical with the AFxx phase (compare Fig. \ref{modesxx}),
reflecting the orbital alternation, while the AF coupling along the $c$-axis
results in the folding of the zone along the $\Gamma-Z$ direction, with
$Z'=(0,0,\pi/2)$ and $S'=(\pi/2,\pi/2,\pi/2)$. One finds one Goldstone mode,
and three other finite-energy modes at the $\Gamma$ point. If no AF coupling
along the $c$-axis is present, similar positive-energy modes describe the
excitation spectrum in the MO{\scriptsize FF} phase in the 2D part of the
BZ (in the region of stability shown in Fig. \ref{mfa2d}), and the
symmetric negative-frequency modes carry then no weight. In contrast, due to
the strong AF interactions in the MO{\scriptsize FFA} phase, the negative
modes give a large energy renormalization due to quantum fluctuations, as
discussed in more detail in Sec. \ref{sec:rpa}.

The spin-wave and spin-and-orbital-wave excitations are well separated along
the $\Gamma-X-S-\Gamma$ path, with a gap of $\sim 0.5J$, as the FM
interactions $\propto J\eta$ are considerably weaker than the orbital
interactions which are $\propto J$. Therefore, the neutron intensity
$\chi({\vec q})$ is found mainly as originating from the lowest energy mode,
$\omega_{\vec k}^{(1)}$, with a small admixture of the higher-energy
spin-and-orbital excitation, $\omega_{\vec k}^{(3)}$. The magnetic
interactions are considerably stronger along the $c$-axis; the modes mix and
the higher-energy excitations, $\omega_{\vec k}^{(n)}$ with $n=3,4$, have
a larger dispersion in the remaining directions with $k_z\neq 0$. Strong
mixing of the modes in this part of the BZ is also visible in the intensity 
distribution, with the modes $n=1$ and $n=3$ contributing with comparable 
intensities (Fig. \ref{modesffa}). The fact that modes labelled as 2 and 4
have zero intensity is due to the path $\Gamma-Z'-S'-\Gamma$ being in the
high-symmetry BZ plane where $k_x=k_y$ so that $\gamma_{-}(\vec{k})=0$.
Then modes 2 and 4 have equal amplitude but are exactly out-of-phase between
$A$ and $B$ sites as well as between $C$ and $D$ sites, and so their neutron
intensities vanish, and only the companion in-phase modes 1 and 3 are
observable by neutrons. Unfortunately, no experimental verification
of these spectra is possible at present, as the spin excitations
measured in neutron scattering for KCuF$_3$ are consistent with the Bethe
ansatz and thus suggest a spin-liquid ground state with strong 1D AF
correlations instead of the A-AF phase with magnetic LRO.\cite{Ten93}

Interestingly, although the order in the $(a,b)$ planes is FM, the energy
of the Goldstone mode increases {\it linearly in all three directions\/} with
increasing ${\vec k}$, and the slopes are proportional to the respective
exchange interactions. This behavior is a manifestation of the A-AF spin
order; a qualitatively similar spectrum is found experimentally in
LaMnO$_3$,\cite{Hir96} where, however, the excitation spectra describe large
spins $S=2$ of Mn$^{3+}$ ions. The rather small dispersion of the spin-wave
part at low energies is due to small values of the exchange constants for
the actual optimal orientation of orbitals found at $J_H/U=0.3$. We note,
however, that the AF interactions along the $c$-axis are much stronger at
$J_H\to 0$ than in the present case. The AF structure along the $c$-axis may
be easily recognized from the symmetric spin-wave mode in the $\Gamma-Z$
direction with respect to $Z'=(0,0,\pi/2)$, while this mode increases all the 
way from the $\Gamma$ to the $X$ point. The fact that only two modes have 
nonzero neutron scattering intensity along $\Gamma-Z'-S'-\Gamma$ is due to 
this BZ path being in the high-symmetry BZ plane, where $k_x=k_y$ and
$\gamma_{-}(\vec{k})=0$. Then two modes have equal amplitude but are exactly 
out-of-phase between $A$ and $B$ sites as well as between $C$ and $D$ sites, 
and so their neutron intensities vanish, while only the companion in-phase 
modes are visible to neutrons. Unlike in the AF phases, the purely orbital 
excitation is here energetically separated from the spin-wave and 
spin-and-orbital-wave modes. The dispersion is quite small and decreases 
with $\eta$.

\subsection{Mixed-orbital AFF phase}
\label{sec:moaffrpa}

The elementary excitations in the MO{\scriptsize AFF} phase may be obtained
using a similar scheme to that used in Sec. \ref{sec:moffarpa} for the
MO{\scriptsize FFA} phase. First of all, one defines new quantum states
which correspond to the minimum of the classical problem. This is realized
by a unitary transformation of the Hamiltonian to the new orbitals
defined for $i\in A$ sublattice as,
\begin{equation}
\label{affstatei}
\left( \begin{array}{c}
 |i\mu_+\rangle   \\
 |i\nu_+\rangle
\end{array} \right) =
\left(\begin{array}{cc}
 \cos\theta_+ &  \sin\theta_+ \\
-\sin\theta_+ &  \cos\theta_+
\end{array} \right)
\left( \begin{array}{c}
 |iz\rangle   \\
 |ix\rangle
\end{array} \right) ,
\end{equation}
and for $j\in B$ sublattice as,
\begin{equation}
\label{affstatej}
\left( \begin{array}{c}
 |j\mu_-\rangle   \\
 |j\nu_-\rangle
\end{array} \right) =
\left(\begin{array}{cc}
 \cos\theta_- & -\sin\theta_- \\
 \sin\theta_- &  \cos\theta_-
\end{array} \right)
\left( \begin{array}{c}
 |jz\rangle   \\
 |jx\rangle
\end{array} \right) .
\end{equation}
By choosing the angles $\theta_+$ and $\theta_-$ at the values which
minimize the classical energy, given by Eqs. (\ref{thetamoaff1}) and
(\ref{thetamoaff1}), we guarantee that $|i\mu_+\rangle$ and $|j\mu_-\rangle$,
respectively, are at each site the orbital state realized in the classical
MO{\scriptsize FFA} phase, and one may easily define the new excitation
operators with respect to the symmetry-breaking which occurs in this phase;
they are analogous to those given in Eqs. (\ref{newspin1})--(\ref{newsping2}).
Next, the Hamiltonian is rotated to the new representation as described in
Sec. \ref{sec:moffarpa}. We do not present an explicit form of the
spin-orbital Hamiltonian (\ref{somcu}) in this case, as it may be obtained
from Eqs. (\ref{hparalong})--(\ref{kk3long}) by replacing the angle $\theta$
by $\theta_+$ and $\theta_-$ for the sublattice $A$ and $B$, respectively.
Furthermore, due to the degeneracy between the MO{\scriptsize AFF} and
MO{\scriptsize FAF} phases, we had to average the crystal-field between the
two sublattices in the actual calculation.

We have verified that the transverse excitations have a similar dependence
on the ${\vec k}$-vector to those found in the MO{\scriptsize FFA} phase,
and we show the representative data in Fig. \ref{modesaff}. For convenience,
we have rotated the BZ and use just the same notation as in Fig.
\ref{modesffa}. The value of the crystal-field $E_z$ is in the present case
effectively smaller by a factor of two in comparison with the
MO{\scriptsize FFA} phase. This asymmetry is a consequence of the choice of
$|x\rangle$ and $|z\rangle$ states as the orbital basis.

One finds again that the spin-wave and spin-and-orbital-wave excitations are
well separated along the $\Gamma-X-S-\Gamma$ path, and the gap between them
has increased to $\sim 1.2J$. We note a stronger renormalization of the
low-energy modes which follows from weakened FM interactions between the
alternating orbitals in the $(b,c)$-planes in the present case as compared
with those within the $(a,b)$-planes in the MO{\scriptsize FFA} phase.
Although the orbital excitations are still well separated from the remaining
transverse modes, their dispersion is larger than that in Fig. \ref{modesffa}.

\section{ Quantum fluctuations }
\label{sec:rpa}

The size of quantum fluctuation corrections to the classical order 
parameters determines the stability of the classical phases. As mentioned 
in Sec. \ref{sec:orbitals}, frustration of magnetic interactions leads in 
spin models to divergent quantum corrections within the LSW theory. Before
calculating these corrections in the present situation, a generalization
of the usual RPA procedure to a system with several excitations is necessary.
Here we present only the relations needed to calculate the quantum 
corrections to the LRO parameter and ground state energy, while more details 
will be reported separately.\cite{rvb99}

For that purpose, let us denote here the local operators constituting
the $so(4)$ Lie algebra at site $i$ as Hubbard operators,
$X_i^{\alpha\beta}=|i\alpha\rangle \langle i\beta|$. Using the unity 
operator, $\sum_{\beta}X_i^{\beta\beta}=\openone$,
the diagonal operator that refers to the state $|i\alpha\rangle$
{\em realized at site $i$ in the classical ground state\/} of the phase
under consideration may be expanded in terms of the excitation operators,
\begin{equation}
\label{expex}
X_i^{\alpha\alpha}=\openone
-\sum_{\beta\neq\alpha}X_i^{\beta\alpha}X_i^{\alpha\beta},
\end{equation}
while the diagonal operators referring to an {\em excited\/} state
$|i \beta\rangle$ are expressed as
\begin{equation}
\label{expexb}
X_i^{\beta\beta}=X_i^{\beta\alpha}X_i^{\alpha\beta}.
\end{equation}
Applying these equations to the $z$-th spin component
$S^z_i=S^z_{ixx}+S^z_{izz}$ of the total spin at site $i$ in one of the AF 
phases with pure orbital character (say AFxx for definiteness), one finds, 
for $i$ in the spin-up sublattice,\cite{noteexp}
\begin{eqnarray}
\label{exps}
S_i^z &=& \case{1}{2} (X_i^{x\uparrow,x\uparrow}
                    - X_i^{x\downarrow,x\downarrow}
                    + X_i^{z\uparrow,z\uparrow}
                    - X_i^{z\downarrow,z\downarrow} ) \nonumber \\
      &=& \case{1}{2} \openone
          - X_i^{x\downarrow,x\uparrow} X_i^{x\uparrow,x\downarrow}
          - X_i^{z\downarrow,x\uparrow} X_i^{x\uparrow,z\downarrow}
                     \nonumber \\
      &=& \case{1}{2} \openone
          - S_{ixx}^- S_{ixx}^+ - K_{izx}^- K_{ixz}^+ .
\end{eqnarray}
Taking the average one obtains, with the MF value 
$\langle S_i^z\rangle=\frac{1}{2}$,
\begin{eqnarray}
\label{expes}
\langle S_i^z\rangle_{\rm RPA}&=&
   \case{1}{2}- \langle S_{ixx}^- S_{ixx}^+\rangle
              - \langle K_{izx}^-K_{ixz}^+\rangle    \nonumber\\
&=&\case{1}{2}- \langle S_{i}^- S_{i}^+\rangle
              - \langle K_{i}^-K_{i}^+\rangle    \nonumber\\
&=&\langle S_i^z\rangle-\delta\langle S_i^z\rangle,
\end{eqnarray}
where the second equality is valid because averages like
$\langle S_{ixx}^-S_{izz}^+\rangle$ are zero since they involve `ghost' 
modes, so that one may formally replace $S_{ixx}^+$ by $S_{ixx}^+ 
+S_{izz}^+=S_i^+$, {\it etcetera\/}. The first contribution 
$\propto\langle S_i^-S_i^+\rangle$ is the usual renormalization due to spin 
waves, while the second term $\propto\langle K_i^-K_i^+\rangle$ stands for 
the reduction of $\langle S_i^z\rangle_{\rm RPA}$ due to 
spin-and-orbital-wave excitations. Both terms involve a local excitation 
preceded by a deexcitation which reproduces the initial local state. 
As expected only the transverse excitations contribute to the spin 
renormalization. Note that, since Eq. (\ref{exps}) is an {\em exact operator 
relation\/}, the present procedure gurantees that Eq. (\ref{expes}) is a
{\em conserving approximation\/} which respects the sum rule for the
occupancies of all states, $\sum_{\beta}\langle X_i^{\beta\beta}\rangle=1$.
The generalization of Eq. (\ref{expes}) to the MO phases using the
operators (\ref{newspin1}), (\ref{newspin2}), or to other order
parameters, like the orbital polarization, is obvious.

The local correlation functions which renormalize the order parameter in 
Eq. (\ref{expex}) are determined in the standard way,\cite{Hal72}
\begin{equation}
\label{theorem}
\langle B_i^{\dagger}A_i\rangle=
   \frac{1}{N}\sum_{\vec k}\int_{-\infty}^{+\infty}d\omega
   {\cal A}_{AB^{\dagger}}({\vec k},\omega)\frac{1}{\exp(\beta\omega)-1},
\end{equation}
where $\beta=1/k_BT$, and
\begin{eqnarray}
\label{weight}
{\cal A}_{AB^{\dagger}}({\vec k},\omega<0)&=&
   2{\rm Im}\langle\langle A_{\vec k}|
      B^{\dagger}_{\vec k}\rangle\rangle_{\omega-i\epsilon}   \nonumber \\
   &=&\sum_{\nu<0}{\cal A}_{AB^{\dagger}}^{(\nu)}({\vec k})
      \delta(\omega-\omega_{\vec k}^{(\nu)})
\end{eqnarray}
is the respective spectral density for the negative frequencies ($\nu<0$),
and ${\cal A}_{AB^{\dagger}}^{(\nu)}({\vec k})$ are the respective spectral
weights. Therefore, the correlation functions at $T=0$ are found by summing
up the total spectral weight at the negative frequencies,
\begin{equation}
\label{average}
\langle B_i^{\dagger}A_i\rangle=\frac{1}{N}\sum_{\vec k}\sum_{\nu<0}
                                {\cal A}_{AB^{\dagger}}^{(\nu)}({\vec k}).
\end{equation}

As we show elsewhere,\cite{rvb99} the Hamiltonian of the spin-orbital model
(\ref{somcu}) may be expanded in RPA in terms of the excitation and
deexcitation operators,
\begin{equation}
\label{exph}
{\cal H}\simeq {\cal H}_{\rm MF}+{\cal H}_{\rm RPA},
\end{equation}
where ${\cal H}_{\rm MF}$ is given by Eq. (\ref{somcumf}), and
\begin{eqnarray}
\label{exprpa}
{\cal H}_{\rm RPA}&=&
   \sum_{i\in A}\sum_{\mu\mu'}X_i^{\mu\alpha}a^{\mu\mu'}_AX_i^{\alpha\mu'}
  +\sum_{j\in B}\sum_{\nu\nu'}X_i^{\nu \beta}a^{\nu\nu'}_BX_i^{ \beta\nu'}
                                                          \nonumber \\
 &+&\sum_{\langle ij\rangle}\sum_{\mu\nu}
                \left( X_i^{\mu\alpha}b_{ij}^{\mu\nu}X_j^{ \beta\nu}
                      +X_i^{\alpha\mu}b_{ij}^{\mu\nu}X_j^{\nu \beta}\right)
                                                          \nonumber \\
 &+&\sum_{\langle ij\rangle}\sum_{\mu\nu}
                \left( X_i^{\alpha\mu}c_{ij}^{\mu\nu}X_j^{ \beta\nu}
                      +X_i^{\mu\alpha}c_{ij}^{\mu\nu}X_j^{\nu \beta}\right)
\end{eqnarray}
for a two-sublattice phase (the generalization to the four-sublattice MO
phases is straightforward). The MF part describes the classical problem which 
was discussed in Sec. \ref{sec:mfa}. The RPA part (\ref{exprpa}) describes 
the many-body problem in a linear approximation, with the fixed indices 
$\alpha$ and $\beta$ referring to the symmetry-broken state at site $i$ and 
$j$, respectively. This expansion leads, after changing the order of 
excitation operators $X_i^{\alpha\beta}$ to normal order, and after making 
straightforward transformations, to a compact expression for the average 
energy contribution per site,
\begin{eqnarray}
\label{erpa}
E_{\rm RPA}&=&\frac{1}{N}\langle {\cal H}_{\rm RPA}\rangle     \nonumber \\
           &=&\frac{1}{4}\left[ - {\rm Tr}\{A\}
      +\sum_{\nu>0}\frac{2}{N}\sum_{\vec k}\omega_{\vec k}^{(\nu)}\right],
\end{eqnarray}
where $A$ is the matrix of positive on-site coefficients $a_A^{\mu\mu'}$, 
$a_B^{\nu\nu'}$, appearing in the first line of Eq. (\ref{exprpa}), and
with the sum running over all modes with positive frequencies (counting 
doubly-degenerate modes twice) in the reduced BZ. This expression is seen 
to be a direct generalization of the familiar result for the HAF, the 
distinction being that more modes contribute here, and so Eq. (\ref{erpa}) 
represents the energy gain ($E_{\rm RPA}<0$) due to the reduction in 
zero-point energy of the propagating modes in comparison with that of the 
local excitations. We use Eq. (\ref{erpa}) to calculate the total energy in 
RPA,
\begin{equation}
\label{etotal}
E=E_{\rm MF}+E_{\rm RPA}.
\end{equation}

Before discussing the renormalization of the order parameter and the
corresponding energies in RPA, we concentrate ourselves on the behavior of
the transverse excitations when the crossover lines between the classical
phases are approached. As already emphasized in Sec. \ref{sec:magnons}, the
spin-wave and spin-and-orbital-wave excitations couple. As a consequence,
the modes in all considered phases {\em soften\/} when the transition lines
between different classical phases, or classical degeneracy point are
approached. This softening is shown for a representative value of $J_H/U=0.3$
in Fig. \ref{swafreal} for the two AF phases. In the AFxx phase the energy 
scales of both excitations are separated for $E_z>4J$, while the 
spin-and-orbital mode moves towards zero energy with decreasing $E_z$, and 
finally becomes soft at the $X$ point, along $\vec{k}=(\pi,0,k_z)$ and along
equivalent lines in the BZ for $E_z\simeq 1.54J$. A similar mode 
softening is found for the AFzz phase at $E_z<0$, with the soft mode along
$\Gamma-X$ and equivalent directions in the BZ at $E_z\simeq -1.84J$. This 
pecular softening along lines and not at points in the BZ shows that the 
modes behave 2D-like instead of 3D-like: constant-frequency surfaces are 
cilinders contracting towards lines, not spheres contracting towards a point.

By making an expansion of Eq. (\ref{afsw}) around the soft-mode lines, one 
finds that the (positive) excitation energies are characterized by 
{\em finite} masses in the perpendicular directions:
\begin{equation}
\label{massx}
\omega_{\rm AFxx}(\vec{k}) \rightarrow \Delta_x
      + B_x \left( {\bar k}_x^4+14{\bar k}_x^2k_y^2+k_y^4 \right)^{1/2}, 
\end{equation}
independently of $k_z$ (here ${\bar k}_x=k_x-\pi$), and 
\begin{equation}
\label{massz}
\omega_{\rm AFzz}(\vec{k}) \rightarrow \Delta_z
      + B_z \left( k_y^2 + 4k_z^2 \right),
\end{equation}
independently of $k_x$, and similarly along the $\Gamma-Y$ direction with 
$k_y$ replaced by $k_x$. As an example we give explicit expressions for the 
AFxx phase at $\eta=0$,
\begin{equation}
\label{massxx}
\Delta_x= \frac{9}{2} \frac{\varepsilon_z}{\varepsilon_z+3}, 
\hskip 1.2cm 
     B_x=\frac{27}{16}\frac{      1      }{\varepsilon_z+3}, 
\end{equation}
where one finds that the gap $\Delta_x\to 0$ when $\varepsilon_z \to 0$,
i.e. upon approaching the $M=(E_z,H_H)=(0,0)$ point at which 
the AF order is changed to the AFzz phase. This illustrates a general
principle: $\Delta_i\to 0$ when the crossover line to another phase is 
approached, and $B_i\neq 0$ when the modes (\ref{massx}) and (\ref{massz}) 
soften, making quantum fluctuation corrections to the order parameter 
to diverge logarithmically, $\langle\delta S\rangle\sim
\int d^3k/\omega(\vec{k})\sim\int d^2k/(\Delta_i+B_ik^2)\sim\ln\Delta_i$.
We emphasize that for the occurrence of this divergence not only the
finiteness of the mass but also the 2D-like nature of the dispersion is
essential. It enables a 3D system to destabilize LRO by what are essentially 
2D fluctuations. So the divergence of the order parameter near the cross-over 
lines in the phase diagram and the associated instability of the classical 
phases, may be regarded as another manifestation of the effective reduction 
of the dimensionality occurring in the spin-orbital model.
We do not present explicitly the softening of the longitudinal modes
which also happens at the transition lines.

A seemingly attractive way to simplify the calculation of the transverse
excitations would be to make a decoupling of the spin-waves and
spin-and-orbital-waves. However, this is equivalent to violating the 
commutation rules between the spin and spin-and-orbital operators in 
Appendix \ref{sec:commute},\cite{Fei98} and this changes the physics. It 
gives the same excitation energies as Eq. (\ref{afsw}), but with 
$P_{\alpha\vec k}=0$; the numerical result is given in Fig. \ref{swafpoor}. 
Of course, the spin-wave excitation does not depend then on the orbital 
splitting $E_z$, and the spin-and-orbital-wave excitation gradually 
approaches the line $\omega_{\vec k}=0$ with decreasing $|E_z|$. It has a 
weak dispersion which depends on $J_H$ and on the value of $|E_z|$, and 
gives an instability at the $\Gamma$ point only, not at lines in the BZ, and 
in the phase diagram well beyond the transition lines of Fig. \ref{mfa3d}, 
i.e., within the MO{\scriptsize FFA} and MO{\scriptsize AFF} phase for 
$E_z<0$ and $E_z>0$, respectively. Such spin-wave and spin-and-orbital-wave 
modes give, of course, much smaller quantum corrections of the order parameter
and energy than the correct RPA spectra of Fig. \ref{swafreal}.\cite{Fei98}

The spin-waves in the MO{\scriptsize FFA} phase, stable at $E_z<0$, soften
with decreasing $\eta$ (\ref{jh}), as shown in Fig. \ref{swmoffa}. At large
$\eta$ the spin-and-orbital-waves at high energies are well separated from
the spin-wave modes. The latter have a rather small dispersion at $J_H/U=0.3$
which follows from relatively weak FM interactions in the $(a,b)$ planes, and
AF interactions along the $c$-axis. The modes start to mix stronger with
decreasing $\eta$, and finally the gap in the spectrum closes below
$\eta=0.1$. The mode softening occurs again along lines in the BZ, namely 
along the $\Gamma-X$ direction. Unfortunately, we could not perform an 
analogous analytic expansion of the energies near the softening point to that 
in the AFxx and AFzz phases, but the numerical results reported here suggest 
a qualitatively similar behavior to these two phases. The MO{\scriptsize AFF} 
phase gives an analogous instability at $E_z>0$.

The soft modes in the excitation spectra give a very strong renormalization
of the order parameter $\langle S^z\rangle_{\rm RPA}$ in RPA (\ref{expes})
near the mode softening, as shown in Fig. \ref{szreal}. The quantum 
corrections {\em exceed\/} the MF values of the order parameter in the AFxx 
and AFzz phases in a region which separates these two types of LRO. Although 
one might expect that another classical phase with mixed orbitals and FM 
planes sets in instead, and the actual instabilities where
$\delta\langle S_z\rangle\to\infty$ are found indeed beyond the transition 
lines to another phase, the lines where 
$\delta\langle S^z\rangle=\langle S^z\rangle$ occur still {\em before\/} the 
phase boundaries in the phase diagram of Fig. \ref{mfa3d} (see Fig. 1 of Ref. 
\onlinecite{Fei97}). This leaves a window where {\em no classical order is 
stable\/} in between the G-AF and A-AF spin structures.

The origin of such a strong renormalization of $\langle S^z\rangle$ may be
better understood by decomposing the quantum corrections into individual
contributions as given in Eq. (\ref{expes}) (see Table I). The leading
correction comes from the local spin fluctuation expressed by
$\langle S^-_iS^+_i\rangle$ and enhanced with respect to the the pure spin
model (HAF), while the spin-and-orbital fluctuation,
$\langle K^-_iK^+_i\rangle$, increases rapidly when the instability lines
$\langle S^z\rangle_{\rm RPA}=0$ are approached. Interestingly, the latter
fluctuation is stronger in the AFxx than in the AFzz phase for the same 
values of $J_H$ and $|E_z|$ which demonstrates that the AFzz phase is more 
robust due to the directionality of the $|z\rangle$ orbitals and the strong 
AF bonds along the $c$-axis. This asymmetry is also visible in Fig. 
\ref{szreal}, where $\langle S^z\rangle_{\rm RPA}$ decreases somewhat faster 
towards zero for $E_z>0$.

In both G-AF phases (AFxx and AFzz) the leading contribution to the
renormalization of $\langle S^z\rangle_{\rm RPA}$ comes from the lower-energy
mode, especially at larger values of $J_H$. In the case of $J_H=0$ one finds, 
however, that the contribution from the lower mode either stays approximately 
constant (in the AFxx phase), or even decreases (in the AFzz phase) when the 
line of the collapsing LRO is approached at $|E_z|\to 0$ (Table I). This 
latter behavior shows again that the coupling between the spin-wave and 
spin-and-orbital-wave excitations is of crucial importance.\cite{Fei98}
This is further illustrated by Fig. \ref{szpoor}, which shows the
renormalization of $\langle S_z \rangle$ as obtained when spin waves and
spin-and-orbital waves are decoupled in the manner discussed above.
One observes that significant reduction of $\langle S_z \rangle$ then sets
in only very close to the actual divergence.

Also the orbital polarization is renormalized by the quantum fluctuations,
but this is a rather mild effect not showing any instability, since this
renormalization involves only the spin-and-orbital and the orbital
excitation but not the spin excitation, which is the one participating most
strongly in the lowest transverse mode that goes soft. This is seen in Fig. 
\ref{nxholes}, where we show $\langle n_x \rangle$, the occupation of the
$|x\rangle$ orbital, again for $J_H/U=0.3$, both at the MF level as well as
including the RPA quantum fluctuations, calculated from an expression
similar to Eq. (\ref{expes}), e.g. in the AFxx phase from
\begin{equation}
\label{expnx}
\langle n_{ix}\rangle = 1-4 \, \langle T_{izx}T_{ixz}\rangle
                        - \langle K^-_i K^+_i\rangle.
\end{equation}
Especially in the MO{\scriptsize FFA} and MO{\scriptsize AFF} phases the
deviation from the classical value of $\theta$ as given by Eq. 
(\ref{thetamoffa}) and by Eqs. (\ref{thetamoaff1}) and (\ref{thetamoaff2}),
respectively, is small. Only in the AFxx phase a significant admixture of 
$|z\rangle$ occupancy could occur close to the regime where this phase 
becomes unstable due to the divergence of $\langle S^z\rangle_{\rm RPA}$.

The reduction of $\langle S^z\rangle_{\rm RPA}$ in the
MO{\scriptsize FFA}/MO{\scriptsize AFF} phases (Table II), described by a
relation similar to Eq. (\ref{expes}), is in general weaker than that in the
G-AF phases. This is understandable, as the quantum fluctuations contribute
here only from a single AF direction, while the FM order in the planes does
not allow for excitations which involve spin flips and stabilizes the LRO
of A-AF type. For fixed $J_H$ one finds increasing quantum corrections
$\delta\langle S^z\rangle$ when the lines of phase transitions towards the
AF phases are approached. These corrections increase faster with increasing
$|E_z|$ in the MO{\scriptsize FFA} phase, as the increasing occupancy of the 
$|z\rangle$-orbital makes the AF interaction stronger there than in the 
MO{\scriptsize AFF} phase, where the occupancy of the $|x\rangle$ orbital 
increases slower roughly by a factor of two. This qualitative difference
between these two A-AF phases may be seen in Fig. \ref{nxholes}. As in the
G-AF phases, we find that the two lower-energy modes give the larger
contribution to the renormalization of the order parameter. The 
spin-and-orbital fluctuation $\langle {\cal K}^-_i{\cal K}^+_i\rangle$ 
remains almost independent of $E_z$, but increases with decreasing values of 
$J_H$. Thus we conclude that the collapse of the LRO in the A-AF (MO) phases 
is primarily due to increasing spin fluctuations,
$\langle {\cal S}^-_i{\cal S}^+_i\rangle$, while the spin-and-orbital
fluctuations become of equal importance only when the multicritical point of
the Kugel-Khomskii model $M=(E_z,J_H)=(0,0)$ is approached.

The representative quantum corrections to the ground state energy are given
in Table III. First of all, these corrections are larger by roughly a factor
of two in the G-AF phases (AFxx and AFzz) than in the A-AF phases
(MO{\scriptsize FFA} and MO{\scriptsize AFF}/MO{\scriptsize FAF}). We believe 
that this is a generic difference between the quantum corrections in the 
A-type and G-type AF phases, with the latter stabilized more due to the spin 
fluctuations contributing at all the bonds. Therefore, the G-AF phases win 
over the A-AF ones near the transition lines, as for example found at 
$J_H/U=2.0$ and $E_z/J=0.2$. However, one should keep in mind that the energy 
alone does not suffice for the stability of a particular phase in RPA, since 
the MF value of the order parameter, $\langle S^z\rangle$, has to remain 
larger than the respective quantum correction, $\delta\langle S^z\rangle$. 
Secondly, the 2D AFxx phase is characterized by larger quantum corrections 
than the strongly anisotropic AFzz phase at the same values of $J_H/U$ and 
$|E_z|/J$. The same observation was made before at the multicritical point
$M=(E_z,J_H)=(0,0)$.\cite{Fei98} This is not surprising since the 2D HAF is 
already quite close to the disordered spin state. We note that the energy 
gain due to quantum fluctuations of $0.423J$ (obtained for the actual 
interactions of $\frac{9}{4}J$ in a 2D HAF) is there considerably smaller 
than the values of $\delta E$ of the order of $0.65J$ reported in Table III.

Finally, we note that the dominating contribution to the quantum corrections
to the energy comes from the transverse excitations. The longitudinal 
excitations do not contribute at all at $J_H/U=0$, where these modes are 
dispersionless. Otherwise, the orbital excitations have always a 
significantly smaller dispersion than the value of the orbital gap in the 
spectrum, and the resulting quantum corrections are therefore almost 
negligible.

\section{ Summary and conclusions }
\label{sec:spinliquid}

Summarizing, we have presented here the case that a generic (Kugel-Khomskii)
model for the dynamics of an orbitally degenerate MHI is characterized by a 
number of peculiar features. In this paper we have followed a semi-classical 
strategy. Assuming that the ground state exhibits some particular classical 
spin- and orbital order, the stability of this order can be
investigated by considering the Gaussian fluctuations around this state.
In this way we find that in various regimes of the zero-temperature
phase-diagram, conventional order is defeated by the quantum fluctuations,
and we expect a qualitative phase diagram as shown in Fig. \ref{artistic}.

In the first place, near the transition lines between the different phases
modes soften, and these soft modes cause the zero-point fluctuations to
diverge. This is not dissimilar from the general theme associated with
the geometrically frustrated quantum spin-models, like the $J_1-J_2-J_3$ 
model.\cite{Cha88} A significant difference is that in the present case the 
source of the problems is distinct: it is associated with the difficulty to 
simultaneously satisfy the requirements for a stable spin- and orbital order. 
The cause of the frustration is dynamical instead of geometrical.

The most interesting feature is the point at the origin of the phase diagram.
On the classical level it is a point in the zero-temperature phase diagram
where a quasi-1D antiferromagnet (MO{\scriptsize FFA} phase), a 2D
antiferromagnet (AFxx phase), and a mildly anisotropic 3D antiferromagnet
(AFzz phase) become degenerate (Fig. \ref{mfa3d}). In fact, these
possibilities make up only an infinitisimal fraction of the total degeneracy
characterizing this special point. In addition, the orbitals can be freely
rotated on every site, if the spins form a 3D antiferromagnet. Likewise, the
phase diagram of Fig. \ref{artistic} is highly incomplete. Next to $E_z$,
there exist an infinity of other axes emerging from this special point, all
corresponding with distinct ways of explicit {\em local} symmetry breaking 
in the orbital sector. One can either call this point an infinite-critical 
point, or a point of perfect dynamical frustration, or a point where local 
symmetry is dynamically generated.

The obvious problem is that the above wisdom applies only when
quantum-mechanics does not play a role. Physical reality is different, and
since the classical limit is pathological, quantum-mechanics is bound to
take over. Although we have not found a way to make the case precise, it
appears to us that the local symmetry referred to in the previous paragraph
exists only in the classical limit. For this to be active on the quantum
level, it should be that the true ground state is also highly degenerate.
Although we did not prove the uniqueness of the quantum ground state, so
much is clear that the classical local symmetry gets lifted at the moment
that quantum fluctuations become significant: the cancellations occur only
if the spins are fully classical. Regardless the nature of the true ground
state, it is generated by a quantum order-out-of-disorder
mechanism.\cite{Chu91}

The first possibility is a straightforward order-out-of-disorder physics:
the quantum fluctuations affect the energies of the various classical states
in different ways, thereby breaking the classical degeneracy. One of the
saddle points might get uniquely favored and this is what is suggested in
Ref. \onlinecite{Kha97}, where it was argued that the AFzz phase becomes the
ground state at the origin of the phase diagram. Although this is a credible
possibility, one would have to demonstrate that the other possibilities are
less favoured, and moreover, we have showed elsewhere\cite{Fei98} that the 
actual calculation by Khaliullin and Oudovenko\cite{Kha97} is flawed. 
The case is still open.

Yet another possibility is unconventional spin- and orbital order which is
in a sense dual to the orbital- and spin (anti)ferromagnetism characterizing
the `classical' order: spin-orbital (resonating) valence bond (R)VB states.
We demonstrated before\cite{Fei97} that these straightforward
generalizations of the spin RVB states, well known from the study of quantum
spin-problems, appear as exceptionally stable. In a next publication we will
further elaborate on these matters.\cite{rvb99}

The status of both proposals is rather unsure: they rely at best on the 
variational principle and the true vacuum can still be completely different. 
In this regard, some recent experiments on the system LiNiO$_2$ are quite
interesting.\cite{Kit98} In this material a Mott-insulator seems to be
realized, characterized by a low spin ($S=1/2$) $e_g$ degenerate Ni(III)
state. One would naively expect this system to be unstable towards a
collective Jahn-Teller distortion, accompanied by spin ordering. This indeed
happens in the closely related system NaNiO$_2$, but in LiNiO$_2$ ordering
phenomena are completely absent,\cite{Hir85} a peculiarity pointed out long
ago.\cite{Bon57} Instead, some quantum-critical
state appears to be present, characterized by power-law behavior of physical
quantities, carrying unusual exponents. Pending the magnitude of the
Li-mediated kinetic exchange ($J_{\rm Li}$), one can view this system as
either disconnected triangular layers of Ni(III) ions (vanishing
$J_{\rm Li}$), or as interpenetrating cubic lattices of these ions which are
described by the Kugel-Khomskii Hamiltonian (large $J_{\rm Li}$).\cite{Fei97}
Hence, the peculiar state seen in the experiments can either originate in
some phenomenon associated with the triangular layers,\cite{whafnium} but
it could also be related to the matters discussed in this paper.

It is easy to settle this issue experimentally. Compare NaNiO$_2$ and
LiNiO$_2$; if the physics of the quantum disorder in the latter has to do
with the (111) layers, one would expect on general grounds that in order to
stabilize an ordered state, the effective dimensionality has to be increased,
of course assuming that the basics of the electronic structure (such like
covalency) do not change appreciably. Hence, in this layer scenario one
would expect stronger layer-layer interactions in NaNiO$_2$ as compared to
LiNiO$_2$, following the standard result of quantum field theory that
fluctuations increase upon lowering dimensionality. This standard wisdom
does not apply to the Kugel-Khomskii model, however. The fluctuations find
their origin in a dynamical frustration, and this frustration is only 
present in three space dimensions. Hence, if the disorder in LiNiO$_2$ is
caused by the physics discussed in this paper, its quantum magnetism should
be rather isotropic in 3D space, while NaNiO$_2$ should be more 2D. It is 
noticed that according to elementary quantum chemistry Li ions should be 
more effective in mediating kinetic exchange than Na ions.

\acknowledgements

We thank P. Horsch, D. I. Khomskii, J. Richter and M. Takano for valuable
discussions. A.M.O. acknowledges the support by the Committee of Scientific
Research (KBN) of Poland, Project No. 2~P03B~175~14.

\end{multicols}
\appendix

\section{ Derivation of the spin-orbital model }
\label{sec:derivation}

The derivation of the effective interactions between two $d^9$ ions at sites
$i$ and $j$ takes the simplest form for a bond $\langle ij\rangle$ oriented
along the $c$-axis. In that case the only nonvanishing hopping element is
that between the two $|z\rangle$ orbitals on the neighboring sites, and
thus the orbital occupancies in the initial and final $d^9_id^9_j$ states
have to be identical (apart from a possible simultaneous and opposite spin
flip at both sites). The possible initial states are described by a direct
product of the total spin state, either a triplet ($S=1$) or a singlet
($S=0$), and the orbital configuration, which takes one of four
possibilities:
$|x_ix_j\rangle$, $|x_iz_j\rangle$, $|z_ix_j\rangle$, or $|z_iz_j\rangle$.
Moreover, the effective interaction vanishes if the holes occupy the
$|x_ix_j\rangle$ configuration. The total spin per two sites is conserved
in the $d^9_id^9_j\rightarrow d^{10}_id^8_j$ excitation process, and
therefore the spin dependence of the resulting second order Hamiltonian
can be expressed in terms of the projection
operators on the total spin states:
$(\frac{3}{4}+\vec{S}_i\cdot\vec{S}_j)$ for the triplet, and
$(\frac{1}{4}-\vec{S}_i\cdot\vec{S}_j)$ for the singlet.

The general form of the effective Hamiltonian may be derived from the formula
which includes all possible virtual transitions to the excited $d^8d^{10}$
configurations,
\begin{equation}
\label{allv}
H_{\langle ij\rangle}=-\sum_{n,\alpha\beta}
\frac{t^2}{\varepsilon_n}Q_{S(i,j)}P_{i\alpha}P_{j\beta},
\end{equation}
where $t$ stands for the $z-z$ hopping along the $c$-axis, $Q_{S(i,j)}$ is
the projection operator on the total spin state, and $P_{i\alpha}$ is the
projection operator on the orbital state $\alpha$ at site $i$, while
$\varepsilon_n$ stands for the excitation energies given by Eqs.
(\ref{specd8}). The orbital projection operators on
$|x\rangle$ and $|z\rangle$ orbital in the initial and final state of
the $d^9$ configuration at site $i$ are, respectively,
\begin{eqnarray}
\label{orbproject}
P_{ix}&=&|ix\rangle\langle ix|=\case{1}{2}+\tau^c_i, \nonumber \\
P_{iz}&=&|iz\rangle\langle iz|=\case{1}{2}-\tau^c_i,
\end{eqnarray}
where $\tau^c_i$ is defined as in Eqs. (\ref{orbop}).

Therefore, one finds from Eq. (\ref{allv}) for a bond
$\langle ij\rangle$ along the $c$-direction,
\begin{eqnarray}
\label{fullij}
H_{\langle ij\rangle}&=&
 - \frac{t^2}{\varepsilon(^3A_2)}
   \left(\vec{S}_i\cdot\vec{S}_j+\frac{3}{4}\right)
   \left(P_{ix}P_{jz}+P_{iz}P_{jx}\right)              \nonumber \\
&+& \frac{t^2}{\varepsilon(^1E_\epsilon)}
   \left(\vec{S}_i\cdot\vec{S}_j-\frac{1}{4}\right)
   \left(P_{ix}P_{jz}+P_{iz}P_{jx}\right)              \nonumber \\
&+&\!\left[\frac{t^2}{\varepsilon(^1E_\theta)}
        +\frac{t^2}{\varepsilon(^1A_1)}\right]\!
   \left(\vec{S}_i\cdot\vec{S}_j-\frac{1}{4}\right) 2P_{iz}P_{jz}.
\end{eqnarray}
While the first two terms in (\ref{fullij}) cancel for the magnetic
interactions in the limit of $\eta\to 0$, the last term favors AF spin
orientation. We recognize that Hamiltonian (\ref{fullij}) describes the
superexchange with the superexchange constant of $4t^2/U$.\cite{And59,Cha77}
However, for convenience we define the energy unit as $J=t^2/U$ in the
present paper. Although the form (\ref{fullij}) might in principle be used
for further analysis, we prefer to make an expansion of the excitation
energies $\varepsilon_n$ in the denominators for small $J_H$, and use
$\eta=J_H/U$ (\ref{jh}) as a parameter which quantifies the Hund's rule
exchange.\cite{noteuj} Using the explicit form of the orbital projection
operators $P_{i\alpha}$ (\ref{orbproject}) this results in the following form
of the effective Hamiltonian for the bond $\langle ij\rangle\parallel c$,
\begin{eqnarray}
\label{expij}
H_{\langle ij\rangle}
&=&J\left[(1+\eta)\left(\vec{S}_i\cdot\vec{S}_j+\case{3}{4}\right)
         - \left(\vec{S}_i\cdot\vec{S}_j-\case{1}{4}\right)\right]
                                                              \nonumber  \\
& &\times\left[
   \left(\tau^c_i+\case{1}{2}\right)\left(\tau^c_j-\case{1}{2}\right)
  +\left(\tau^c_i-\case{1}{2}\right)\left(\tau^c_j+\case{1}{2}\right)\right]
                                                              \nonumber  \\
&+&4J\left(1-\case{1}{2}\eta\right)
     \left(\vec{S}_i\cdot\vec{S}_j-\case{1}{4}\right)
     \left(\tau^c_i-\case{1}{2}\right)\left(\tau^c_j-\case{1}{2}\right),
\end{eqnarray}
which may be further simplified to the form
\begin{eqnarray}
\label{effij}
H_{\langle ij\rangle}
&=&J\left\{\left(4\vec{S}_i\cdot\vec{S}_j+1\right)
         \left(\tau^c_i-\case{1}{2}\right)\left(\tau^c_j-\case{1}{2}\right)
         + \tau^c_i + \tau^c_j -1\right.               \nonumber  \\
&+& \left.\eta\left(\vec{S}_i\cdot\vec{S}_j\right)
         (\tau^c_i+\tau^c_j-1)\right.                  \nonumber  \\
&+&\left.\frac{1}{2}\eta\left[
   \left(\tau^c_i-\case{1}{2}\right)\left(\tau^c_j-\case{1}{2}\right)
            +3\left(\tau^c_i\tau^c_j-\case{1}{4}\right)\right]\right\}.
\end{eqnarray}
The first line represents the AF superexchange interactions $\propto J$,
while the other two lines describe the weaker FM interactons $\propto J\eta$,
and stand for the corrections due to the multiplet splittings of the $d^8$
excited states.

It is straightforward to verify that the above form of the effective
Hamiltonian simplifies in the limit of occupied $|z\rangle$ orbitals to
\begin{equation}
\label{effijz}
H_{\langle ij\rangle}=4J\left(1-\case{1}{2}\eta\right)
\left(\vec{S}_i\cdot\vec{S}_j-\case{1}{4}\right),
\nonumber \\
\end{equation}
and one recognizes the same constant $-\frac{1}{4}$, and the same
superexchange interaction $4J=4t^2/U$ as in the $t-J$ model at
half-filling.\cite{Cha77} However, the effective superexchange is somewhat
reduced by the factor $(1-\frac{1}{2}\eta)$ in the presence of the Hund's
rule interaction.

The effective interactions on the bonds within the $(a,b)$ planes may be now
obtained by rotating Eq. (\ref{expij}) the orbital operators $\tau^c_i$ on
Eq. (\ref{expij}) by $\pi/2$ to the cubic axes $a$ and $b$ which generates
the orbital operators $\tau^a_i$ and $\tau^b_i$ (\ref{orbop}), respectively.
This results in a nontrivial coupling between the orbital and spin degrees
of freedom, as given in Eq. (\ref{somj}). We note that in the case of
a single $s$-orbital per site, it would suffice to rotate instead the simpler
projected form (\ref{effijz}), which would give the same superexchange
interaction in any direction.

\section{ Commutation rules in the $so(4)$ algebra for the spin-orbital
          model }
\label{sec:commute}

In order to illustrate the full algebraic structure of our problem, we
present here the $so(4)$ commutators between the various excitation
operators which are needed for calculating the excitation spectra in Sec.
\ref{sec:magnons}. As the operators defined on different sites commute,
we only specify the on-site commutators.

The spin operators fulfill the usual relations for each orbital $\alpha=x,z$,
\begin{eqnarray}
{[S_{i\alpha\alpha}^+,S_{i\alpha\alpha}^z]}&=&
                    - S_{i\alpha\alpha}^+ ,            \nonumber \\
{[S_{i\alpha\alpha}^+,S_{i\alpha\alpha}^-]}&=&
                     2S_{i\alpha\alpha}^z .
\label{scom1}
\end{eqnarray}
Their commutators with the other operators which describe either
spin-and-orbital (transverse), or orbital (longitudinal, i.e., excitonic)
excitations are responsible for the coupling between spin- and
spin-and-orbital excitations (here $\alpha\neq\beta$),
\begin{eqnarray}
{[S_{i\alpha\alpha}^+,K_{i\alpha\beta }^z]}&=&
        - \case{1}{2} K_{i\alpha\beta }^+ ,                    \nonumber \\
{[S_{i\alpha\alpha}^+,K_{i\beta \alpha}^z]}&=&
        - \case{1}{2} K_{i\beta \alpha}^+ ,                    \nonumber \\
{[S_{i\alpha\alpha}^+,K_{i\alpha\beta }^-]}&=&
                     (K_{i\alpha\beta }^z+T_{i\alpha\beta }) , \nonumber \\
{[S_{i\alpha\alpha}^+,K_{i\beta \alpha}^-]}&=&
                     (K_{i\beta \alpha}^z-T_{i\beta \alpha}) , \nonumber \\
{[S_{i\alpha\alpha}^+,T_{i\alpha\beta }  ]}&=&
          \case{1}{2} K_{i\alpha\beta }^+ ,                    \nonumber \\
{[S_{i\alpha\alpha}^+,T_{i\beta \alpha}  ]}&=&
        - \case{1}{2} K_{i\beta \alpha}^+ ,
\label{scom2}
\end{eqnarray}
while they commute with the orbital-polarization operator,
\begin{equation}
[S_{i\alpha\alpha}^+,n_{i-}] = 0.
\label{scomn}
\end{equation}

The operators for spin-and-orbital excitations have the following commutators:
(i) with the spin operators,
\begin{eqnarray}
{[K_{i\alpha\beta }^+,S_{i\alpha\alpha}^z]}&=&
        - \case{1}{2} K_{i\alpha\beta }^+ ,                     \nonumber \\
{[K_{i\alpha\beta }^+,S_{i\beta \beta }^z]}&=&
        - \case{1}{2} K_{i\alpha\beta }^+ ,                     \nonumber \\
{[K_{i\alpha\beta }^+,S_{i\alpha\alpha}^-]}&=&
                      K_{i\alpha\beta }^z-T_{i\alpha\beta } ,   \nonumber \\
{[K_{i\alpha\beta }^+,S_{i\beta \beta }^-]}&=&
                      K_{i\beta \alpha}^z+T_{i\beta \alpha} ,
\label{ocom1}
\end{eqnarray}
(ii) with the spin-and-orbital operators,
\begin{eqnarray}
{[K_{i\alpha\beta }^+,K_{i\alpha\beta }^z]}&=& 0,               \nonumber \\
{[K_{i\alpha\beta }^+,K_{i\beta \alpha}^z]}&=&
       - \case{1}{2} (S_{i\alpha\alpha}^++S_{i\beta \beta }^+), \nonumber \\
{[K_{i\alpha\beta }^+,K_{i\alpha\beta }^-]}&=& 0,               \nonumber \\
{[K_{i\alpha\beta }^+,K_{i\beta \alpha}^-]}&=&
      \case{1}{2} (n_{i\alpha}-n_{i\beta })
                     +S_{i\alpha\alpha}^z+S_{i\beta \beta }^z , \nonumber \\
{[K_{i\alpha\beta }^+,T_{i\alpha\beta }  ]}&=& 0,               \nonumber \\
{[K_{i\alpha\beta }^+,T_{i\beta \alpha}  ]}&=&
    - \case{1}{2} (S_{i\alpha\alpha}^+-S_{i\beta \beta }^+) ,
\label{ocom2}
\end{eqnarray}
and (iii) with the orbital-polarization operator,
\begin{eqnarray}
{[K_{ixz }^+,n_{i-}]} &=& - K_{ixz }^+ ,                        \nonumber \\
{[K_{izx }^+,n_{i-}]} &=& + K_{izx }^+ .
\label{ocomn}
\end{eqnarray}

The relevant excitonic operators in the symmetry-broken state (\ref{excop})
commute with the above spin-transverse operators, $S_{i\alpha\alpha}^+$ and
$K_{i\alpha\beta }^+$, and give the following commutators with the
remaining spin-longitudinal operators,
\begin{eqnarray}
{[T_{i\alpha\beta \sigma},S_{i\alpha\alpha}^z]}&=&
    - \case{1}{2} \lambda_{\sigma} T_{i\alpha\beta \sigma},   \nonumber \\
{[T_{i\alpha\beta \sigma},S_{i\beta \beta }^z]}&=&
    + \case{1}{2} \lambda_{\sigma} T_{i\alpha\beta \sigma},   \nonumber \\
{[T_{i\alpha\beta \sigma},K_{i\alpha\beta }^z]}&=& 0,        \nonumber \\
{[T_{i\alpha\beta \sigma},K_{i\beta \alpha}^z]}&=&
     \case{1}{2} (S_{i\alpha\alpha}^++S_{i\beta \beta }^+)
    +\case{1}{4}\lambda_{\sigma} (n_{i\alpha}-n_{i\beta }),   \nonumber \\
{[T_{i\alpha\beta \sigma},T_{i\alpha\beta }  ]}&=& 0,        \nonumber \\
{[T_{i\alpha\beta \sigma},T_{i\beta \alpha}  ]}&=&
   \case{1}{2}\lambda_{\sigma}
         (S_{i\alpha\alpha}^++S_{i\beta \beta }^+)
  +\case{1}{4} (n_{i\alpha}-n_{i\beta }),                     \nonumber \\
{[T_{ixz \sigma},n_{i-}]}&=& -T_{ixz \sigma},                 \nonumber \\
{[T_{izx \sigma},n_{i-}]}&=& +T_{izx \sigma},
\label{excom}
\end{eqnarray}
where $\lambda_{\sigma}=\pm 1$ for $\sigma=\uparrow,\downarrow$. Therefore,
the subset of longitudinal operators $\{T_{i\alpha\beta\sigma}\}$ generates
the excitations which do not couple to the transverse excitations.

\section{ Green function equations for spin and orbital excitations }
\label{sec:dynama}

Here we present the dynamical matrices obtained for the phases with LRO
for the spin-orbital model (\ref{somj}). It is easy to verify that the
presented dynamical matrices have an RPA structure and thus describe
symmetric spectra with respect to $\omega=0$.

Let us start with the G-AF phases with either $|x\rangle$ or $|z\rangle$
orbitals occupied. The spin and spin-and-orbital excitations are determined
from Eqs. (\ref{gfafxx1}) and (\ref{gfafxx2}) for the AFxx phase, and from
Eqs. (\ref{gfafzz1}) and (\ref{gfafzz2}) for the AFzz phase. After using the
translational symmetry and performing the familiar RPA decoupling
procedure,\cite{Zub60,Hal72}
\begin{equation}
\langle\langle {\cal A}_i {\cal B}_j|...\rangle\rangle \simeq
\langle {\cal A}_i\rangle \langle\langle {\cal B}_j|...\rangle\rangle +
\langle {\cal B}_j\rangle \langle\langle {\cal A}_i|...\rangle\rangle ,
\label{deco}
\end{equation}
where $i$ and $j$ refere to {\em different\/} sites,
one finds a system
of linear equations for the excitation energies. A straightforward but
somewhat lengthy calculation shows that the same matrix with different
coefficients describes the elementary excitations for both AF phases,
\begin{equation}
\label{gfeq}
\left(\begin{array}{cccc}
 \lambda_{\alpha}-\overline{\omega}_{\vec k} & 0 & Q_{\alpha\vec k}
                                            & P_{\alpha\vec k}   \\
    0 & \tau_{\alpha}-\overline{\omega}_{\vec k} & P_{\alpha\vec k}
                                            & R_{\vec k}         \\
-Q_{\alpha\vec k} & -P_{\alpha\vec k}
                  & -\lambda_{\alpha}-\overline{\omega}_{\vec k}&0 \\
      -P_{\alpha\vec k} & -R_{\vec k} & 0
                        & -\tau_{\alpha}-\overline{\omega}_{\vec k}
\end{array} \right)
\left( \begin{array}{c}
 \langle\langle S^+_{{\vec k}xx}|\cdots\rangle\rangle_A   \\
 \langle\langle K^+_{{\vec k}xz}|\cdots\rangle\rangle_A   \\
 \langle\langle S^-_{{\vec k}xx}|\cdots\rangle\rangle_B   \\
 \langle\langle K^-_{{\vec k}xz}|\cdots\rangle\rangle_B
\end{array} \right) = 0,
\end{equation}
where $\overline{\omega}_{\vec k}$ is the frequency in units of $J$, i.e.,
$\overline{\omega}_{\vec k} = \omega_{\vec k}/J$. The constants,
$\lambda_{\alpha}$ and $\tau_{\alpha}$, and the ${\vec k}$-dependent
functions $P_{\alpha\vec k}$ and $Q_{\alpha\vec k}$ depend on the considered 
AF phase and are specified in Sec. \ref{sec:magnons}, while
$R_{\vec k}=\case{3}{2}\gamma_{+}(\vec k)$. The solution for the
eigenenergies is given by Eq. (\ref{afsw}). As discussed in Section 
\ref{sec:genrpa}, the same $4\times 4$ matrix equation written down in Eq. 
(\ref{gfeq}) for $\langle\langle S_{{\vec k}xx}^+|...\rangle\rangle_A$, 
{\it etcetera,\/} describing the modes generated by the spin-raising 
operators, is also valid for the Green functions
$\langle\langle S_{{\vec k}xx}^-|...\rangle\rangle_A$ {\it etc.\/},
describing the modes generated by the spin-lowering operators
$\{S_{ixx}^-,K_{ixz}^-,S_{jxx}^-,K_{jxz}^-\}$, with $i\in A$ and $j\in B$,
and all transverse modes are doubly degenerate.

The orbital (longitudinal) excitations correspond to exciting an electron
from one orbital to the other without changing the spin direction. If $A$
($B$) is an up (down) sublattice in the N\'eel state, the basis operators
which define the modes are $\uparrow$-spin ($\downarrow$-spin) orbital
excitations, as introduced in Sec. \ref{sec:magnons}. One finds the
following eigenvalue problem using the RPA,
\begin{equation}
\label{gfeqor}
\left(\begin{array}{cccc}
 u_{\alpha}-\overline{\zeta}_{\vec k} & 0 & +\rho_{\alpha\vec k}
                                          & +\rho_{\alpha\vec k} \\
0 & -u_{\alpha}-\overline{\zeta}_{\vec k} & -\rho_{\alpha\vec k}
                                          & -\rho_{\alpha\vec k} \\
-\rho_{\alpha\vec k} & -\rho_{\alpha\vec k}
                     & -u_{\alpha}-\overline{\zeta}_{\vec k} & 0 \\
+\rho_{\alpha\vec k} & +\rho_{\alpha\vec k} & 0
                     & u_{\alpha}-\overline{\zeta}_{\vec k}
\end{array} \right)
\left( \begin{array}{c}
 \langle\langle T_{{\vec k}xz\uparrow  }|\cdots\rangle\rangle_A   \\
 \langle\langle T_{{\vec k}zx\uparrow  }|\cdots\rangle\rangle_A   \\
 \langle\langle T_{{\vec k}xz\downarrow}|\cdots\rangle\rangle_B   \\
 \langle\langle T_{{\vec k}zx\downarrow}|\cdots\rangle\rangle_B
\end{array} \right) = 0,
\end{equation}
where again $\overline{\zeta}_{\vec k}$ is in units of $J$, i.e.,
$\overline{\zeta}_{\vec k}=\zeta_{\vec k}/J$, and the quantities
$u_{\alpha}$ and $\rho_{\alpha\vec k}$ depend on the considered G-AF phase.

The classical A-AF ground state is discussed here on the example of the
MO{\scriptsize FFA} phase. It consists of four sublattices: two sublattices
($A$ and $B$) due to different orbital order in the $(a,b)$ planes (see Fig.
\ref{allmfa}), and two other ($C$ and $D$) due to spins which alternate
along the $c$-direction. Therefore, one finds an $(8\times 8)$-matrix which
determines the energies of the elementary excitations. If the operators
transformed to ${\vec k}$-space are ordered
as ${\cal S}_{A\mu\mu}^+$, ${\cal S}_{B\mu\mu}^+$,
   ${\cal K}_{A\mu\nu}^+$, ${\cal K}_{B\mu\nu}^+$,
   ${\cal S}_{C\mu\mu}^+$, ${\cal S}_{D\mu\mu}^+$,
   ${\cal K}_{C\mu\nu}^+$, ${\cal K}_{D\mu\nu}^+$,
one finds a general structure of the eigenvalue problem,
\begin{equation}
\left(\begin{array}{cc}
  {\cal A}-\overline{\omega}_{\vec k} {\cal I}  &  {\cal B}   \\
 -{\cal B}  & -{\cal A}-\overline{\omega}_{\vec k} {\cal I}
\end{array} \right) = 0,
\label{moffarpa}
\end{equation}
where ${\cal A}$ and ${\cal B}$ are $(4\times 4)$ symmetric matrices,
${\cal I}$ is the $(4\times 4)$ identity matrix, and 
$\overline{\omega}_{\vec k}=\omega_{\vec k}/J$. Using the averages of the 
diagonal operators in the classical ground state,
\begin{eqnarray}
 \langle {\cal S}_{A\mu\mu}^z\rangle =
 \langle {\cal S}_{B\mu\mu}^z\rangle&=&
-\langle {\cal S}_{C\mu\mu}^z\rangle =
-\langle {\cal S}_{D\mu\mu}^z\rangle = \case{1}{2},     \\
 \langle {\cal N}_{i-}\rangle       &=&\case{1}{2},
\label{avzz}
\end{eqnarray}
one finds the following elements of matrix ${\cal A}$,
\begin{eqnarray}
{\cal A}_{11}={\cal A}_{22}&=&-\case{1}{2}(1-\case{1}{2}\eta)
          (1-2\cos 2\theta)^2 + 2(2-\eta)\cos^4 \theta
        +\case{1}{2}\eta(\case{3}{2}+\sin^2 2\theta),                 \\
{\cal A}_{12}&=&\left[\case{1}{2}(1-\case{1}{2}\eta)
         (1-2\cos 2\theta)^2
         -\eta(\case{3}{4}+\sin^2 2\theta)\right]\gamma_+(\vec k),    \\
{\cal A}_{13}=-{\cal A}_{24}&=&-\case{1}{2}(1-\case{1}{2}\eta)
         \sin 2\theta(2-\cos 2\theta)
        -\case{1}{4}(3+\case{11}{2}\eta)\sin 4\theta
        -\case{1}{2}\varepsilon_z\sin 2\theta,                        \\
{\cal A}_{14}&=&-(1-\case{1}{2}\eta-2\cos 2\theta)\sin 2\theta\gamma_+(\vec k)
        +\case{\sqrt{3}}{2}\left[1-(2-\eta)\cos 2\theta\right]\gamma_-(\vec k),
                                                                      \\
{\cal A}_{23}&=&+(1-\case{1}{2}\eta-2\cos 2\theta)\sin 2\theta\gamma_+(\vec k)
        +\case{\sqrt{3}}{2}\left[1-(2-\eta)\cos 2\theta\right]\gamma_-(\vec k),
                                                                      \\
{\cal A}_{33}={\cal A}_{44}&=&
        -(1-\case{1}{2}\eta)(1-2\cos 2\theta)+\case{1}{2}\eta
        -\case{1}{2}(1+2\eta)(1+2\sin^2 2\theta)
        -\varepsilon_z \cos 2\theta,                                  \\
{\cal A}_{34}&=&\case{1}{2}(1+2\cos 4\theta)\gamma_+(\vec k),
\label{amat}
\end{eqnarray}
and the following nonzero elements of matrix ${\cal B}$,
\begin{eqnarray}
{\cal B}_{11}={\cal B}_{22}&=&\left[
          (1-\case{1}{2}\eta)(1+\cos 2\theta)-\case{1}{4}\eta\right]
          (1+\cos 2\theta)\gamma_z(\vec k),         \\
{\cal B}_{33}={\cal B}_{44}&=&
          \sin^2 2\theta \; \gamma_z(\vec k),         \\
{\cal B}_{13}={\cal B}_{31}=-{\cal B}_{24}=-{\cal B}_{42}&=&-\left(
          1-\case{1}{2}\eta+\cos 2\theta\right)\sin 2\theta\;\gamma_z(\vec k).
\label{bmat}
\end{eqnarray}

The longitudinal excitations in the A-AF phases were obtained by solving the
respective Green function equations for the excitation operators
(\ref{excop1}). After transforming these equations to ${\vec k}$-space,
and taking the following sequence of excitation operators,
${\cal T}_{A\mu\nu  \uparrow}$, ${\cal T}_{B\mu\nu  \uparrow}$,
${\cal T}_{A\nu\mu  \uparrow}$, ${\cal T}_{B\nu\mu  \uparrow}$,
${\cal T}_{C\mu\nu\downarrow}$, ${\cal T}_{D\mu\nu\downarrow}$,
${\cal T}_{C\nu\mu\downarrow}$, ${\cal T}_{D\nu\mu\downarrow}$,
one finds an eigenvalue problem of the form,
\begin{equation}
\left(\begin{array}{cccc}
 {\cal P}-\omega &  {\cal R}        &    +{\cal Q}     &   +{\cal Q}     \\
   -{\cal R}     & -{\cal P}-\omega &    -{\cal Q}     &   -{\cal Q}     \\
   -{\cal Q}     &    -{\cal Q}     & -{\cal P}-\omega &   -{\cal R}     \\
   +{\cal Q}     &    +{\cal Q}     &     {\cal R}     & {\cal P}-\omega
\end{array} \right) = 0,
\label{matrixmoffa}
\end{equation}
where ${\cal P}$, ${\cal R}$, and ${\cal Q}$ are symmetric matrices.
Their nonvanishing elements are defined as follows:
\begin{eqnarray}
\label{pmat}
{\cal P}_{11}={\cal P}_{22}&=&
        \case{1}{2}(1-\case{1}{2}\eta)
        [1-2\cos 4\theta+2\cos 2\theta(2+\cos 2\theta)]        \nonumber \\
     &+&\case{3}{4}\eta \cos 2\theta
      -\case{3}{2}(1+2\eta)\cos 4\theta-\varepsilon_z\cos 2\theta,       \\
{\cal P}_{12}={\cal P}_{21}&=&
        \case{1}{2}(1+\eta)(1-2\cos 2\theta)\gamma_+(\vec k),            \\
\label{rmat}
{\cal R}_{12}={\cal R}_{21}&=&
        \case{1}{2}(1+\eta)(1-2\cos 2\theta)\gamma_+(\vec k),            \\
\label{qmat}
{\cal Q}_{11}={\cal Q}_{22}&=&\eta\sin^2 2\theta\gamma_z(\vec k).
\end{eqnarray}
As in the AF phases, the coupling between the sublattices $A$ and $C$ and
between $B$ and $D$, respectively, is proportional to the weak FM component,
$\eta$. The mechanism of this coupling is explained in Fig. \ref{orbex}.

\section{ Neutron intensities in transverse excitations }
\label{sec:neutrons}

In this Appendix we explain the intensities $\chi(\omega)$ in neutron
scattering seen in the presence of orbital degrees of freedom. One can
start from the general expression for the cross section for pure magnetic
scattering,\cite{deGen}
\begin{eqnarray}
\frac{d^2\sigma}{d\Omega d\omega}&\propto &
\frac{k_1}{k_0}\sum_{ij} f_j^*({\vec q})f_i({\vec q})    \nonumber \\
&\times& \frac{1}{2\pi} \int dt e^{-i\omega t}
\langle {\vec S}_{i\perp}(0) {\vec S}_{j\perp}(t) \rangle
e^{-i{\vec q}({\vec R}_i-{\vec R}_j)},
\label{crosssec}
\end{eqnarray}
where $k_0$ and $k_1$ are the initial and final momenta, while ${\vec q}$ is
the momentum transfer. The spin components at site $i$ and $j$ are
perpendicular to ${\vec q}$. By integrating over time $t$ one finds that the
neutron cross section (\ref{crosssec}) is related to the imaginary part of
the spin-spin Green function,
\begin{eqnarray}
\frac{d^2\sigma}{d\Omega d\omega}&\propto &
\frac{k_1}{k_0}\sum_{ij} f_j^*({\vec q})f_i({\vec q})
e^{-i{\vec q}({\vec R}_i-{\vec R}_j)}                    \nonumber \\
& &\times\frac{1}{2\pi} 2{\rm Im} \left\{\sum_{\alpha}\langle\langle
S^{\alpha}_{j\perp}|S^{\alpha}_{i\perp}\rangle\rangle_{-\omega} \right\}
\Theta(\omega),
\label{crosssecgf}
\end{eqnarray}
where $\Theta(\omega)=1$ for $\omega>0$, and $\Theta(\omega)=0$ for 
$\omega<0$, and we took the limit of temperature $T\to 0$. In order to 
extract the perpendicular component of the spin-spin correlation function 
from the Green functions,
$\langle\langle S^{\alpha}_{j\perp}|
                S^{\alpha}_{i\perp}\rangle\rangle_{-\omega}$,
we use the identity,
\begin{equation}
S^{\alpha}_{i\perp}=\sum_{\beta}S^{\beta}_{i}
\left(\delta_{\alpha\beta}-\frac{q^{\alpha}q^{\beta}}{q^2}\right).
\label{spinperp}
\end{equation}
The components of the Green functions in ${\vec q}$-space, $\langle\langle
S^{\alpha}_{\vec q}|S^{\beta}_{-{\vec q}}\rangle\rangle_{-\omega}$, are found
using the following properties of the transverse spin-spin functions,
\begin{eqnarray}
 {\rm Im}\langle\langle S^{\alpha}_{\vec q}|
                   S^{ \beta}_{-{\vec q}}\rangle\rangle_{-\omega}&=&
-{\rm Im}\langle\langle S^{ \beta}_{\vec q}|
                   S^{\alpha}_{-{\vec q}}\rangle\rangle_{ \omega}, \nonumber \\
\langle\langle S^+_{\vec q}|S^+_{-{\vec q}}\rangle\rangle_{\omega}&=&
\langle\langle S^-_{\vec q}|S^-_{-{\vec q}}\rangle\rangle_{\omega}=0,
\label{gfrelas}
\end{eqnarray}
and 
$\langle\langle S^{\alpha}_{\vec q}|S^z_{-{\vec q}}\rangle\rangle_{\omega}=0$
for the wave-vectors ${\vec q}\neq {\vec Q}$, where ${\vec Q}$ is the nesting
vector. One finds that the neutron cross section normalized per one site may
be written as follows,
\begin{equation}
\frac{1}{N}\frac{d^2\sigma}{d\Omega d\omega} \propto
\frac{1}{8\pi}\frac{k_1}{k_0}\frac{1}{N}\sum_{ij}
f_j^*({\vec q})f_i({\vec q})\chi({\vec q}),
\label{crosssecchi}
\end{equation}
where $\chi({\vec q})$ is the neutron scattering intensity which includes
the geometrical factor which originates from Eq. (\ref{spinperp}). It is
proportional to a linear combination of the diagonal and off-diagonal
elements of the Green function, and one finds for a two-sublattice magnetic
structure, as for example in AFxx and AFzz phases,
\begin{eqnarray}
\chi({\vec q})&=&\left(1+\frac{q^2_z}{q^2}\right) 2{\rm Im}\left[
 G_{AA}({\vec q},-\omega)+G_{BB}({\vec q},-\omega)\right.  \nonumber \\
& &\left.\hskip .9cm
+G_{AB}({\vec q},-\omega)+G_{BA}({\vec q},-\omega)\right]\Theta(\omega),
\label{chidef}
\end{eqnarray}
with the element $G_{AA}({\vec q},-\omega)$ standing for the transverse Green 
function, $\langle\langle S^{+}_{A,{\vec q}}|
S^{-}_{A,{-{\vec q}}}\rangle\rangle_{-\omega}$, {\it etcetera\/}, and the 
indices $A$ and $B$ refer to two sublattices. The explicit formula in terms 
of the spectral intensities ${\cal A}_{mn}^{(\nu)}({\vec q})$ is given by
\begin{eqnarray}
\chi({\vec q})&=&\left(1+\frac{q^2_z}{q^2}\right)\sum_{\nu(>0)}
\left[ {\cal A}_{AA}^{(\nu)}({\vec q})+{\cal A}_{BB}^{(\nu)}({\vec q})\right.
                                                    \nonumber \\
& &\left. \hskip .5cm
+{\cal A}_{AB}^{(\nu)}({\vec q})+{\cal A}_{BA}^{(\nu)}({\vec q})\right]
\delta(\omega-\omega^{(\nu)}_{\vec q}).
\label{chia}
\end{eqnarray}
We have used Eq. (\ref{chia}) to determine the contributions to the neutron
cross-section due to different excitations, as analyzed in Sec.
\ref{sec:magnons} and presented in Figs. \ref{modesxx}--\ref{modesaff}.
The generalization to the case of four-sublattice structures found in the
MO{\scriptsize FFA} and MO{\scriptsize AFF} phases is straightforward.

\begin{multicols}{2} 

\eject
\widetext

\begin{table}
\caption{Individual contributions to quantum corrections
$\langle\delta S^z\rangle$ of the AF order parameter in AFxx ($E_z>0$) and
AFzz ($E_z<0$) phases due to: spin-wave ($\langle S^{-}S^{+}\rangle$),
spin-and-orbital-wave ($\langle K^{-}K^{+}\rangle$), and the leading
contribution from low-energy mode, $\langle\delta S^z\rangle_1$. The values
of the order parameter in RPA are given by $\langle S^z\rangle_{\rm RPA}$.}
\vskip .5cm
\begin{tabular}{ccccccc}
$J_H/U$    & $E_z/J$ &
 $\langle S^{-}S^{+} \rangle$ & $\langle K^{-}K^{+} \rangle$ &
 $\langle\delta S^z\rangle_1$ &
 $\langle\delta S^z\rangle$ & $\langle S^z\rangle_{\rm RPA}$ \\
\hline
   0.0 & -3.0 & 0.2680 & 0.0117 & 0.2731 & 0.2797 & 0.2203  \\
   0.0 & -2.0 & 0.2733 & 0.0187 & 0.2606 & 0.2920 & 0.2080  \\
   0.0 & -1.0 & 0.2839 & 0.0368 & 0.2146 & 0.3207 & 0.1793  \\
   0.0 &  1.0 & 0.2645 & 0.0901 & 0.2440 & 0.3546 & 0.1454  \\
   0.0 &  2.0 & 0.2416 & 0.0516 & 0.2426 & 0.2932 & 0.2068  \\
   0.0 &  3.0 & 0.2298 & 0.0352 & 0.2455 & 0.2650 & 0.2350  \\
\hline
   0.1 & -3.0 & 0.2919 & 0.0140 & 0.2963 & 0.3059 & 0.1941  \\
   0.1 & -2.0 & 0.2995 & 0.0245 & 0.2757 & 0.3240 & 0.1760  \\
   0.1 & -1.0 & 0.3188 & 0.0612 & 0.2339 & 0.3800 & 0.1200  \\
   0.1 &  1.0 & 0.2925 & 0.1461 & 0.2864 & 0.4387 & 0.0613  \\
   0.1 &  2.0 & 0.2519 & 0.0665 & 0.2493 & 0.3183 & 0.1817  \\
   0.1 &  3.0 & 0.2352 & 0.0421 & 0.2519 & 0.2773 & 0.2227  \\
\hline
   0.2 & -3.0 & 0.3270 & 0.0174 & 0.3291 & 0.3445 & 0.1555  \\
   0.2 & -2.0 & 0.3398 & 0.0351 & 0.3023 & 0.3750 & 0.1250  \\
   0.2 &  2.0 & 0.2687 & 0.0928 & 0.2647 & 0.3615 & 0.1385  \\
   0.2 &  3.0 & 0.2428 & 0.0521 & 0.2593 & 0.2950 & 0.2050  \\
   0.2 & 10.0 & 0.2071 & 0.0092 & 0.2077 & 0.2163 & 0.2837  \\
\hline
   0.3 & -3.0 & 0.3861 & 0.0232 & 0.3834 & 0.4093 & 0.0907  \\
   0.3 & -2.0 & 0.4215 & 0.0601 & 0.3720 & 0.4816 & 0.0184  \\
   0.3 &  2.0 & 0.3026 & 0.1530 & 0.3179 & 0.4556 & 0.0444  \\
   0.3 &  3.0 & 0.2545 & 0.0680 & 0.2706 & 0.3224 & 0.1776  \\
   0.3 & 10.0 & 0.2076 & 0.0097 & 0.2083 & 0.2173 & 0.2827  \\
\end{tabular}
\label{table1}
\end{table}

\begin{table}
\caption{Individual contributions to the quantum corrections of the magnetic
order parameter $\langle\delta {\cal S}^z\rangle$ in MO phases due to
spin-wave, $\langle {\cal S}^{-}{\cal S}^{+}\rangle$, and due to
spin-and-orbital-wave excitations, $\langle {\cal K}^{-}{\cal K}^{+}\rangle$,
and due to individual modes as labelled in Figs. \protect\ref{modesffa} and
\protect\ref{modesaff}, $\langle\delta {\cal S}^z\rangle_n$, respectively.
The values of the renormalized order parameter in RPA are given by
$\langle {\cal S}^z\rangle_{\rm RPA}$.}
\vskip .5cm
\begin{tabular}{cccccccccccc}
& $J_H/U$   & $E_z/J$ &
 $\langle {\cal S}^{-}{\cal S}^{+}\rangle$ &
 $\langle {\cal K}^{-}{\cal K}^{+}\rangle$ &
 $\langle\delta{\cal S}^z\rangle_1$ & $\langle\delta{\cal S}^z\rangle_2$ &
 $\langle\delta{\cal S}^z\rangle_3$ & $\langle\delta{\cal S}^z\rangle_4$ &
 $\langle\delta{\cal S}^z\rangle$ & $\langle{\cal S}^z\rangle_{\rm RPA}$ & \\
\hline
&  0.2 &  0.0 & 0.1350 & 0.0508 & 0.0114 & 0.0344  &
                0.0709 & 0.0691 & 0.1858 & 0.3142  &  \\
\hline
&  0.3 & -2.0 & 0.2138 & 0.0323 & 0.0673 & 0.0646  &
                0.0585 & 0.0557 & 0.2461 & 0.2539  &  \\
&  0.3 & -1.0 & 0.1338 & 0.0336 & 0.0411 & 0.0025  &
                0.0547 & 0.0691 & 0.1674 & 0.3326  &  \\
&  0.3 &  0.0 & 0.0918 & 0.0354 & 0.0122 & 0.0241  &
                0.0425 & 0.0485 & 0.1273 & 0.3727  &  \\
&  0.3 &  1.0 & 0.1095 & 0.0323 & 0.0285 & 0.0041  &
                0.0684 & 0.0408 & 0.1418 & 0.3582  &  \\
&  0.3 &  2.0 & 0.1330 & 0.0328 & 0.0327 & 0.0076  &
                0.0754 & 0.0502 & 0.1658 & 0.3342  &  \\
&  0.3 &  3.0 & 0.1664 & 0.0329 & 0.0465 & 0.0146  &
                0.0738 & 0.0644 & 0.1993 & 0.3007  &  \\
\hline
&  0.4 & -3.0 & 0.2144 & 0.0232 & 0.0876 & 0.0958  &
                0.0294 & 0.0249 & 0.2376 & 0.2624  &  \\
&  0.4 & -2.0 & 0.1373 & 0.0258 & 0.0552 & 0.0145  &
                0.0453 & 0.0482 & 0.1631 & 0.3369  &  \\
&  0.4 & -1.0 & 0.0928 & 0.0269 & 0.0370 & 0.0020  &
                0.0302 & 0.0505 & 0.1197 & 0.3803  &  \\
&  0.4 &  0.0 & 0.0647 & 0.0274 & 0.0224 & 0.0080  &
                0.0257 & 0.0360 & 0.0921 & 0.4079  &  \\
&  0.4 &  1.0 & 0.0776 & 0.0254 & 0.0258 & 0.0038  &
                0.0494 & 0.0240 & 0.1030 & 0.3970  &  \\
&  0.4 &  2.0 & 0.0924 & 0.0258 & 0.0292 & 0.0063  &
                0.0552 & 0.0276 & 0.1182 & 0.3818  &  \\
&  0.4 &  3.0 & 0.1117 & 0.0259 & 0.0363 & 0.0104  &
                0.0590 & 0.0319 & 0.1376 & 0.3624  &  \\
\end{tabular}
\label{table2}
\end{table}

\eject
\narrowtext

\begin{table}
\caption{
The mean-field energy, $E_{\rm MF}$, the quantum energy correction due to
transverse modes and due to longitudinal modes, $\delta E_t$ and
$\delta E_l$, respectively, and the ground state energy in RPA,
$E_{\rm RPA}$ (all in the units of $J$). The labels FFA and AFF indicate
the way of staggering of FM planes in the MO phases with A-AF order. }
\vskip .5cm
\begin{tabular}{ccccccc}
$J_H/U$  &  $E_z/J$ & $E_{\rm MF}$ & $\delta E_t$
                                   & $\delta E_l$ & $E_{\rm RPA}$ & phase \\
\hline
   0.0 & -2.0 & -4.0000 & 0.6440 & 0.0    & -4.6440 & AFzz \\
   0.0 & -1.0 & -3.5000 & 0.6700 & 0.0    & -4.1700 & AFzz \\
   0.0 &  1.0 & -3.5000 & 0.7073 & 0.0    & -4.2073 & AFxx \\
   0.0 &  2.0 & -4.0000 & 0.6399 & 0.0    & -4.6399 & AFxx \\
\hline
   0.1 & -2.0 & -3.9250 & 0.6354 & 0.0008 & -4.5612 & AFzz \\
   0.1 & -1.0 & -3.4250 & 0.6735 & 0.0021 & -4.1006 & AFzz \\
   0.1 &  1.0 & -3.4250 & 0.7344 & 0.0020 & -4.1614 & AFxx \\
   0.1 &  2.0 & -3.9250 & 0.6384 & 0.0008 & -4.5642 & AFxx \\
\hline
   0.2 & -3.0 & -4.3500 & 0.6082 & 0.0024 & -4.9606 & AFzz \\
   0.2 & -2.0 & -3.8500 & 0.6328 & 0.0042 & -4.4870 & AFzz \\
   0.2 & -1.0 & -3.4769 & 0.3964 & 0.0009 & -3.8742 &  FFA \\
   0.2 &  0.0 & -3.2558 & 0.2992 & 0.0028 & -3.5577 &  FFA \\
   0.2 &  1.0 & -3.3543 & 0.3437 & 0.0010 & -3.6990 &  AFF \\
   0.2 &  2.0 & -3.4769 & 0.3962 & 0.0005 & -3.8738 &  AFF \\
   0.2 &  2.0 & -3.8500 & 0.6472 & 0.0041 & -4.5013 & AFxx \\
\hline
   0.3 & -3.0 & -4.2750 & 0.6052 & 0.0062 & -4.8864 & AFzz \\
   0.3 & -3.0 & -4.2272 & 0.5252 & 0.0194 & -4.7717 &  FFA \\
   0.3 & -2.0 & -3.7750 & 0.6419 & 0.0134 & -4.4303 & AFzz \\
   0.3 & -2.0 & -3.8651 & 0.3944 & 0.0037 & -4.2632 &  FFA \\
   0.3 & -1.0 & -3.5892 & 0.3040 & 0.0019 & -3.8951 &  FFA \\
   0.3 &  0.0 & -3.3996 & 0.2335 & 0.0054 & -3.6384 &  FFA \\
   0.3 &  1.0 & -3.4836 & 0.2664 & 0.0031 & -3.7531 &  AFF \\
   0.3 &  2.0 & -3.5892 & 0.3038 & 0.0016 & -3.8947 &  AFF \\
   0.3 &  2.0 & -3.7750 & 0.6768 & 0.0134 & -4.4652 & AFxx \\
   0.3 &  3.0 & -3.7164 & 0.3459 & 0.0015 & -4.0638 &  AFF \\
   0.3 &  3.0 & -4.2750 & 0.5773 & 0.0063 & -4.8586 & AFxx \\
   0.3 & 10.0 & -7.7750 & 0.4048 & 0.0014 & -8.1812 & AFxx \\
\end{tabular}
\label{table3}
\end{table}


\begin{figure}
\caption
{Virtual transitions $d_i^9d_j^9\rightarrow d_i^{10}d_j^8$ which lead to a
spin-flip and generate effective interactions for a bond
$\langle ij\rangle\parallel c$-axis, with the excitation energies at $E_z=0$.
For two holes in different orbitals (a), either the triplet $^3A_2$ or the
interorbital singlet $^1E_{\theta}$ occurs as an intermediate $d^8$
configuration, while if both holes are in $|z\rangle$ orbitals (b), two other
singlets, $^1E_{\epsilon}$ and $^1A_1$, with double occupancy of $|z\rangle$
orbital, contribute. The latter processes are possible either from $i$ to $j$
or from $j$ to $i$.}
\label{virtual}
\end{figure}

\begin{figure}
\caption
{Energies of the virtual excitations $\varepsilon_i/U$ shown in Fig.
\protect{\ref{virtual}} as functions of $E_z/J$ for $J_H/U=0.3$.
The lowest triplet $|^3A_2\rangle$ state is indicated by full circles, and
the singlet states ($|^1E\rangle$ and $|^1A_1\rangle$) by full lines.}
\label{msd8}
\end{figure}

\begin{figure}
\caption
{Schematic representation of orbital and magnetic long-range order within
the $(a,b)$ planes of AFxx, AFzz, MO{\protect{\scriptsize FFA}}, and
MO{\protect{\scriptsize AFF}} phases, respectively. The shadded parts of
different orbitals are oriented along the $c$-axis. The spins (arrows) in
the next $(a,b)$ plane in the $c$-direction are AF to those below them in
AFzz and MO{\protect{\scriptsize FFA}} phases, and FM in
MO{\protect{\scriptsize AFF}} phase. In the AFxx phase there is no magnetic
coupling to the next plane along the $c$-axis, but this degeneracy is
removed in MO{\protect{\scriptsize AAF}} phase, where a small $|z\rangle$
component promotes a FM coupling. }
\label{allmfa}
\end{figure}

\begin{figure}
\caption
{Mean-field phase diagram of the 3D spin-orbital model (\protect{\ref{somcu}})
in the $(E_z,J_H)$ plane ($\beta=1$). The lines separate the classical
states shown in Fig. \protect{\ref{allmfa}}; the transition from AFxx to
MO{\protect{\scriptsize AFF}} phase is second order (dashed line), while
the remaining transitions are first order (full lines). }
\label{mfa3d}
\end{figure}

\begin{figure}
\caption
{Mean-field phase diagrams of the spin-orbital model (\protect{\ref{somcu}})
in the $(E_z,J_H)$ plane for different values of hopping along $c$-axis: (a)
$\beta=1.414$, and (b) $\beta=0.707$. The magnetic phases and lines as in
Fig. \protect{\ref{mfa3d}}.}
\label{beta}
\end{figure}

\begin{figure}
\caption
{Mean-field phase diagram of the spin-orbital model (\protect{\ref{somcu}})
in the $(E_z,J_H)$ plane in two dimensions ($\beta=0$). Full lines separate 
the classical states AFxx, AFzz, and MO{\protect{\scriptsize FF}} shown in 
Fig. \protect{\ref{allmfa}}, while the spin order in the 
MO{\protect{\scriptsize AA}} phase is AF, and the orbitals are in between 
those in AFxx and MO{\protect{\scriptsize FF}} phase. }
\label{mfa2d}
\end{figure}

\begin{figure}
\caption
{Schematic representation of the mixed orbitals in $(a,b)$ planes of the
MO{\protect{\scriptsize FF}} phase in a 2D model: (a) the orbitals with 
their phases, and (b) the resulting distortion in the oxygen lattice, 
stabilized by the orbital ordering.}
\label{disto}
\end{figure}

\begin{figure}
\caption
{Schematic propagation of the orbital (excitonic) excitation (a).
If $J_H=0$, an orbital excitation can propagate only to state (b) and is
accompanied by a spin-flip (top), while $J_H>0$ allows also the spin-flip
in the intermediate $d_i^8$ state, and thus the propagation without
spin-flip (c) becomes possible (bottom).}
\label{orbex}
\end{figure}

\begin{figure}
\caption
{Lower panel: spin-wave and spin-and-orbital-wave transverse excitations
(full lines) and longitudinal excitations (dashed lines) in AFxx phase;
upper panel: neutron intensities of the transverse excitations.
Parameters: $E_z/J=3.0$ and $J_H/U=0.3$. }
\label{modesxx}
\end{figure}

\begin{figure}
\caption
{The same as in Fig. \protect\ref{modesxx}, but for the AFzz phase,
as obtained for $E_z/J=-3.0$ and $J_H/U=0.3$. }
\label{modeszz}
\end{figure}

\begin{figure}
\caption
{The same as in Fig. \protect\ref{modesxx}, but for the
MO{\protect{\scriptsize FFA}} phase, as obtained for $E_z/J=-1.0$ and
$J_H/U=0.3$. Different modes are labelled by the increasing indices
$i=1,\dots,4$ with increasing energy.}
\label{modesffa}
\end{figure}

\begin{figure}
\caption
{The same as in Fig. \protect\ref{modesxx}, but for the
MO{\protect{\scriptsize AFF}} phase, as obtained for $E_z/J=1.0$ and
$J_H/U=0.3$. }
\label{modesaff}
\end{figure}

\begin{figure}
\caption
{Spin-wave and spin-and-orbital-wave excitations in the G-AF phases: AFxx
(left) and AFzz (right), in the main directions of the 3D BZ for a few
values of $E_z$ (in the units of $J$), and for $J_H/U=0.3$. The lower-energy
mode becomes soft for $E_z/J<1.54$ ($E_z/J>-1.84$) in the AFxx (AFzz) phase.}
\label{swafreal}
\end{figure}

\begin{figure}
\caption
{The same as in Fig. \protect{\ref{swafreal}}, but without the coupling
between the spin-wave and spin-and-orbital-wave excitations in both G-AF
phases: AFxx (left) and AFzz (right). }
\label{swafpoor}
\end{figure}

\begin{figure}
\caption
{Transverse (full lines) and longitudinal (dashed lines) excitations in 
MO{\protect{\scriptsize FFA}} phase in the
main directions of the 3D BZ for a few values of $J_H/U$, and for
$E_z/J=-0.5$. The lower-energy mode becomes soft for $J_H/U<0.06$ .}
\label{swmoffa}
\end{figure}

\begin{figure}
\caption
{Renormalization of the magnetic LRO parameter $\langle S^z_i\rangle$ by
quantum fluctuations as obtained in RPA in:
(a) AFzz (left) and AFxx (right) phases as functions of $E_z/J$ for
$J_H/U=0.1$ and $0.3$;
(b) MO{\protect{\scriptsize FFA}} phase as function of $J_H/U$ for
$E_z/J=0.5$, -0.5 and -1.5. }
\label{szreal}
\end{figure}

\begin{figure}
\caption
{Renormalization of the magnetic LRO parameter $\langle S^z_i\rangle$ by
quantum fluctuations obtained for the G-AF phases as in Fig.
\protect{\ref{szreal}}(a), but for decoupled spin-wave and
spin-and-orbital-wave excitations shown in Fig. \protect{\ref{swafpoor}}.}
\label{szpoor}
\end{figure}

\begin{figure}
\caption
{Average density of $|x\rangle$-holes $\langle n_x\rangle$ as obtained for 
$J_H/U=0.3$ in MF approximation (dashed lines) and with the quantum 
corrections calculated in RPA (full lines). The splitting of lines for 
$E_z/J>0$ corresponds to the MO{\protect{\scriptsize AFF}} phase with two 
different hole densities $\langle n_x\rangle_A\neq \langle n_x\rangle_B$ on 
the ions belonging to two sublattices (see Fig. \protect{\ref{allmfa}}). }
\label{nxholes}
\end{figure}

\begin{figure}
\caption
{Schematic phase diagram of the spin-orbital model including quantum
fluctuations. The {\em spin liquid\/} phase is expected to separate the AF
phases with different types of magnetic LRO: G-AF phases with either
$d_{x^2-y^2}$ ($|xx\rangle$) or $d_{3z^2-r^2}$ ($|zz\rangle$) orbital
occupied on both sublattices from the A-AF phases with mixed orbitals (MO)
ordered on two sublattices.}
\label{artistic}
\end{figure}

\end{multicols} 


\begin{references}
\bibitem{notecov} The covalency due to $p-d$ hybridization is responsible
                   for the admixture of the $d^{10}\underline{L}$
                   configuration and leads to some quantitative
                   changes of the parameters in the present effective
                   model which describes only $d$ orbitals at transition
                   metal ions.
\bibitem{noteuj} For convenience, we defined $U$ as the average excitation
                   energy of the $d^8$ configuration. Therefore, $U$ is not
                   the interorbital Coulomb element, and with the present 
                   definition of $J_H$ as equal to the singlet-triplet
                   splitting, the interorbital interaction between holes
                   of opposite (equal) spins is $U-J_H/2$ ($U-J_H$).
\bibitem{noted8} The Hamiltonian given in Ref. \protect{\onlinecite{Zaa93}}
                 does not contain the pair-hopping terms\protect{\cite{Ole83}}
                 in Eq. (\protect{\ref{hint}}), and gives thus
                 a different multiplet structure from that obtained in Eqs.
                 (\protect{\ref{specd8}}).
\bibitem{notebugs}
                A simplified multiplet structure used in Refs.
                \protect{\onlinecite{Kug73}} and \protect{\onlinecite{Kug82}}
                leads to a different expression. We also note that this
                earlier proposed form does not observe the cubic symmetry
                in the orbital space.
\bibitem{noteexp} 
                The expansion (\protect\ref{exps}) is exact and holds for 
                the global spin operator $S^z_i$, while a similar relation 
                holds for the spin operators projected on one orbital state,
                and in the AFxx phase $S^z_{ixx}=\case{1}{2}\openone 
                -S_{ixx}^-S_{ixx}^+ -\case{1}{2}K_{izx}^-K_{ixz}^+
                -\case{1}{2}T_{izx\uparrow}^-T_{ixz\uparrow}^+$, if $i\in A$.
                Eq. (\protect\ref{exps}) was also used in Ref.
                \protect\onlinecite{Fei98} to calculate the quantum correction
                to the order parameter $\delta\langle S^z\rangle$ in the AFxx 
                and AFzz phase at $J_H=0$, with the coefficient in front of 
                $\langle S_i^-\sigma_i^-S_i^+\sigma_i^+\rangle$ being 
                $\case{1}{4}$. It follows from a relation between the 
                present operators $K_{i\alpha\beta}^{\pm}$ and the composite 
                spin-and-orbital operators, 
                $K_i^{\pm}=\case{1}{2}S_i^{\pm}\sigma_i^{\pm}$, and was
                misprinted in Eq. (15) of Ref. \protect\onlinecite{Fei98}.
\end{references}
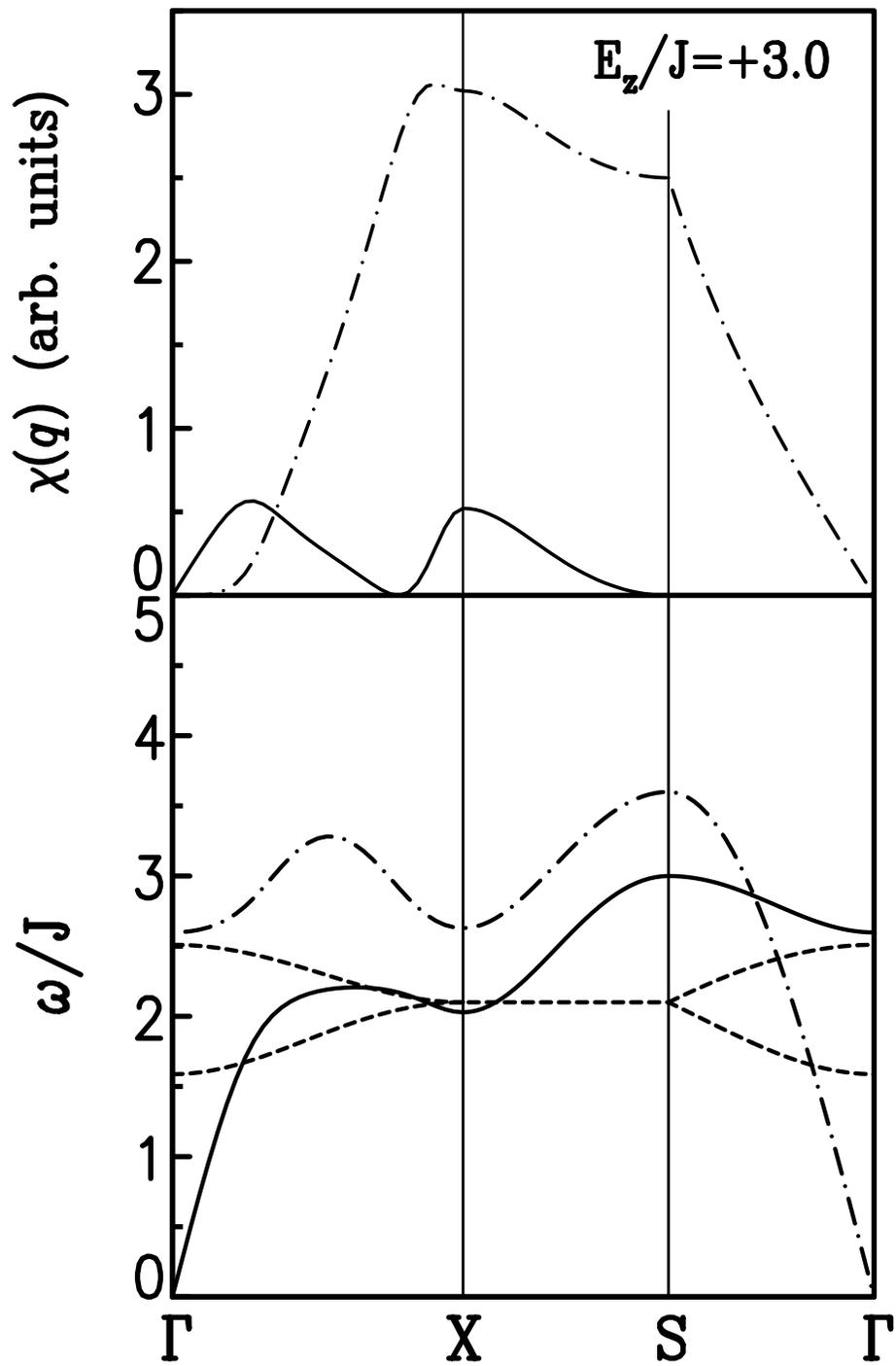
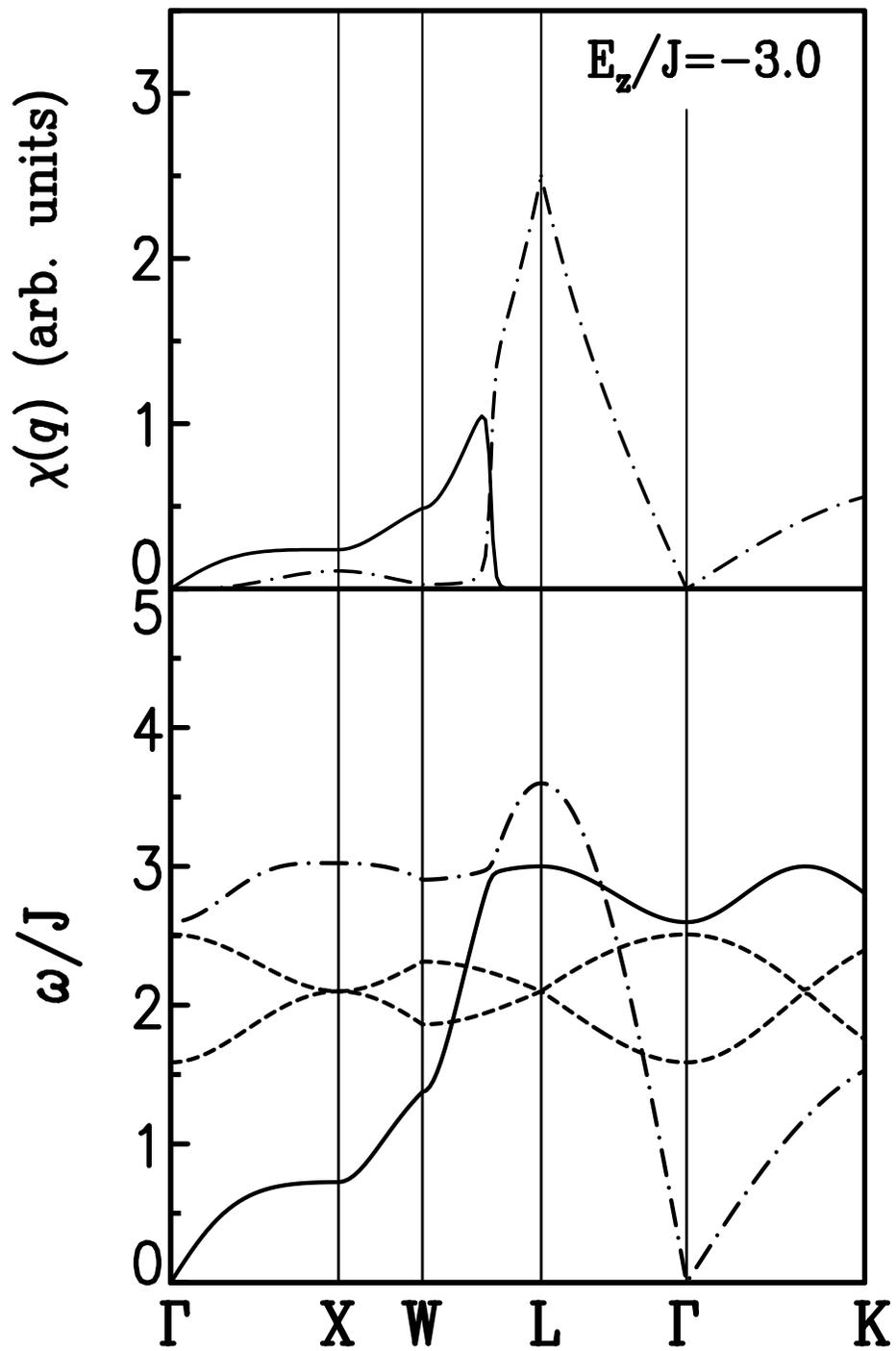
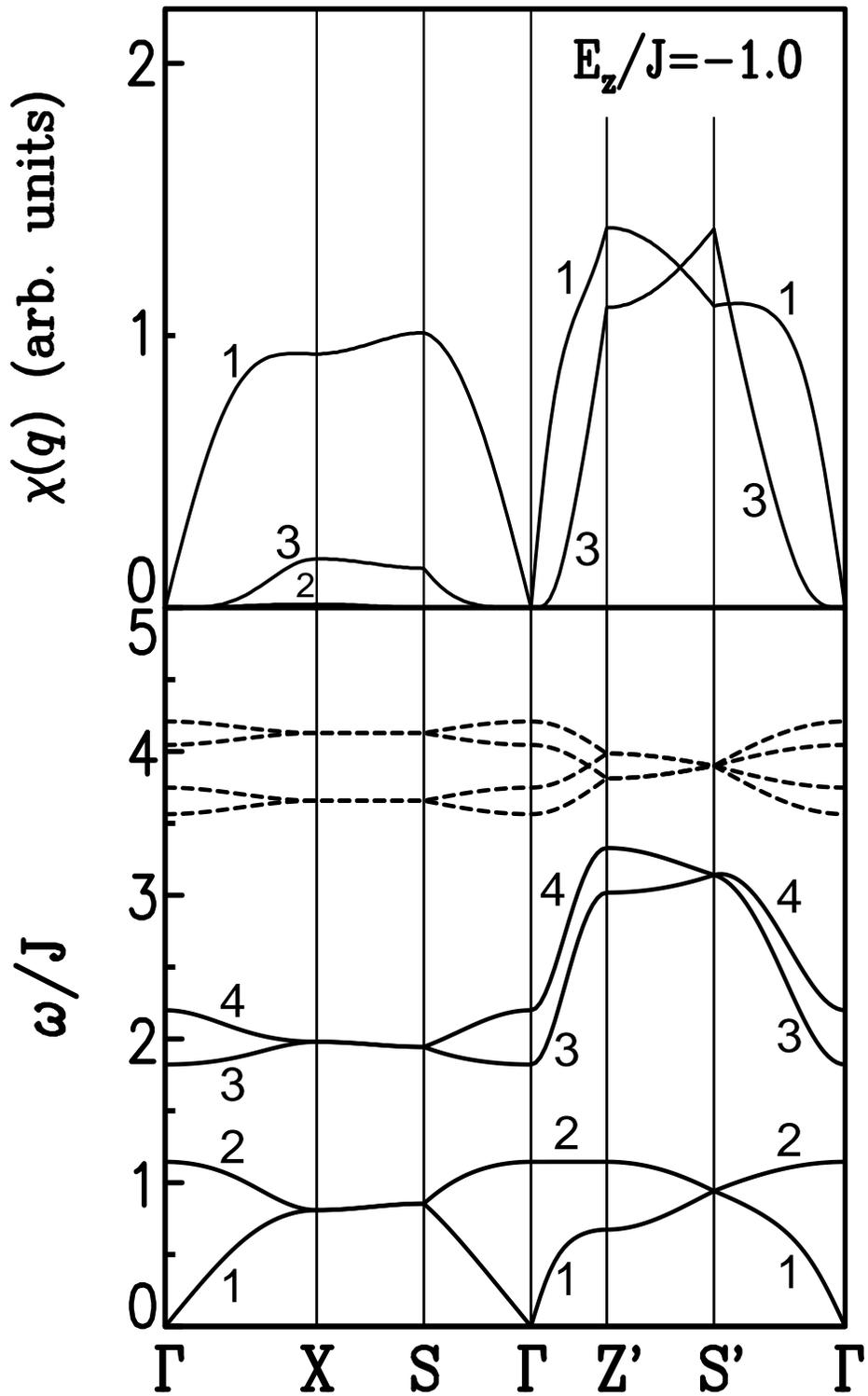
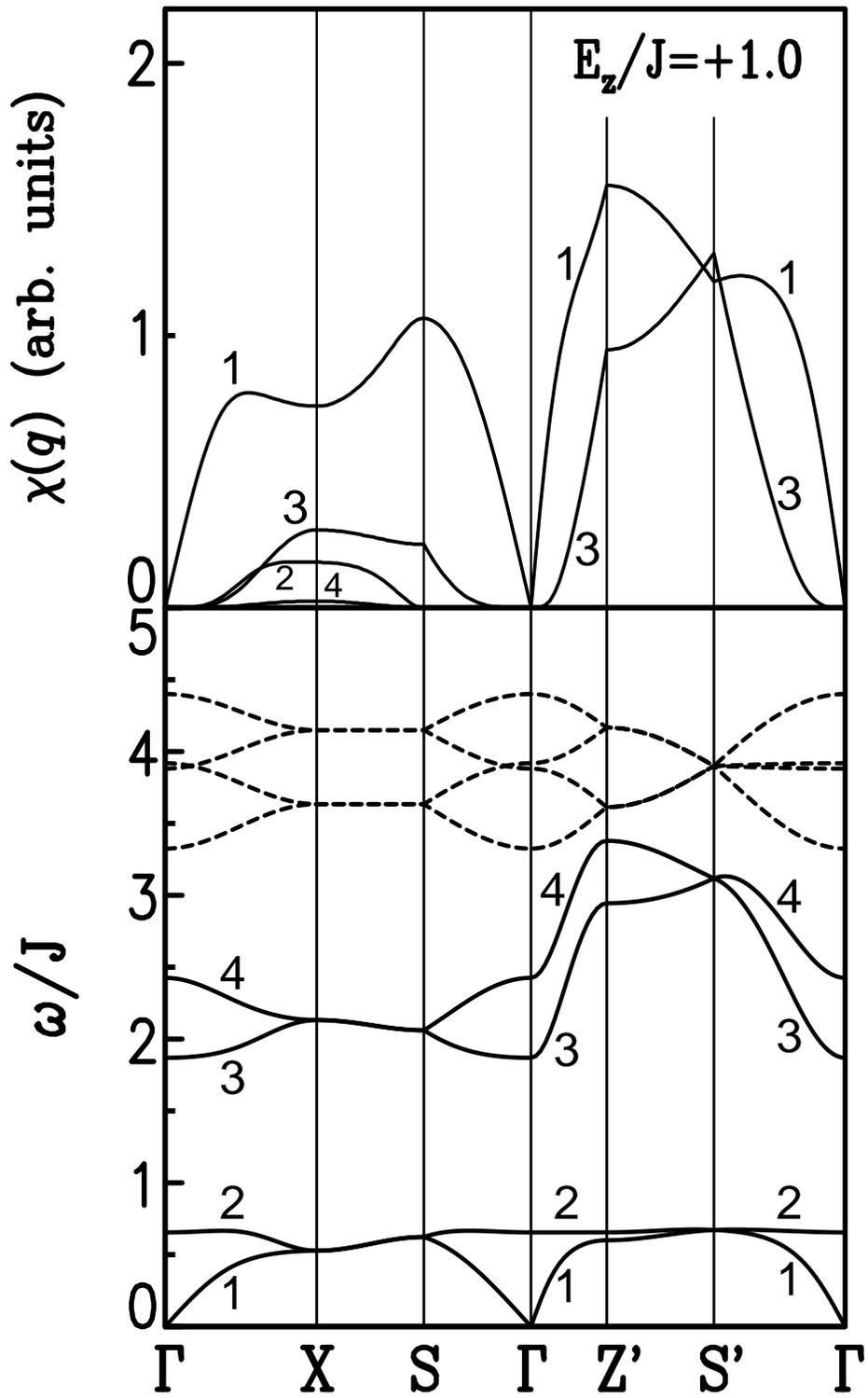
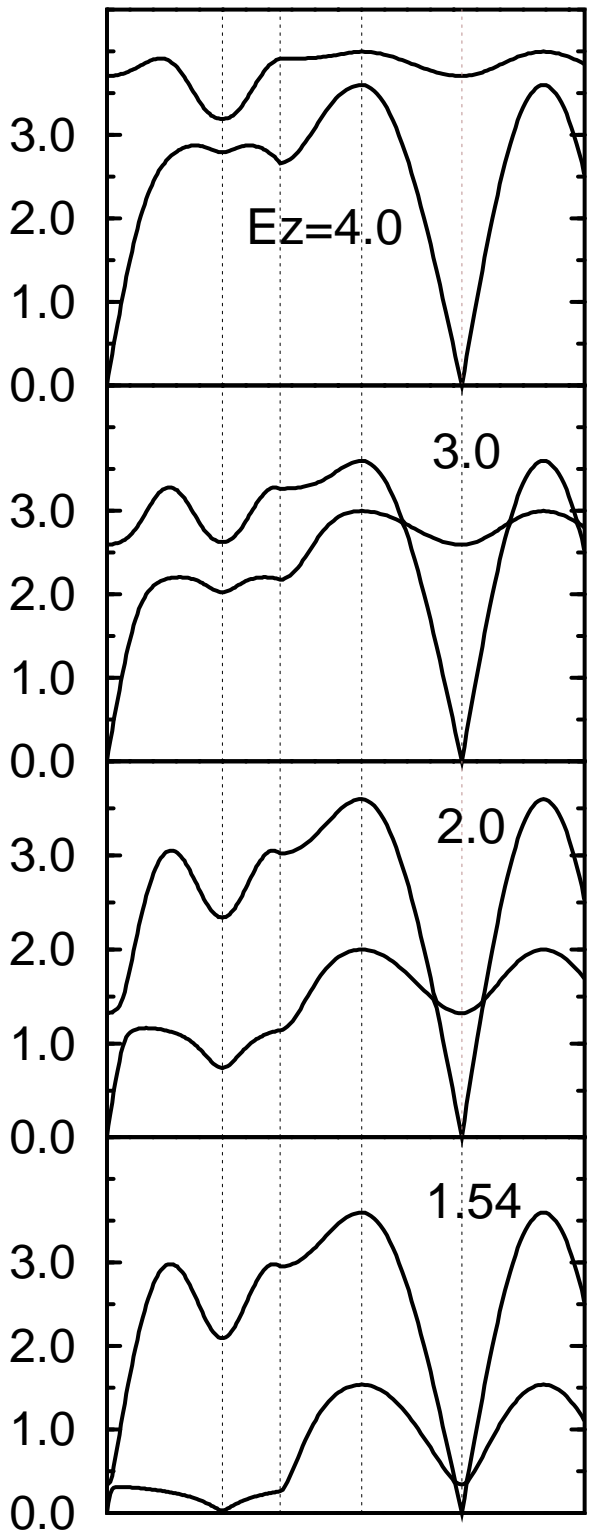
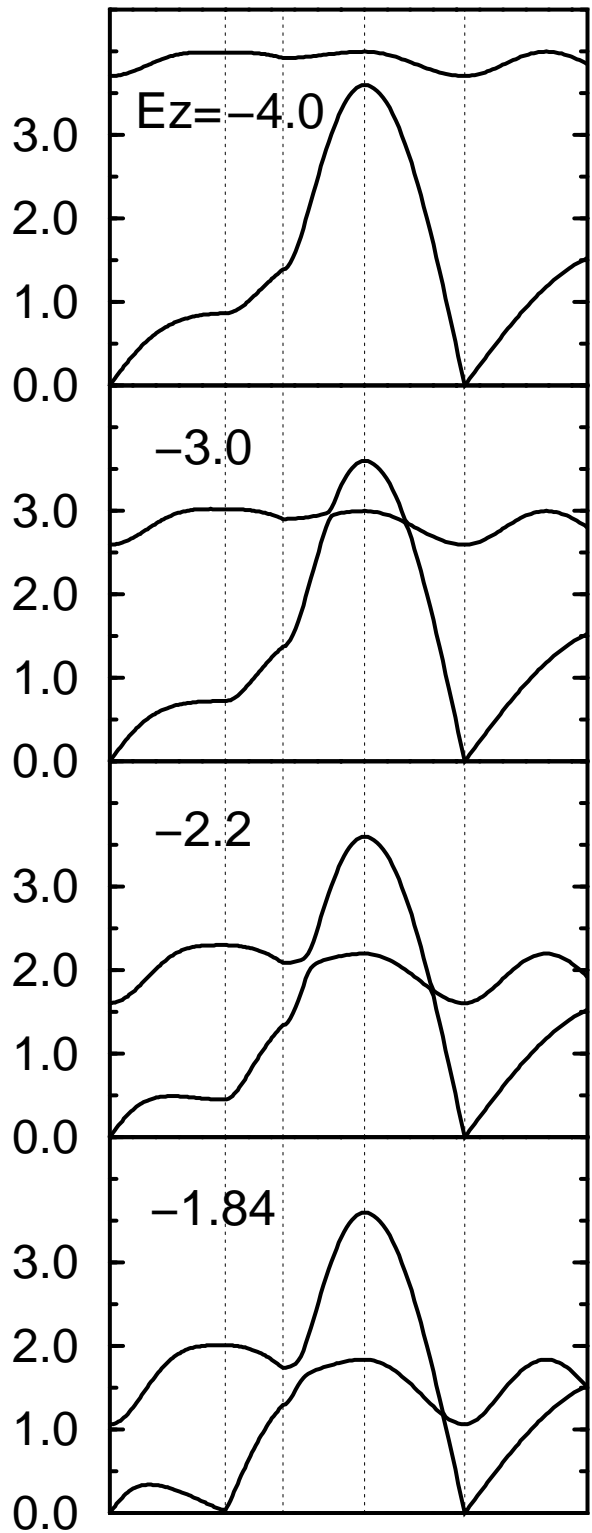
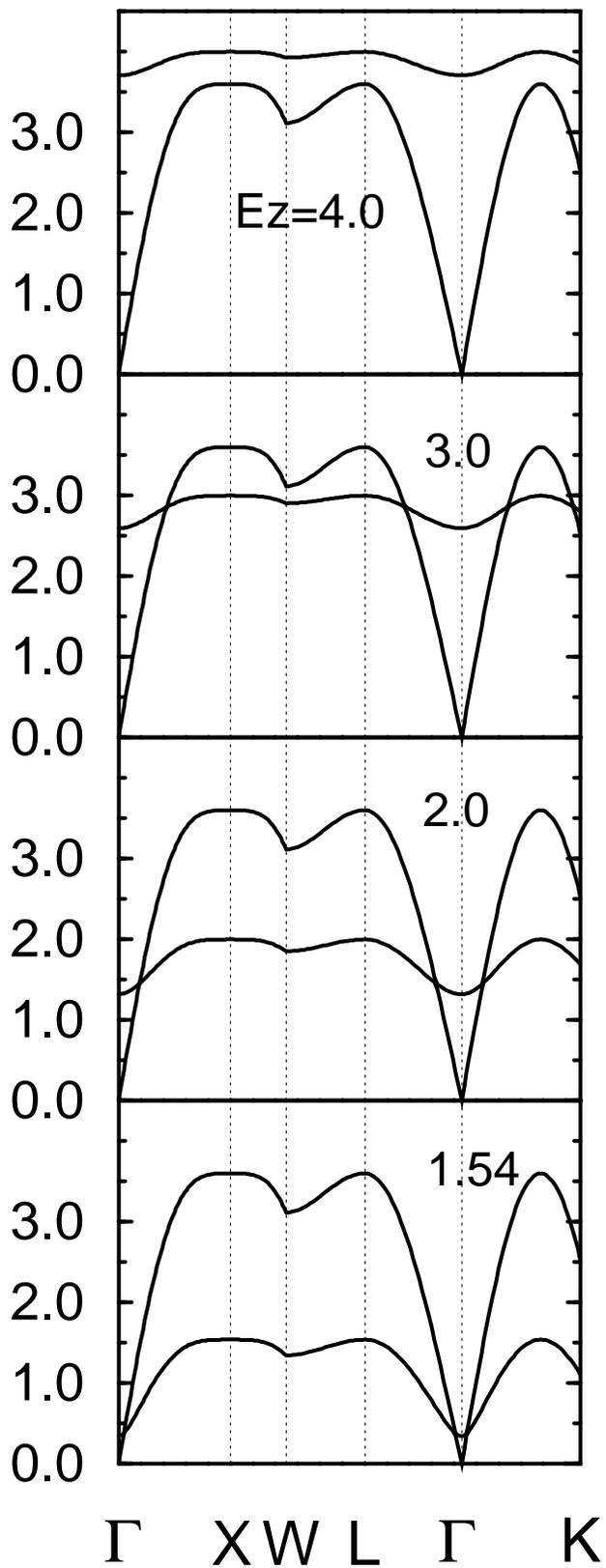
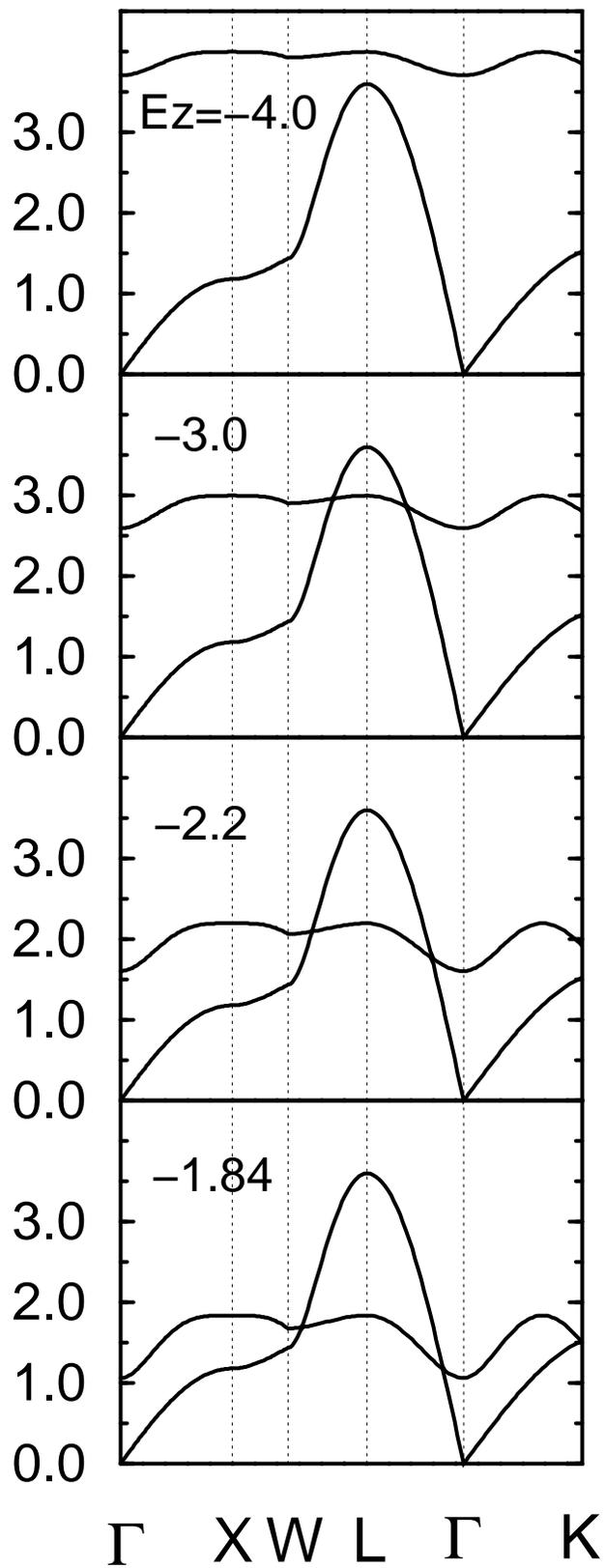
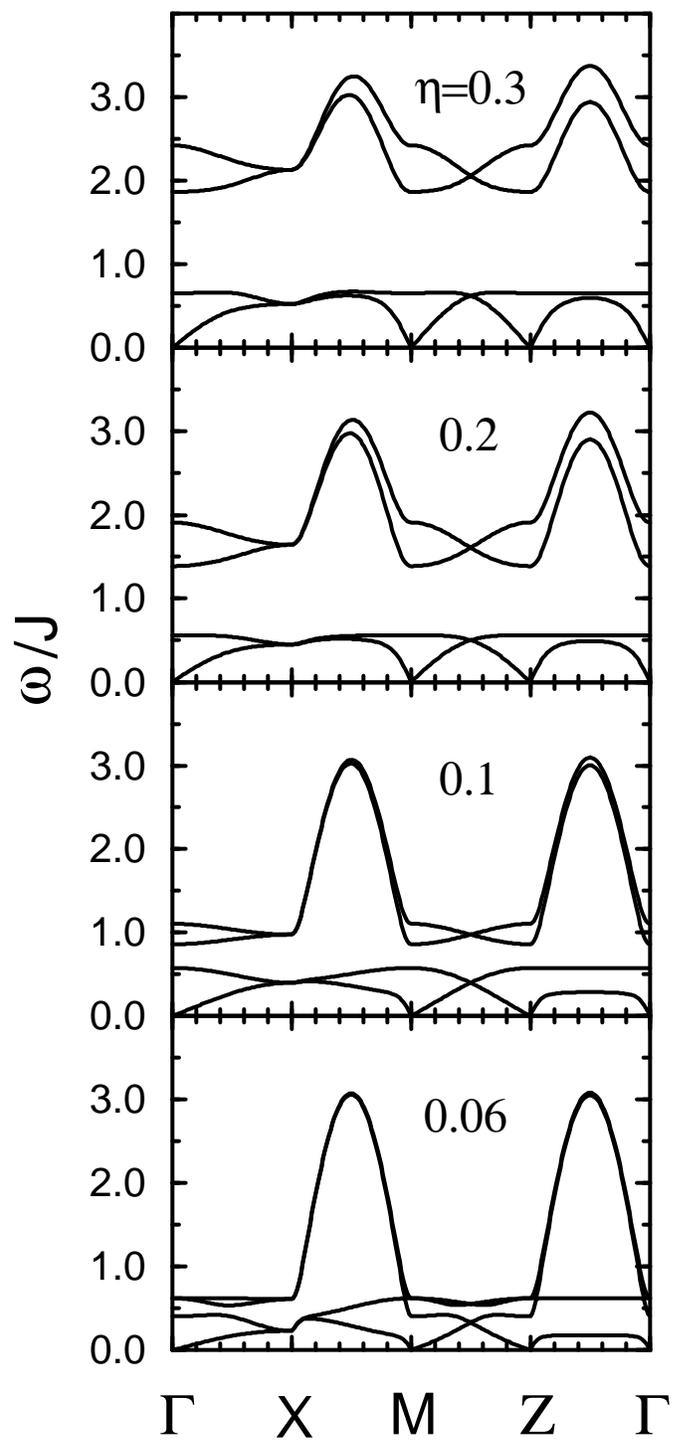
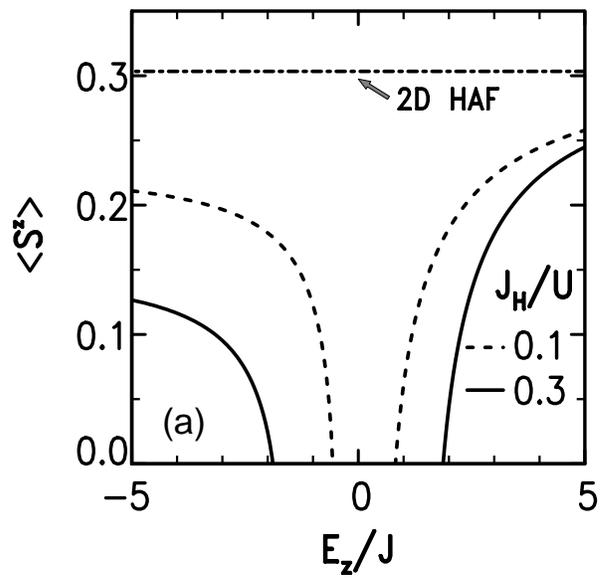
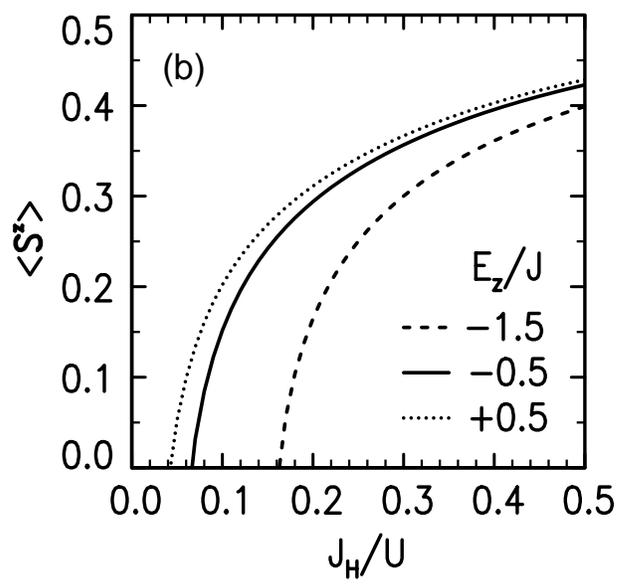
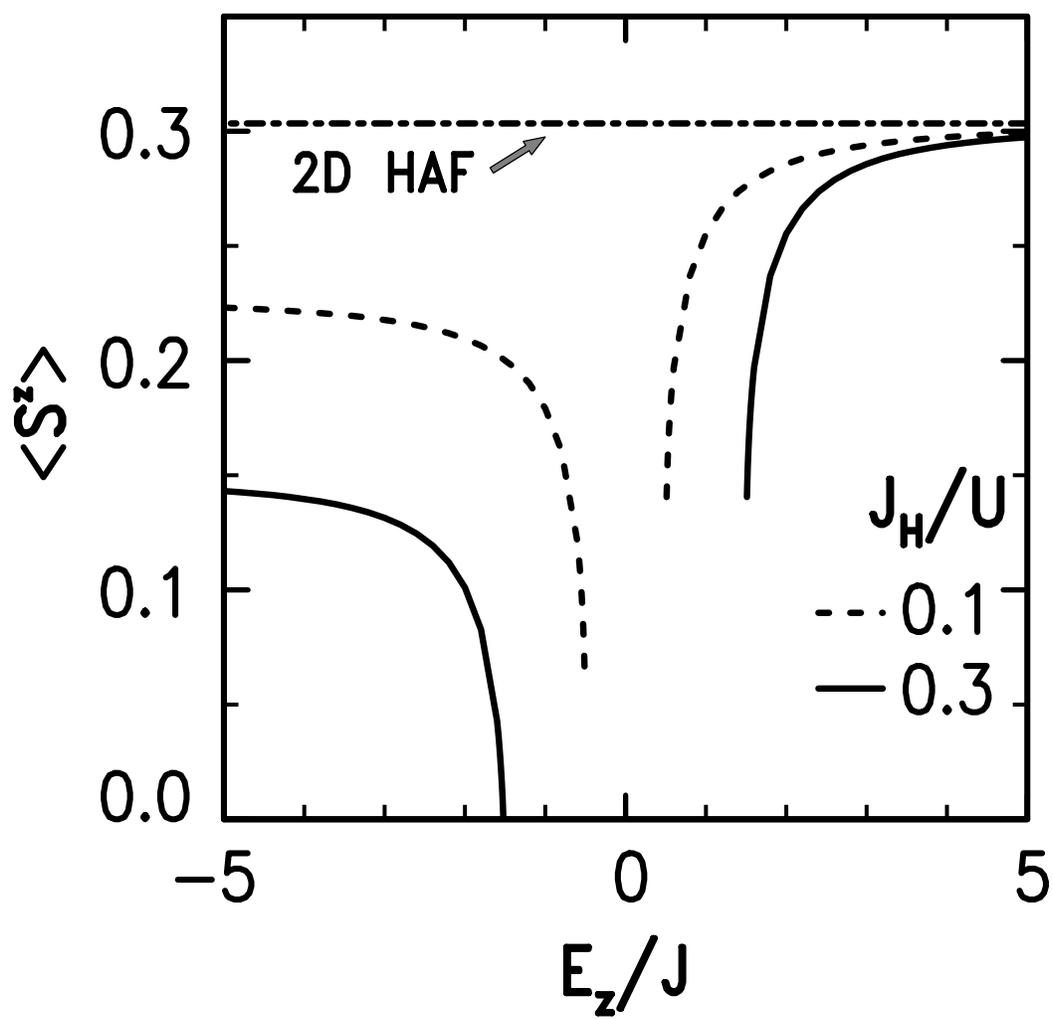
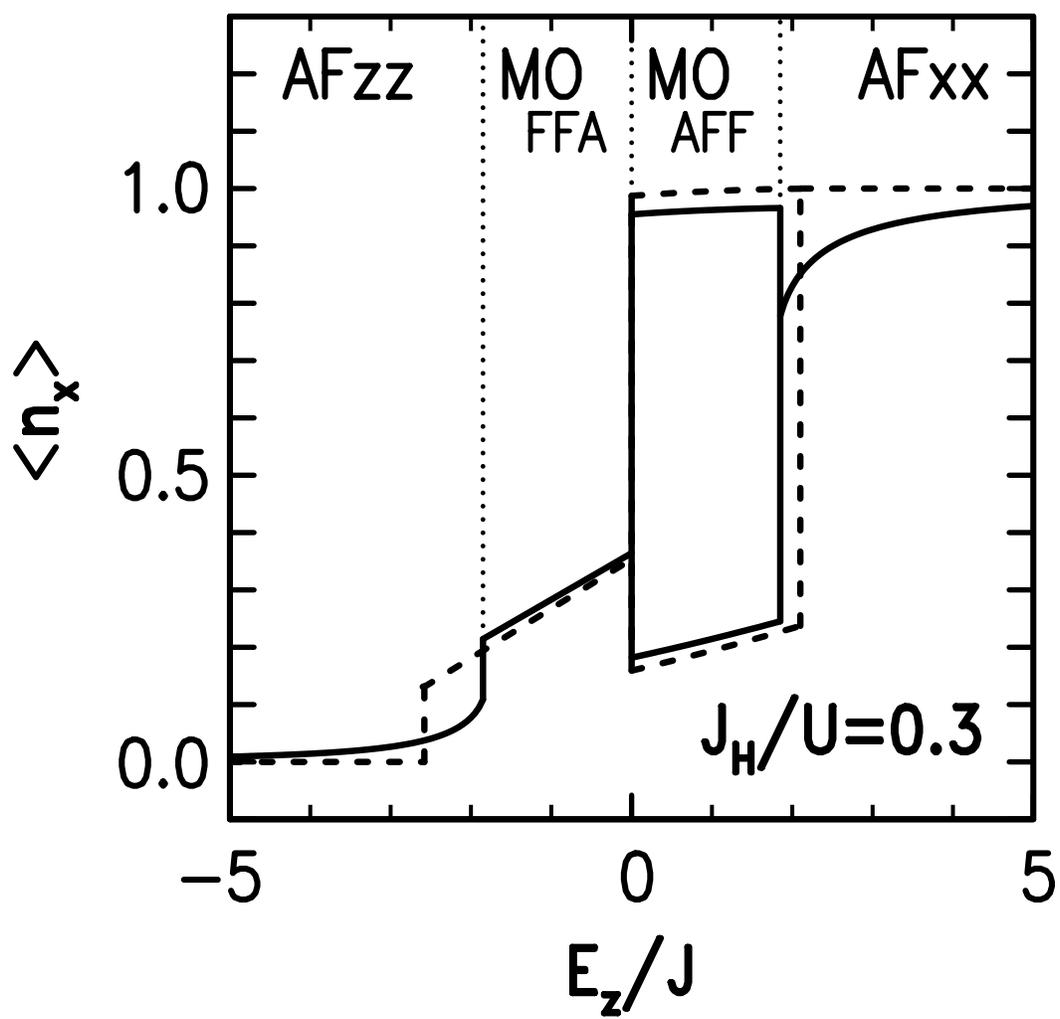
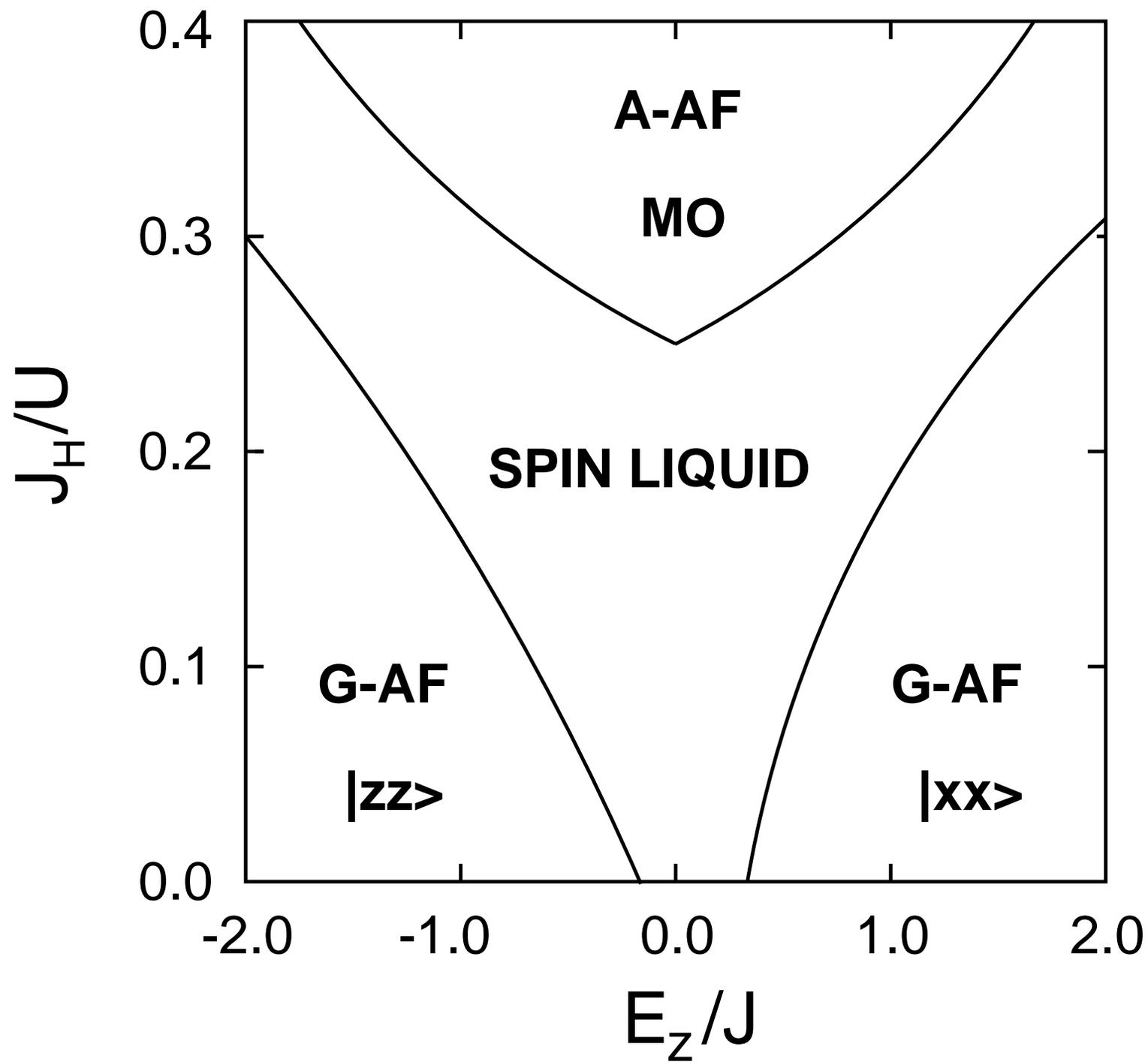
\end{document}